\documentclass[twocolumn,tighten,times]{aastex63}

\usepackage{amsmath}

\newcommand{\avg}[1]{\left\langle{#1}\right\rangle}
\newcommand{\pd}[2]{\frac{\partial{#1}}{\partial{#2}}}

\makeatletter\def\@keys@name{\textit{Unified Astronomy Thesaurus concepts:}\/~\mbox{}}\makeatother
\iftrack
\renewcommand{\added}[1]{{\bf #1}}
\renewcommand{\deleted}[1]{}
\renewcommand{\replaced}[2]{{\bf #2}}
\fi
\shorttitle{COMAP Early Science: V. $z\sim3$}
\shortauthors{Chung et al.}
\graphicspath{{./}}

\NewPageAfterKeywords

\begin{document}

\title{COMAP Early Science: V. Constraints and Forecasts at $z\sim3$}

\correspondingauthor{Dongwoo T.~Chung}
\email{dongwooc@cita.utoronto.ca}

\author[0000-0003-2618-6504]{Dongwoo T.~Chung}
\affiliation{Canadian Institute for Theoretical Astrophysics, University of Toronto, 60 St. George Street, Toronto, ON M5S 3H8, Canada}
\affiliation{Dunlap Institute for Astronomy and Astrophysics, University of Toronto, 50 St. George Street, Toronto, ON M5S 3H4, Canada}

\author[0000-0001-8382-5275]{Patrick C.~Breysse}
\affiliation{Center for Cosmology and Particle Physics, Department of Physics, New York University, 726 Broadway, New York, NY, 10003, USA}
\author[0000-0002-8214-8265]{Kieran A. Cleary}
\affiliation{California Institute of Technology, 1200 E. California Blvd., Pasadena, CA 91125, USA}
\author[0000-0003-3420-7766]{H\aa vard T.~Ihle}
\affiliation{Institute of Theoretical Astrophysics, University of Oslo, P.O. Box 1029 Blindern, N-0315 Oslo, Norway}
\author[0000-0002-8800-5740]{Hamsa Padmanabhan}
\affiliation{D\'epartement de Physique Th\'{e}orique, Universit\'{e} de Gen\`{e}ve, 24 Quai Ernest-Ansermet, CH-1211 Gen\`{e}ve 4, Switzerland}
\author[0000-0003-0209-4816]{Marta B.~Silva}
\affiliation{Institute of Theoretical Astrophysics, University of Oslo, P.O. Box 1029 Blindern, N-0315 Oslo, Norway}

\author[0000-0003-2358-9949]{J.~Richard Bond}
\affiliation{Canadian Institute for Theoretical Astrophysics, University of Toronto, 60 St. George Street, Toronto, ON M5S 3H8, Canada}
\author{Jowita Borowska}
\affiliation{Institute of Theoretical Astrophysics, University of Oslo, P.O. Box 1029 Blindern, N-0315 Oslo, Norway}
\author{Morgan Catha}
\affiliation{Owens Valley Radio Observatory, California Institute of Technology, Big Pine, CA 93513, USA}
\author{Sarah E.~Church}
\affiliation{Kavli Institute for Particle Astrophysics and Cosmology \& Physics Department, Stanford University, Stanford, CA 94305, USA}
\author[0000-0002-5223-8315]{Delaney A.~Dunne}
\affiliation{California Institute of Technology, 1200 E. California Blvd., Pasadena, CA 91125, USA}
\author[0000-0003-2332-5281]{Hans Kristian Eriksen}
\affiliation{Institute of Theoretical Astrophysics, University of Oslo, P.O. Box 1029 Blindern, N-0315 Oslo, Norway}
\author[0000-0001-8896-3159]{Marie Kristine Foss}
\affiliation{Institute of Theoretical Astrophysics, University of Oslo, P.O. Box 1029 Blindern, N-0315 Oslo, Norway}
\author{Todd Gaier}
\affiliation{Jet Propulsion Laboratory, California Institute of Technology, 4800 Oak Grove Drive, Pasadena, CA 91109, USA}
\author{Joshua Ott Gundersen}
\affiliation{Department of Physics, University of Miami, 1320 Campo Sano Avenue, Coral Gables, FL 33146, USA}
\author[0000-0001-7911-5553]{Stuart E.~Harper}
\affiliation{Jodrell Bank Centre for Astrophysics, Alan Turing Building, Department of Physics and Astronomy, School of Natural Sciences, The University of Manchester, Oxford Road, Manchester, M13 9PL, U.K.}
\author[0000-0001-6159-9174]{Andrew I.~Harris}
\affiliation{Department of Astronomy, University of Maryland, College Park, MD 20742, USA}
\author[0000-0001-7449-4638]{Brandon Hensley}
\affiliation{Department of Astrophysical Sciences, Princeton University, Princeton, NJ 08544, USA}
\author{Richard Hobbs}
\affiliation{Owens Valley Radio Observatory, California Institute of Technology, Big Pine, CA 93513, USA}
\author[0000-0001-5211-1958]{Laura C.~Keating}
\affiliation{Leibniz-Institut f{\"u}r Astrophysik Potsdam (AIP), An der Sternwarte 16, D-14482 Potsdam, Germany}
\author[0000-0002-4274-9373]{Junhan Kim}
\affiliation{California Institute of Technology, 1200 E. California Blvd., Pasadena, CA 91125, USA}
\author{James W.~Lamb}
\affiliation{Owens Valley Radio Observatory, California Institute of Technology, Big Pine, CA 93513, USA}
\author{Charles R.~Lawrence}
\affiliation{Jet Propulsion Laboratory, California Institute of Technology, 4800 Oak Grove Drive, Pasadena, CA 91109, USA}
\author{Jonas Gahr Sturtzel Lunde}
\affiliation{Institute of Theoretical Astrophysics, University of Oslo, P.O. Box 1029 Blindern, N-0315 Oslo, Norway}
\author{Norman Murray}
\affiliation{Canadian Institute for Theoretical Astrophysics, University of Toronto, 60 St. George Street, Toronto, ON M5S 3H8, Canada}
\author[0000-0001-5213-6231]{Timothy J.~Pearson}
\affiliation{California Institute of Technology, 1200 E. California Blvd., Pasadena, CA 91125, USA}
\author[0000-0001-7612-2379]{Liju Philip}
\affiliation{Jet Propulsion Laboratory, California Institute of Technology, 4800 Oak Grove Drive, Pasadena, CA 91109, USA}
\author{Maren Rasmussen}
\affiliation{Institute of Theoretical Astrophysics, University of Oslo, P.O. Box 1029 Blindern, N-0315 Oslo, Norway}
\author{Anthony C.~S.~Readhead}
\affiliation{California Institute of Technology, 1200 E. California Blvd., Pasadena, CA 91125, USA}
\author[0000-0002-1667-3897]{Thomas J. Rennie}
\affiliation{Jodrell Bank Centre for Astrophysics, Alan Turing Building, Department of Physics and Astronomy, School of Natural Sciences, The University of Manchester, Oxford Road, Manchester, M13 9PL, U.K.}
\author[0000-0001-5301-1377]{Nils-Ole Stutzer}
\affiliation{Institute of Theoretical Astrophysics, University of Oslo, P.O. Box 1029 Blindern, N-0315 Oslo, Norway}
\author[0000-0001-8526-3464]{Bade D.~Uzgil}
\affiliation{California Institute of Technology, 1200 E. California Blvd., Pasadena, CA 91125, USA}
\author[0000-0003-0545-4872]{Marco P.~Viero}
\affiliation{California Institute of Technology, 1200 E. California Blvd., Pasadena, CA 91125, USA}
\author[0000-0002-5437-6121]{Duncan J.~Watts}
\affiliation{Institute of Theoretical Astrophysics, University of Oslo, P.O. Box 1029 Blindern, N-0315 Oslo, Norway}
\author[0000-0003-2229-011X]{Risa H.~Wechsler}
\affiliation{Kavli Institute for Particle Astrophysics and Cosmology \& Physics Department, Stanford University, Stanford, CA 94305, USA}
\affiliation{SLAC National Accelerator Laboratory, Menlo Park, CA 94025, USA}
\author[0000-0003-3821-7275]{Ingunn Kathrine Wehus}
\affiliation{Institute of Theoretical Astrophysics, University of Oslo, P.O. Box 1029 Blindern, N-0315 Oslo, Norway}
\author{David P.~Woody}
\affiliation{Owens Valley Radio Observatory, California Institute of Technology, Big Pine, CA 93513, USA}

\collaboration{37}{(COMAP Collaboration)\vspace{-0.42cm}}



\begin{abstract}
We present the current state of models for the $z\sim3$ carbon monoxide (CO) line-intensity signal targeted by the CO Mapping Array Project (COMAP) Pathfinder in the context of its early science results. Our fiducial model, relating dark matter halo properties to CO luminosities, informs parameter priors with empirical models of the galaxy--halo connection and previous CO(1--0) observations. The Pathfinder early science data spanning wavenumbers $k=0.051$--$0.62\,$Mpc$^{-1}$ represent the first direct 3D constraint on the clustering component of the CO(1--0) power spectrum. Our 95\% upper limit on the redshift-space clustering amplitude $A_\text{clust}\lesssim70\,\mu$K$^2$ greatly improves on the indirect upper limit of $420\,\mu$K$^2$ reported from the CO Power Spectrum Survey (COPSS) measurement at $k\sim1\,$Mpc$^{-1}$. The COMAP limit excludes a subset of models from previous literature, and constrains interpretation of the COPSS results, demonstrating the complementary nature of COMAP and interferometric CO surveys. Using line bias expectations from our priors, we also constrain the squared mean line intensity--bias product, $\avg{Tb}^2\lesssim50\,\mu$K$^2$, and the cosmic molecular gas density, $\rho_\text{H2}<2.5\times10^8\,M_\odot\,$Mpc$^{-3}$ (95\% upper limits). Based on early instrument performance and our current CO signal estimates, we forecast that the five-year Pathfinder campaign will detect the CO power spectrum with overall signal-to-noise of 9--17. Between then and now, we also expect to detect the CO--galaxy cross-spectrum using overlapping galaxy survey data, enabling enhanced inferences of cosmic star-formation and galaxy-evolution history.
\end{abstract}

\keywords{\href{http://astrothesaurus.org/uat/262}{CO line emission (262)}; \href{http://astrothesaurus.org/uat/336}{Cosmological evolution (336)}; \href{http://astrothesaurus.org/uat/734}{High-redshift galaxies (734)}; \href{http://astrothesaurus.org/uat/1073}{Molecular gas (1073)}; \href{http://astrothesaurus.org/uat/1338}{Radio astronomy (1338)}}


\section{Introduction} \label{sec:intro}

Line-intensity mapping (LIM) surveys propose to map 3D fluctuations in integrated redshifted spectral line emission across large cosmological volumes (cf.~reviews by~\citealt{Kovetz17} and~\citealt{Kovetz2019}). These survey designs generally focus on statistical measurements of the line emitters as a whole, including faint populations of galaxies that cannot be detected in isolation but may be inferred in aggregate. Investigating what measurements and inferences this perspective enables, previous literature has studied the potential of high-redshift LIM with carbon monoxide (CO) lines for over a decade (see, e.g.:~\citealt{2008A&A...489..489R,visbal_loeb_10,2011JCAP...08..010V,Lidz11,Pullen13,li_etal_16,Padmanabhan18,MoradinezhadKeating19,Sun19,Yang21,Yang21b}).

The history of direct power spectrum measurements of CO intensity is somewhat shorter, as surveys like the CO Power Spectrum Survey (COPSS;~\citealt{keating_etal_16}) and the mm-wave Intensity Mapping Experiment (mmIME;~\citealt{mmIME-ACA}) have only begun to publish results relatively recently. Both of these surveys leverage community instruments to make interferometric measurements of the CO line-intensity field over fields of $\sim10$--$100$ square arcminutes ($\sim10^{-3}$--$10^{-2}$\,deg$^2$), with both claiming measurements of CO power slightly beyond a $2\sigma$ level of significance. In addition to the CO auto-spectrum measurements from COPSS and mmIME,~\cite{Keenan21} demonstrated the feasibility of cross-correlation between COPSS data and galaxy surveys, placing an upper limit on the CO--galaxy cross-spectrum. However, neither COPSS nor mmIME probe sufficiently large scales to constrain CO fluctuations shaped by clustering, instead measuring the shot noise chiefly expected to arise from the stochastic bright end of the luminosity function (LF).

Results from the first observing season (Y1) of the CO Mapping Array Project (COMAP) Pathfinder~\citep{es_I}, the first instrument specifically designed for single-dish CO line-intensity mapping, provide the first direct constraints on the clustering component of the high-redshift CO line-intensity power spectrum. COMAP Pathfinder observations at 26--34\,GHz measure CO(1--0) (rest frequency 115.27~GHz) at $z=2.4$--3.4 in three fields of 4 deg$^2$, allowing characterization of larger transverse scales than with previous interferometric LIM surveys. Other papers associated with these results\footnote{Beyond CO LIM, see also early results from continuum observations comprising the COMAP Galactic Plane Survey~\citep{es_VI}.} describe the instrument~\citep{es_II}, data processing and mapmaking procedures~\citep{es_III}, and power spectrum methodology and results~\citep{es_IV}. This paper aims to convert these measurements into astrophysical inferences and consider forecasts for the remainder of the initial five-year Pathfinder campaign, with a separate paper by~\cite{es_VII} considering potential realizations of COMAP beyond the Pathfinder.

\added{LIM surveys like COMAP tend to assume approximate isotropy in the spatial structure of line emission across the survey volume, and so report the spherically averaged power spectrum across comoving wavenumber $k$. The clustering component of this power spectrum, proportional to the first moment of the LF, traces the large-scale structure underlying the CO emitters and dominates at low $k$. The (independent) shot-noise component, proportional to the second moment of the LF, describes scale-independent fluctuations arising from the Poisson statistics of line emitters and dominates at high $k$. Due to how each component relates to the CO LF, shot-noise measurements probe CO stochasticity and bright emitters, but clustering measurements are less skewed towards the bright end of the LF, meaning they will be more sensitive to the properties of faint emitters. We expect COMAP Pathfinder results to be significant in this context because our high-sensitivity instrumentation is purpose-built for CO LIM and enables access to large angular scales unavailable to COPSS or mmIME, thus allowing measurements of the power spectrum at clustering-dominated ranges of $k$ previously inaccessible to CO LIM surveys.

} While current COMAP Pathfinder measurements are consistent with white noise and thus provide an upper limit for the spherically averaged CO power spectrum $P(k)$ at $k\sim10^{-1}$\,Mpc$^{-1}$, several years remain in the observing campaign, 
during which we anticipate a detection based on previous models in the literature. Furthermore, members of the COMAP collaboration have worked on updating our own fiducial CO models and expectations for LF and molecular gas density constraints at the conclusion of the COMAP Pathfinder survey. 
In this context this paper aims to answer questions about the COMAP Pathfinder campaign following Y1 and early science verification:
\begin{itemize}
    \item What inferences do our early science verification data enable about the $z\sim3$ CO(1--0) power spectrum, and molecular gas abundance?
    \item Given early science sensitivities and updated $z\sim3$ models, what are our present expectations for constraints on these same quantities, and others like the CO LF, at the end of five years of COMAP Pathfinder observations?
\end{itemize}

We organize the paper as follows. In~\autoref{sec:model} we outline our fiducial model for CO emission at $z\sim3$, chiefly in comparison to the model of~\cite{li_etal_16} and to observational results. Then, in~\autoref{sec:ULinterp} we consider implications of the current COMAP Pathfinder $P(k)$ limit in relation to other models and observational results in the literature. Finally, in~\autoref{sec:forecasts} we outline simulated constraints of our new CO model based on expected five-year results from the COMAP Pathfinder, before outlining overall conclusions in~\autoref{sec:conclusions}.

Unless otherwise stated, we assume base-10 logarithms, and a $\Lambda$CDM cosmology with parameters $\Omega_m = 0.286$, $\Omega_\Lambda = 0.714$, $\Omega_b =0.047$, $H_0=100h$\,km\,s$^{-1}$\,Mpc$^{-1}$ with $h=0.7$, $\sigma_8 =0.82$, and $n_s =0.96$, to maintain consistency with previous COMAP simulations~\citep{li_etal_16,Ihle19}. The cosmology is also broadly consistent with nine-year WMAP results~\citep{WMAP9}. Distances carry an implicit $h^{-1}$ dependence throughout, which propagates through masses (all based on virial halo masses, proportional to $h^{-1}$) and volume densities ($\propto h^3$).
\section{Devising a Model for CO at Redshift 3}
\label{sec:model}
\added{As we discussed in~\hyperref[sec:intro]{the Introduction}, substantial astrophysical literature over the past decade has formulated models for the line emission fluctuations probed by CO LIM, some of which we will consider later in this paper in relation to COMAP Pathfinder observations. While a detailed examination of these prior models is not within the scope of the present work, we distinguish two general approaches to dark matter halo-based models of CO fluctuations, as follows.

\begin{enumerate}
    \item \emph{Indirect models of the halo--CO connection via intermediate properties:} These models connect the mass distribution of dark matter halos to a different property like star-formation rate (SFR) or infrared luminosity (usually as a proxy for SFR), often but not necessarily through simulations or abundance matching. This intermediate property then relates to CO luminosity via simulations or fits to data. This group includes the models of~\cite{Pullen13}, which make use of an empirical relation from~\cite{Wang10} between CO luminosity and bolometric far-infrared luminosity, 
    and the model of~\cite{li_etal_16}, which instead uses a CO--SFR relation derived in part from empirical correlations described by the review of~\cite{CW13}. (This group therefore also includes the modification of this last model by~\cite{mmIME-ACA}, which replaces the~\cite{CW13} relation with results from~\cite{Kamenetzky16}.)
    \item \emph{Direct models of the halo--CO connection, motivated by observed CO emitter abundances:} The recent emergence of direct constraints on the high-redshift CO LF now enables direct abundance matching of such measurements to the halo mass distribution to obtain a halo model for CO emission. \cite{Padmanabhan18} first undertook formulation of a model of this kind applicable at $z\sim3$ (using data compiled across a wide redshift range), and the present work now aims to follow this approach also (albeit only for CO(1--0) at $z\sim3$).
\end{enumerate}

}Previous forecasting efforts for COMAP have used the fiducial model of~\citet{li_etal_16}. Since then, we have gained new insight into CO(1--0) emitters at high redshift through two important surveys: the CO Luminosity Density at High Redshift survey (COLDz;~\citealt{COLDzLF}), which provide the strongest constraints on the CO(1--0) LF at $z=2$--3 to date; and COPSS~\citep{keating_etal_16}, which made a tentative detection of shot noise power from small-scale CO fluctuations. Other surveys such as the aforementioned mmIME, the ALMA SPECtroscopic Survey (ASPECS) in the Hubble Ultra-Deep Field, and the Plateau de Bure High-$z$ Blue-Sequence Survey 2 (PHIBBS2) lend insight into emission in higher-$J$ CO lines at these redshifts~\citep{mmIME-ACA,ASPECS-LPLF2,PHIBBS2Lenkic}.

Here we present a new fiducial model that takes into account the COPSS and COLDz measurements---as well as priors from empirical models of the halo mass--\replaced{star-formation rate (SFR)}{SFR} relation and the SFR--CO luminosity relation already used in~\citet{li_etal_16}---and uses a double power-law parameterization modified from~\citet{Padmanabhan18} (removing redshift dependence\footnote{Whereas~\cite{Padmanabhan18} sought to model CO over a broad redshift range of $z\sim0$--$3$, we concentrate on a narrower range where redshift evolution is expected to be much less significant. Therefore, a redshift-dependent parameterization would complicate the model for little benefit.}). The new parameterization models the halo mass--CO luminosity relation with greater flexibility and {directness} compared to~\citet{li_etal_16}.

We first provide an overview of the parameterization in~\autoref{sec:paramet}, then present priors on the model parameters in~\autoref{sec:priors}. An additional aspect of our model is a basic treatment of line broadening, as described in~\autoref{sec:doppler}, which is highly approximate but acceptable when considering the sensitivity expected especially from our early data. Only after laying this groundwork can we discuss our procedure for inferring parameter constraints from COMAP Pathfinder measurements in~\autoref{sec:forecasts}, through Markov Chain Monte Carlo (MCMC) runs using our fiducial parameterization and priors to inform forward models of one- and two-point statistics.

\subsection{Fiducial Parameterization of the Halo--CO Connection}
\label{sec:paramet}

The double power-law parameterization of the halo mass--CO luminosity relation $L(M_h)$ approximates the composition of a series of scaling relations connecting halo mass $M_h$ to CO luminosity $L$ or $L_\text{CO}$ similar to the series considered by~\cite{li_etal_16}.
\begin{itemize}
    \item As in~\citet{li_etal_16}, we consider a single power law relating IR luminosity and CO luminosity:
    \begin{equation}\log{L_\text{IR}}=\alpha\log{L'_\text{CO}}+\beta,\label{eq:CO_IR}\end{equation}
    where for CO(1--0),
    \begin{equation}\frac{L_\text{CO}}{L_\odot}=4.9\times10^{-5}\frac{L'_\text{CO}}{\text{K km s}^{-1}\text{ pc}^2}.\label{eq:Lprime_to_L}\end{equation}\added{This form of scaling relation is commonly fitted to observational data in the literature (see, e.g., the reviews of~\citealt{SVB05} and~\citealt{CW13}). \cite{li_etal_16} found values of $\alpha$ anywhere between 1.00 and 1.37 from the high-redshift CO studies available at the time of that work.}
    \item We also relate IR luminosity to star-formation rate as in~\citet{li_etal_16}:
    \begin{equation}\frac{{\rm SFR}}{M_\odot\,\text{yr}^{-1}} = \delta_\text{MF}\times10^{-10}\left(\frac{L_\text{IR}}{L_\odot}\right).\end{equation}
    for some coefficient $\delta_\text{MF}$ whose value depends on the initial mass function (IMF)\replaced{; t}{. T}his is set to 1 in~\citet{li_etal_16}\added{ but we refer the reader to reviews by~\cite{Calzetti13},~\cite{Casey14}, and~\cite{madau_dickinson_14} for more information about the IMF-dependence of this SFR calibration in the style of~\cite{Kennicutt98b}}.
    \item The UniverseMachine (UM) framework of \citet{Behroozi19} models the average star-formation rate for a star-forming galaxy hosted in a halo with maximum circular velocity at peak halo mass $v_{M_\text{peak}}$ as
    \begin{align}\frac{\left\langle\operatorname{SFR}_\text{SF}\right\rangle(v_{M_\text{peak}})}{M_\odot\,\text{yr}^{-1}}&=\epsilon\left[\frac{1}{v^{\alpha_\text{UM}}+v^{\beta_\text{UM}}}\right.\nonumber\\&\qquad\quad\left.+\gamma\exp{\left(-\frac{\log^2{v}}{2\delta^2}\right)}\right],\end{align}
    where
    \begin{equation}v=\frac{v_{M_\text{peak}}}{V\text{ [km\,s}^{-1}]}.\end{equation}
    (Note that we have added ``UM'' subscripts to $\alpha$ and $\beta$ from~\citealt{Behroozi19} denoting that these are UM parameters, to avoid confusion with $\alpha$ and $\beta$ from~\citealt{li_etal_16}.) This is a double power law with a Gaussian component added to it. However, here we assume the effect of the Gaussian component is negligible (i.e., $\gamma\approx0$) and consider only the double power-law component.
    The above equations are for the star-forming galaxy population rather than the quenched population, but according to the model of~\cite{Behroozi19} the latter is a small enough portion of galaxies at the redshifts we probe that we do not consider it for this exercise.
    \item \citet{Behroozi19} also provide a relation (although approximate) between peak halo mass (which for these redshifts is essentially the same as halo mass) and $v_{M_\text{peak}}$:
    \begin{equation}v_{M_\text{peak}}(M_h) = (200\text{ km\,s}^{-1})\left[\frac{M_h}{M_\text{200 km/s}(a)}\right]^{0.3},\end{equation}
    where $a=1/(1+z)$ is the scale factor at redshift $z$, and
    \begin{equation}M_\text{200 km/s} = \frac{1.64\times10^{12}\,M_\odot}{(a/0.378)^{-0.142}+(a/0.378)^{-1.79}}.\label{eq:M200}\end{equation}
\end{itemize}
Across all of these relations, we can in principle list the independent parameters $\{\alpha,\beta,\delta_\text{MF},\epsilon,\alpha_\text{UM},\beta_\text{UM},V\}$. However, many of these are degenerate in the context of CO LIM data, and for our analyses it makes more sense to deal with combinations of these parameters, in a simplified re-parameterization.

If we make the assumption that $\alpha$ is close to unity\added{, or at least $\alpha\not\gg1$ and $\alpha\not\ll1$}---which seems a justifiable one, given that the prior on this parameter in~\citet{li_etal_16} was $\alpha=1.17\pm0.37$---then we can collapse all of the above scaling relations into a single $L'_\text{CO}(M_h)$ relation (which then exactly corresponds to the intrinsic $L_\text{CO}$ via~\autoref{eq:Lprime_to_L}) with four free parameters:
\begin{equation}\frac{L'_\text{CO}(M_h)}{\text{K km s}^{-1}\text{ pc}^2} = \frac{C}{(M_h/M)^A+(M_h/M)^B}\equiv\frac{C}{m^A+m^B}.\label{eq:Lprime_of_M}\end{equation}
Additionally, we assume that there is some log-normal scatter $\sigma$ (in units of dex) about this relation, which is taken to be the (linear) mean at fixed halo mass.\added{

Note also that $\alpha$ does not need to be exactly one, i.e., the CO--IR or CO--SFR relation does not need to be exactly linear, for~\autoref{eq:Lprime_of_M} to be a reasonable approximation to (even if not an exact description of) the true composition of Equations~\ref{eq:CO_IR} through~\ref{eq:M200}. With $\alpha=1$ the true composition will clearly deviate somewhat from our simplification, but principally around $M_h=M$ and not so much for $M_h\ll M$ or $M_h\gg M$. In any case, our simplification is flexible enough to reflect highly nonlinear trends of CO emission against other properties. This is important as simulations and observational analyses continue to explore the complex environmental factors and physical processes that drive CO emissivity (or lack thereof) relative to H$_2$ content in molecular clouds and galaxies, and thus determine the physical interpretation of CO LIM observations~\citep{Li18,Gong20,Inoguchi20,Keating20,Madden20,Seifried20,Breysse21}.}

\subsection{Priors from Previous Models and Observations}
\label{sec:priors}
We want to formulate a set of priors for our model parameters for two reasons. The first is that they serve as a range of fiducial expectations for forecasting the CO signal at this relatively speculative time. The other is that they will serve as the ground level for Bayesian inferences from COMAP data.

\begin{figure*}
    \centering
    \includegraphics[width=0.27\linewidth]{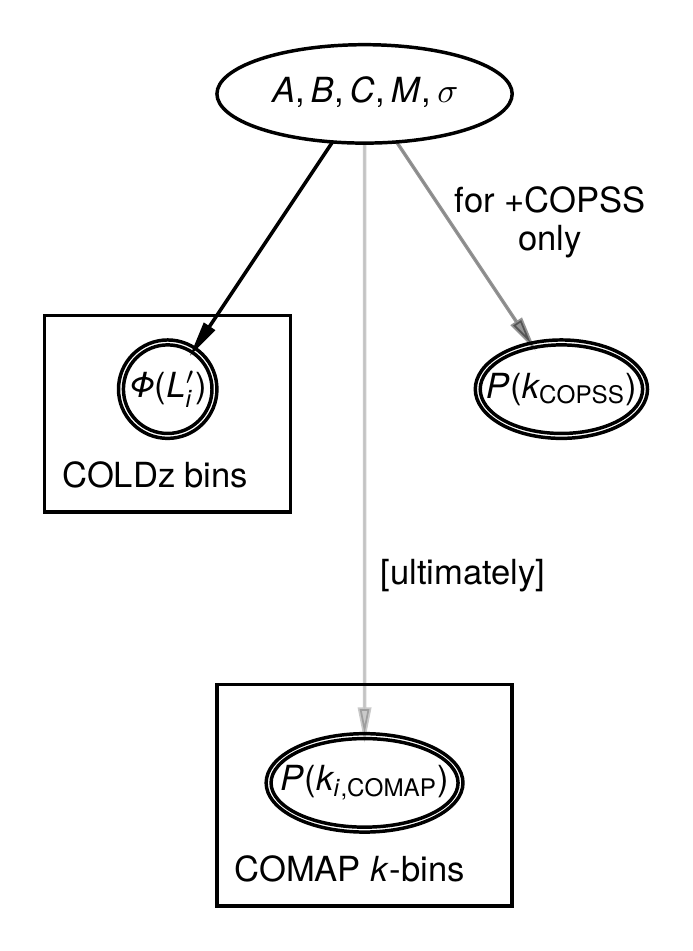}
    \parbox[b]{0.64\linewidth}{\emph{Step 1: Initial priors devised} (\autoref{sec:initialpriors})
    \begin{itemize}
        \item ``flat'' (conservative, uninformative)
        \item ``UM'' (based on empirical fits and models)
        \item ``P18'' (high-information, strong assumptions about CO redshift evolution)
    \end{itemize}
    \emph{Step 2: Condition priors on current observations}
    \begin{itemize}
        \item Likelihood functions based on COLDz LF alone (``+COLDz'') or COPSS $P(k)$ constraint also (``+COLDz+COPSS'')---cf.~\autoref{sec:priorobs}
        \item Infer updated priors via MCMC (\autoref{sec:finalpriors})
    \end{itemize}
    \emph{Step 3: Use resulting posteriors as new data-driven priors}
    \begin{itemize}
        \item \{flat,UM,P18\}+\{COLDz,COLDz+COPSS\} priors can now be meaningfully conditioned on COMAP data once they reach sufficient sensitivity
        \item Can also be used to generate best estimate models for COMAP forecasting
    \end{itemize}}
    \caption{Simplified, annotated graphical representation of the derivation of our model priors for the $z\sim3$ CO(1--0) $L(M_h)$ relation, which is considered in much greater detail throughout~\autoref{sec:model_appendix}.}
    \label{fig:daftpgm}
\end{figure*}

The details of these priors are somewhat ancillary to the primary results of this work, and so are discussed largely in~\autoref{sec:model_appendix}. However, we present a broad overview in~\autoref{fig:daftpgm}. In short, we begin with one of three possible sets of initial priors on our CO model parameters (``flat'', ``UM'', ``P18''), then condition these priors on either the COLDz LF constraints alone or both the COLDz constraints and the COPSS $P(k)$ measurement. These posterior distributions, obtained via MCMC, then act as data-driven priors for COMAP, and can be conditioned on COMAP data at some later date to yield updated posterior distributions.

\begin{figure}[ht!]
    \centering
    \includegraphics[width=0.96\linewidth]{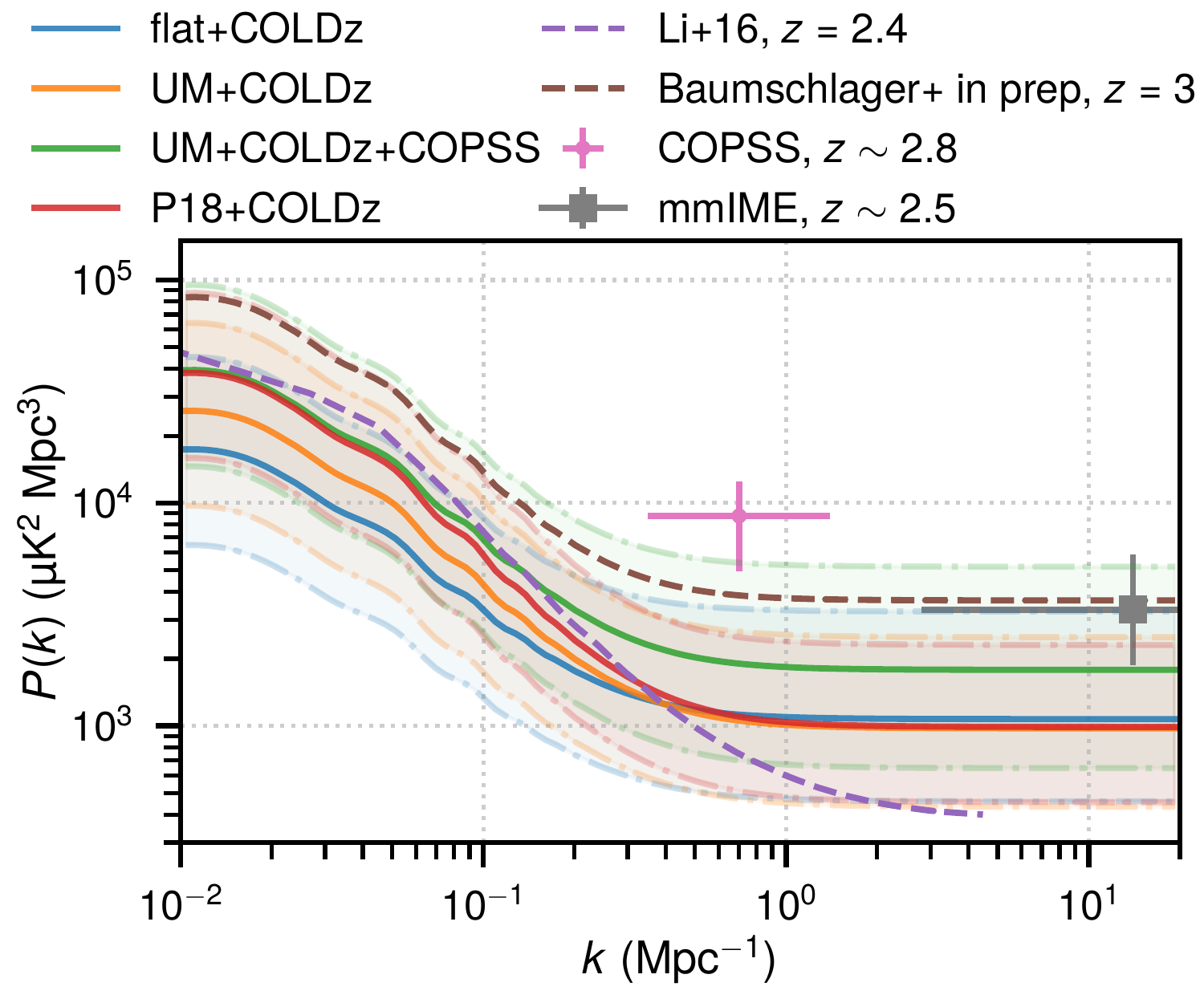}
    \caption{Predictions for the real-space $P(k)$ of the CO(1--0) line-intensity field at $z\sim2$--3. Alongside our data-driven priors and their 68\% credibility intervals (solid lines and shaded areas bounded by dashed-dotted lines), we also show predictions from~\cite{li_etal_16} and the \texttt{TNG300\_2} model of Baumschlager et al.~(in prep.; used by~\citealt{silva_etal_21} for forecasting COMAP--HETDEX synergies) as well as results from COPSS~\citep{keating_etal_16} and mmIME~\citep{mmIME-ACA}.}
    \label{fig:posteriorPk}
\end{figure}

The data-driven priors can also be used to generate analytic estimates for the real-space $P(k)$ at $z\sim2.4$, the central redshift for the COLDz observations. The \texttt{lim} package\footnote{\url{https://github.com/pcbreysse/lim/tree/pcbreysse}} can generate $P(k)$ estimates for every sample of each MCMC. We use a minimum halo mass of $10^9\,M_\odot$ for CO emission in these calculations, but as our models strongly favor a steep super-linear faint-end power law for $L(M_h)$ (i.e., $A<-1$), shifting the minimum halo mass up to $10^{10}\,M_\odot$ has minimal effect on our predictions, including for $P(k)$. We therefore use the higher minimum mass for the remainder of this work, as it matches the value used in our previous fiducial model devised by~\cite{li_etal_16} and as the cosmological simulations we use for simulated COMAP inferences in~\autoref{sec:sim_inf} will only resolve halos with mass $\sim10^{10}\,M_\odot$ (reproducing correct statistics for halos with $M_h>2.5\times10^{10}\,M_\odot$).

We plot the 68\% credibility intervals in~\autoref{fig:posteriorPk} alongside both the model of~\cite{li_etal_16} that previously acted as the fiducial model for COMAP simulations, and observational LIM results from COPSS and mmIME. The COPSS result is in some tension ($\approx2$--$3\sigma$) with our other priors, as are the mmIME estimates ($\approx1$--$2\sigma$). One proposition by~\cite{mmIME-ACA} was that clustering could contribute significantly to the COPSS measurement and thus the best estimate for the shot-noise power spectrum should be adjusted down to $2.0^{+1.1}_{-1.2}\times10^3h^{-3}$\,$\mu$K$^2$\,Mpc$^3$. We discuss the clustering-versus-shot noise balance for the CO power spectrum further in~\autoref{sec:ESconstraints_Aclust}, in the context of current COMAP constraints.

There is one caveat related to this tension that we should consider about our data-driven priors. At present, surveys like COLDz principally constrain CO emitter abundances around or above the knee of the LF, and do not meaningfully constrain the faint-end slope of the LF. The COLDz data prefer neither a positive faint-end slope that would suggest fewer faint CO emitters, nor a negative one that would suggest more faint emitters. Splitting the difference necessarily results in a highly tempered estimate of the total abundance of CO emitters and thus a highly tempered estimate of the total CO power spectrum.

This tempered nature affects not only comparisons of our data-driven priors with observational results, but also in comparisons with previous models informing some of our priors. The best-fit model of~\cite{Padmanabhan18}, for instance, also uses observational data to drive an abundance-matched $L(M_h)$ model. At $z\sim3$ the principal driver is the COPSS data from~\cite{keating_etal_16} but in the form of constraints on the Schechter parameterization of the CO LF. The prior on the faint-end slope that~\cite{keating_etal_16} used is loose but asymmetric and does prefer negative value, and their overall estimate of the LF knee lies higher in both abundance and luminosity than the COLDz constraints. Thus, these data drive the original model $P(k)$ of~\cite{Padmanabhan18} orders of magnitude above our P18+COLDz model $P(k)$, which the COLDz data temper significantly.

One tempting resolution of the tension between our COLDz-driven priors and the COPSS and mmIME results, then, is in interpretation of results from CO line searches like COLDz as a kind of lower bound when considering quantities that involve the faint end of the LF, including the CO power spectrum. We find this idea mirrored in the interpretation of ASPECS data by~\cite{ASPECS-LPPk}, who quoted a lower limit on the mean total CO line temperature based on the individual line detections from that survey. For ASPECS CO(2--1) detections,~\cite{ASPECS-LPPk} were able to use CO--galaxy cross statistics with external optically selected spectroscopic redshifts to constrain the faint-end slope of the $z\sim1$ CO LF. However, they elected not to claim similar constraints for CO(3--2) at $z\sim2.5$ due to potential unreliability of such constraints given the percentage of ASPECS detections without matching optical counterparts. Therefore, any clustering amplitude constraint from direct detections depends strongly on the selection characteristics. Since CO LIM surveys trade this dependence away for the price of potential systematics and contamination, the discrepancy between COLDz and COPSS could be considered a natural result of these caveats.

It is however possible that the resolution of any tension specifically involving the shot noise-dominated measurement of COPSS actually lies in a lower-abundance faint end of the CO LF. If re-weighted based on the COPSS measurement, the COLDz LF Schechter parameter posterior would actually weakly prefer larger positive values of the faint-end log-slope of the LF. The Schechter function as used by~\cite{COLDzLF} models the CO emitter number density per log-luminosity bin as proportional to a power-law \replaced{$L^\alpha$}{$L^{\alpha_\text{LF}}$} times an exponential cutoff $\propto\exp{(-L/L_*)}$. Then the shot noise is proportional to the average integrated squared luminosity of the emitters, which is roughly proportional to \replaced{$\Gamma(\alpha+2)$}{$\Gamma(\alpha_\text{LF}+2)$}. This function reaches a local minimum at \replaced{$\alpha\approx-0.54$}{$\alpha_\text{LF}\approx-0.54$} but will be greater for lower \emph{or} higher values of \replaced{$\alpha$}{$\alpha_\text{LF}$}. We can make sense of this physically: a CO LF with fewer faint emitters and more emitters near or above the knee\replaced{ has}{---e.g., from low-mass halos hosting low-metallicity systems with high CO dissociation rates---leads to} enhanced contrast of CO line-intensity fluctuations at small scales, and thus a greater shot-noise amplitude of the CO power spectrum. Without a clustering measurement like COMAP, independent of both direct-detection surveys like COLDz and shot-noise LIM surveys like COPSS, we have limited ability to bound the faint end of the CO LF from either below or above.

Ultimately, we drive our fiducial UM+COLDz+COPSS estimate with the best and most relevant observational data available, but this model is conservative by nature of the COLDz data (which hold far higher total statistical weight than the COPSS data). In future forecasting and forward models, we should be entirely open to the possibility that faint CO emitters are far more abundant---and thus that the integrated cosmic average CO intensity is considerably higher---than direct CO line searches suggest at the time of writing.

\begin{deluxetable*}{rccccc}
    \tablecaption{CO Model Point Estimates Based on Data-driven Priors\label{tab:priorpoints}}
    \tablehead{\colhead{}& \multicolumn{5}{c}{Point Estimates for:} \\
        \colhead{Data-driven Prior} &\colhead{$A$}&\colhead{$B$}&\colhead{$\log{C}$}&\colhead{$\log{\frac{M}{M_\odot}}$}&\colhead{$\sigma$}}
    \startdata
        ``flat+COLDz'' & $-3.7$ & 7.0 & 11.1 & 12.5 & 0.36\\
        ``UM+COLDz'' & $-2.75$ & 0.05 & 10.61 & 12.3 & 0.42 \\
        \textbf{``UM+COLDz+COPSS''} & $-2.85$ & $-0.42$ & 10.63 & 12.3 & 0.42 \\
        ``P18+COLDz'' & $-2.4$ & $-0.5$ & 10.45 & 12.21 & 0.36
    \enddata
    \tablecomments{Values are determined at $z\sim2.8$ to match the median $P(k)$ and LF values from each data-driven prior. We indicate our fiducial choice in boldface.}
\end{deluxetable*}
For now, reverting to the COMAP central redshift of $z\sim2.8$, we can identify specific parameter values to approximately match the median $P(k)$ and $\phi(L')$ values for each set of priors (as shown in~\autoref{fig:posteriorPk} and~\autoref{sec:model_appendix}). This assumes that the CO signal is relatively insensitive to cosmology and redshift (within the COMAP survey range), which is true when compared to our model uncertainties. We show the parameter point estimates corresponding to each data-driven prior in~\autoref{tab:priorpoints}.

\subsection{Incorporating Line Broadening}
\label{sec:doppler}

Not only are CO emitters not point sources, but their extent in a data cube does not correspond to their extent in physical or comoving space. Some of this is due to instrumental resolution, but some of this is due to observations being in redshift space rather than in real space. One key effect to consider is the peculiar velocities of the gas within each galaxy---due both to overall galactic rotation and to turbulent gas motion separate from this rotation---which results in Doppler broadening of the CO line emission.

\cite{linebroad} provide some methods to account for line broadening, providing an empirical line-width model for CO(1--0) under the assumption that CO emitters are rotation-dominated, mostly disc-like sources. The inclination angle $i$ of each emitter's axis of rotation relative to the observer line of sight is assumed to be random and independent, with a uniform distribution of $\cos{i}\in(0,1)$. Using this model, we set the full width at half maximum (FWHM) of the CO line profile for a host halo of virial mass $M_h$ to the circular velocity of the halo at the median inclination angle of $i=\pi/3$. In this work, we use either numerical calculations based on an analytic model or approximate $N$-body simulations using the peak--patch method~\citep{Stein18} that we consider further in~\autoref{sec:sim_inf}. In both cases, the halo maximum circular velocity is unavailable and we use the virial velocity $v_\text{vir}$ instead. \cite{linebroad} preferred the former but compared using one versus the other and found the choice to not affect results significantly.

The CO line FWHM estimated from the host halo's virial velocity and randomized inclination is
\begin{align}
    v(M_h,z,i)&=v_\text{vir}(M_h,z)\left[\frac{\sin{i}}{\sin{(\pi/3)}}\right]\nonumber\\
    &=\left(\frac{\Delta_c}{2}\right)^{1/6}[GM_hH(z)]^{1/3}\left(\frac{\sin{i}}{\sqrt{3}/2}\right)\nonumber\\
    &\approx35\,\text{km\,s}^{-1}\left(\frac{\sin{i}}{0.866}\right)\left(\frac{\Delta_c}{200}\right)^{1/6}\times\nonumber\\
    &\qquad\left(\frac{M_h}{10^{10}\,M_\odot}\frac{H(z)}{100\,\text{km\,s}^{-1}\,\text{Mpc}^{-1}}\right)^{1/3}.
\end{align}
Here $\Delta_c$ is the spherical overdensity within the virial radius of the halo, relative to the critical density of our cosmology. The value used by~\cite{linebroad} is 180, whereas 200 is also common (being historically considered canonical for a cosmology with critical matter density---cf.~\citealt{White01,White02}). This difference in $\Delta_c$ is of minimal concern as the resulting difference in $v(M_h)$ is only a few percent.

When a forecast of \emph{only} the spherically-averaged $P(k)$ is required, a single Gaussian filter with an effective velocity scale $v_\text{eff}$ is sufficient to describe the smearing of the total CO line-intensity cube. This comes at the cost of some accuracy, but will bring significantly improved computational speed in any contexts where the approximation is applicable. Including adjustments for random inclinations, the appropriate effective velocity given by Equation~46 of~\cite{linebroad} is
\begin{equation}
    v_\text{eff}=\frac{1}{2}\left(\frac{\avg{L^2v_\text{vir}}}{\avg{L^2}}+\frac{4}{\pi\sqrt{3}}\frac{\avg{L^2}}{\avg{L^2v_\text{vir}^{-1}}}\right),\label{eq:veff_success}
\end{equation}
where $\avg{x}\equiv\int dM_h\,(dn/dM_h)\,x$.

As~\cite{linebroad} make clear, stark shortcomings in approximating the effect of line broadening with only $v_\text{eff}$ exist in the context of projections made in the present work for future analyses, which will not only deal with $P(k)$, but also the voxel intensity distribution (VID). Therefore, in mocks of the CO line-intensity field using approximate $N$-body simulations, we bin halos by virial velocity and broaden the CO emission from each bin by its median velocity. We use the two-tier approach outlined in~\cite{linebroad} which ignores line broadening for halos below a certain mass whose line profiles are not possible to resolve with the COMAP Pathfinder science channelisation of 32\,MHz (equivalent to $\approx320$\,km\,s$^{-1}$ in velocity space for 30\,GHz observations). To recap the procedure in full:
\begin{itemize}
    \item Divide the halos into a low-mass subset with $M_h<{10}^{11}\,M_\odot$ and a high-mass subset with $M_h>10^{11}\,M_\odot$. The cut point is equivalent to $v_\text{vir}\approx107$\,km\,s$^{-1}$, so the low-mass subset includes all halos whose CO line widths should span less than one-third of a COMAP science voxel.
    \item Generate a CO cube from the low-mass subset without applying any Gaussian filters.
    \item Divide the high-mass subset into 16 equally spaced linear bins in virial velocity.
    \item For each bin, generate a CO cube with a Gaussian filter applied to approximate line broadening. The median virial velocity across all halos within the bin sets the Gaussian width. This results in 16 CO cubes, one for each velocity bin.
    \item Sum all 17 CO cubes, including the low-mass CO cube, for the final simulated product.
\end{itemize}
Simulations by~\cite{linebroad} show that this approach keeps $P(k)$ within 10\% of the reference simulation (using 64 bins in halo circular velocity) and the VID approximately within Poisson error of the reference simulation. The increase in time for the CO cube computation is around a factor of 30, but the computation is still sufficiently fast when considering the other steps involved in simulations such as power spectrum evaluation. Thus, this will be our approach to simulating line broadening for anything more complicated than simple $P(k)$ forecasts. 

Note that we did not apply this correction above when constraining our priors with observational results. First, the COLDz dataset used is of distributions of discrete emitters and our COLDz-based likelihood does not need models of any line profiles. Second, the UM+COLDz+COPSS calculation needs in principle to correct for the effect of line broadening on the CO(1--0) power spectrum, especially as it will attenuate the apparent power spectrum less for wavenumbers where COMAP measures $P(k)$ compared to COPSS. However, even for COPSS the effect at $k\sim1h$\,Mpc$^{-1}$ is typically $\sim30$\% and thus is subdominant to the overall uncertainty in the $\sim2\sigma$ COPSS $P(k)$ result. Therefore, we err on the conservative side and do not correct for line broadening in devising our priors.
\section{Implications of COMAP Early Science Power Spectrum Measurements}
\label{sec:ULinterp}

\begin{figure*}
    \centering
    \includegraphics[width=0.96\linewidth]{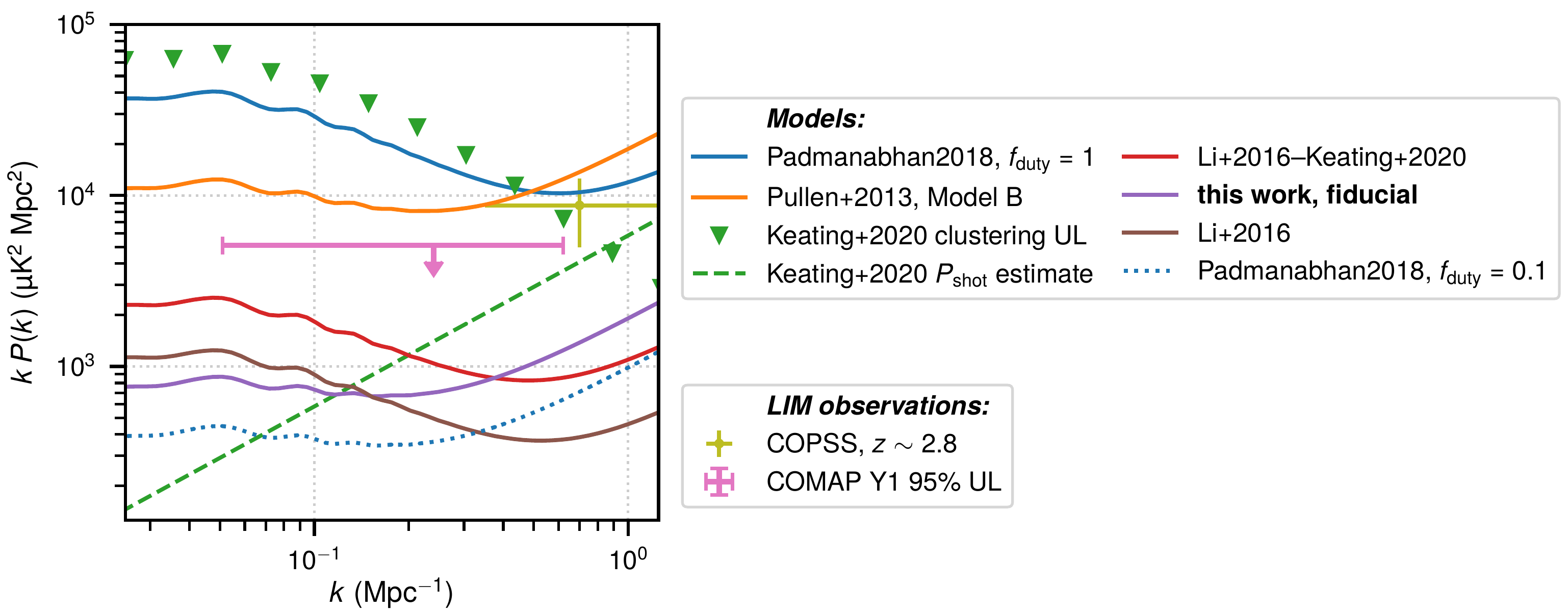}
    \caption{COMAP Pathfinder early science constraint (pink) on the redshift-space CO(1--0) power spectrum at $z\sim3$, alongside model predictions from this work (UM+COLDz+COPSS) and those recalculated based on $L(M_h)$ relations from~\cite{Padmanabhan18},~\cite{Pullen13}, and~\cite{li_etal_16}, as well as a variation on the latter from~\cite{mmIME-ACA}. We also show interpretations of the COPSS result both as a direct $P(k)$ measurement~\citep[yellow error bars;][]{keating_etal_16} and as a constraint on clustering (triangles) and shot-noise amplitudes~\citep[dashed line;][]{mmIME-ACA}.}
    \label{fig:PkSNR_Y1}
\end{figure*}

The present state of COMAP observations do not yet allow for the kinds of analyses that we forecast in~\autoref{sec:forecasts}. However, the $P(k)$ result\footnote{Strictly speaking, as~\cite{es_IV} note in their Section 3.1, the result is based on a pseudo-power spectrum measurement and may have some residual mode-mixing bias. However, their Figure 1 also shows that this mode-mixing bias likely is a small effect (5--30\%) that enhances the pseudo-spectrum relative to the true signal. The measurement obtained from this pseudo-spectrum result should thus still be a valid, if possibly conservative, upper limit on the true CO power spectrum.} obtained by~\cite{es_IV} already has constraining power that strongly complements the COPSS result. \added{We do note that since the original analysis of~\cite{keating_etal_16}, which essentially assumed the measurement to be entirely shot noise-dominated,~\cite{mmIME-ACA} re-analysed the COPSS power spectrum result as a combination of clustering and shot-noise components, setting an upper limit on clustering and revising down the best estimate for shot noise. Nonetheless, the COPSS measurement still reflects a $k$-range where the majority of the signal is shot noise, whereas the majority of any potential power spectrum measurement in the COMAP $k$-range would be from clustering.

}Coadding constant-elevation scan (CES) data across all fields, the all-scale measurement is $P(k=0.051$--$0.62\,\text{Mpc}^{-1})=(-2.7\pm1.7)\times10^4\,\mu$K$^2$\,Mpc$^3$. Asserting $P(k)>0$ on top of this measurement, we obtain a 95\% upper limit of $k\,P(k)<5.1\times10^3$\,$\mu$K$^2$\,Mpc$^2$ at $k=0.24$\,Mpc$^{-1}$, shown in~\autoref{fig:PkSNR_Y1}.

Note that the bulk of the present sensitivity derives from Field~1 CES data, which alone yield $P(k=0.051$--$0.62\,\text{Mpc}^{-1})=(-4.6\pm2.2)\times10^4\,\mu$K$^2$\,Mpc$^3$, or a 95\% upper limit of $k\,P(k)<5.4\times10^3$\,$\mu$K$^2$\,Mpc$^2$ at $k=0.24$\,Mpc$^{-1}$ when requiring $P(k)>0$.

The current COMAP constraint already excludes the predictions of~\cite{Padmanabhan18} (assuming a CO emission duty cycle\replaced{ }{---i.e., the fraction of time that any given galaxy is CO-luminous---}of 1) and Model B of~\cite{Pullen13} at 95\% confidence\footnote{We also exclude but do not consider other models in the literature that are based on outdated assumptions, which subsequent works often supersede. For instance, COPSS data also excluded predictions from the model of~\cite{Lidz11} (even in the pilot analysis done by~\citealt{COPSSPilot}). However, the~\cite{Lidz11} model had already been reformed at $z\sim3$ into Model A of~\cite{Pullen13} with a revised halo mass--SFR scaling that was more applicable at these redshifts.}, and overall constrains the clustering component of the power spectrum better than the COPSS re-analysis of~\cite{mmIME-ACA} by roughly an order of magnitude. We first consider the model exclusions in~\autoref{sec:exclusions} before considering clustering constraints in more detail in~\autoref{sec:ESconstraints_Aclust} and translating these into molecular gas constraints in~\autoref{sec:ESconstraints_rhoH2}.

\subsection{Excluded Models}
\label{sec:exclusions}

Model~B of~\cite{Pullen13} was in principle one of the models already excluded by the COPSS measurement of~\cite{keating_etal_16}, but we exclude it in the clustering regime whereas the COPSS results excluded it in the shot-noise regime ($k=0.5h$--$10h$\,Mpc$^{-1}$). This is a meaningful distinction particularly for this model, as~\cite{Pullen13} implement a duty cycle $f_\text{duty}$ for CO-bright activity, with which the shot noise scales inversely. Therefore, the constraint of~\cite{keating_etal_16} can only be on some combination of the halo mass--CO luminosity scaling and $f_\text{duty}$, encapsulating the shot-noise amplitude. The correction required for Model B of~\cite{Pullen13} to be made consistent with the COPSS measurement could thus be either a different CO--SFR scaling from what~\cite{Pullen13} used---which was a fit to local and high-redshift galaxies by~\cite{Wang10}---or a different value of $f_\text{duty}$.

In a typical halo model, the clustering amplitude scales directly with $f_\text{duty}$. However, Model B of~\cite{Pullen13} derives the cosmic average CO temperature $\avg{T}$ from using the~\cite{Wang10} CO--SFR relation to directly scale the integrated SFR density obtained via Schechter fits to the SFR function tabulated by~\cite{Smit12}. As~\cite{Pullen13} assume that the duty cycle for CO-bright activity matches the duty cycle for star-formation activity, $f_\text{duty}$ does not modify $\avg{T}$ for Model B of~\cite{Pullen13}, and thus should not modify the lower-$k$ power spectrum values that we constrain.

Our results thus suggest that the~\cite{Wang10} CO--SFR relation is not globally applicable to galaxies at $z\sim3$, in the sense that it cannot be used to connect the SFR functions of~\cite{Smit12} to CO luminosity at this redshift range. Indeed, while the~\cite{Wang10} relation suggests $\mathrm{SFR}\sim L_\text{CO}^{1.67}$, this is much steeper than the general correlation at high redshift inferred from data reviewed by~\cite{CW13}, which includes some data not available at the time of~\cite{Wang10}.

Also of interest is our exclusion of the model of~\cite{Padmanabhan18} with $f_\text{duty}=1$, which explicitly folded the~\cite{keating_etal_16} result into its derivation. In comparison to other models, this model predicts a higher clustering amplitude relative to the shot-noise amplitude. Without other significant data available to drive the abundance matching carried out at $z\sim3$ by~\cite{Padmanabhan18}, it was perfectly reasonable for the resulting model to account for the COPSS result through a very high overall power spectrum prediction---including a high clustering amplitude---as opposed to additional parameterization of stochasticity to further decouple the shot-noise and clustering amplitudes. This once again highlights the value of having COMAP data to separately constrain the power spectrum at lower $k$.

Note that \cite{Pullen13} and~\cite{Padmanabhan18} each present an alternate model that we do not exclude. Model A of~\cite{Pullen13} is based on a less empirical, more indirect set of assumptions to connect halo and galaxy properties, and more similar (both qualitatively and quantitatively) to our fiducial models or that of~\cite{li_etal_16}. Meanwhile,~\cite{Padmanabhan18} shows $P(k)$ curves for both $f_\text{duty}=1$ and $f_\text{duty}=0.1$. We do not exclude the latter variation on this model in principle, although as~\autoref{fig:PkSNR_Y1} shows that this variation then predicts shot noise well below our UM+COLDz+COPSS model's expectation as well as the COPSS measurement alone. \cite{Padmanabhan18} also notes that $f_\text{duty}=1$ is somewhat better supported in observational data. The tension between these two extremes (and their implications for the ratio between the clustering and shot-noise components of the power spectrum) could be feasibly bridged by a mass-dependent $f_\text{duty}$ that falls from 1 with higher mass, as is the case for the empirical models of~\cite{Yang21}.

\subsection{Constraints on CO Power Spectrum Clustering and Shot-noise Amplitudes}
\label{sec:ESconstraints_Aclust}

At this early stage of the Pathfinder campaign, COMAP data will not yet place significant constraints on the parameters of our $L(M_h)$ model devised in~\autoref{sec:model}. However, we show that the upper limit does place meaningful constraints on the integrated clustering and shot-noise amplitudes for the CO power spectrum. Furthermore, by leveraging our model priors from~\autoref{sec:model}, we can obtain an upper limit on the mean temperature $\avg{T}$ at $z\sim2.8$ from our clustering amplitude constraint, from which we derive limits on H$_2$ mass density in~\autoref{sec:ESconstraints_rhoH2}.


In real comoving space, we would model the power spectrum as
\begin{equation}
    P(k) = A_\text{clust}P_m(k) + P_\text{shot}.
\end{equation}
This is to say that the total $P(k)$ is the sum of a clustering component, the matter power spectrum $P_m(k)$ scaled by a clustering amplitude $A_\text{clust}$, and a shot-noise component $P_\text{shot}$. This neglects any possible scale-dependent bias or one-halo terms but is sufficient for our purposes.

We should then be able to consider likelihood contours and constraints for $A_\text{clust}$ and $P_\text{shot}$ based on our observational data, both in isolation and in combination with the COPSS $P(k)$ measurements from~\cite{keating_etal_16}. This mirrors the COPSS re-analysis performed by~\cite{mmIME-ACA}.

For the real-space $P(k)$, we would have $A_\text{clust}=\avg{Tb}^2$, or the square of the mean line temperature--bias product across the luminosity function:
\begin{equation}
    \avg{Tb}\propto\int dM_h\,\frac{dn}{dM_h}L(M_h)b(M_h),
\end{equation}
with appropriate conversions applied to convert luminosity density to brightness temperature. Without $b(M_h)$ in the integrand, the analogous integral would yield the mean CO brightness temperature $\avg{T}$; the line luminosity-averaged bias is then $b\equiv \avg{Tb}/\avg{T}$.

However, redshift-space distortions from the coherent infall of galaxies into large-scale structure~\citep{Kaiser1987,Hamilton1998} enhance the clustering component such that $A_\text{clust}\approx \avg{T}^2(b^2+2b/3+1/5)$ for small $k$ (and $\Omega_m(z)\approx1$, which is the case at $z\sim3$). Furthermore, as explained in~\autoref{sec:doppler}, line broadening introduces $k$-dependent attenuation, largely of the shot noise. In the context of $P(k)$, the parameter $v_\text{eff}$ described there is sufficient to encapsulate the overall effect.

Given our limited knowledge of line bias and line broadening for CO at high redshift, we consider two different ways to present constraints on the power spectrum clustering and shot-noise amplitudes.
\begin{itemize}
\item In the first method we carry out a $b$-agnostic, $v_\text{eff}$-agnostic calculation of constraints on $A_\text{clust}$ and $P_\text{shot}$. We make no assumptions about values of $b$, instead constraining the overall observed amplitude $A_\text{clust}$ that scales the matter power spectrum. We also ignore line broadening altogether and make no attempt to compensate for its effect on our data. Thus we assume that the shot-noise component looks the same in real and redshift space (before transfer functions, for which we do compensate). This is closest to the analyses of~\cite{mmIME-ACA} and~\cite{Keenan21}, neither of which correct CO auto-spectra for line broadening or account for linear redshift distortions.
\item In the second method we constrain $\avg{Tb}^2$ and $P_\text{shot}$ in a $b$-informed\footnote{Despite making assumptions around $b$, we do not attempt to directly evaluate constraints in the 2D parameter space of $\avg{T}$--$P_\text{shot}$. Such an approach will involve a scaling of $A_\text{clust}/\avg{T}^2=b^2+2b/3+1/5$ to estimate the clustering component of the power spectrum for a given value of $\avg{T}$, whereas our approach only involves scaling by $A_\text{clust}/\avg{Tb}^2\approx1+2/(3b)+1/(5b^2)$ for a given value of $\avg{Tb}$. Any unreliability in determining $b$ will result in far greater relative error in the former than in the latter for plausible values of $b$ in our models.} and $v_\text{eff}$-informed analysis, incorporating line broadening as well as expectations for line bias based on our UM+COLDz priors. From our UM+COLDz MCMC distribution, we obtain average values of $b$ and $v_\text{eff}$ across these dimensions\footnote{We derive these values from the UM+COLDz priors to avoid double-counting any information from COPSS in our analysis, but the resulting fits hold equally well for the UM+COLDz+COPSS MCMC samples.} and define reasonable two-dimensional polynomial fits to those average values, as described in~\autoref{sec:vbfits}. This allows us to directly calculate $P(k)$ including redshift-space distortions and line broadening, which should be appropriate to fit simultaneously to COPSS data and to the COMAP data that has been corrected in $(k_\parallel,k_\perp)$ space (before spherical averaging) to account for beam, filtering, and spectral effects.
\end{itemize}

\begin{figure}[t!]
    \centering
    \includegraphics[width=0.96\linewidth]{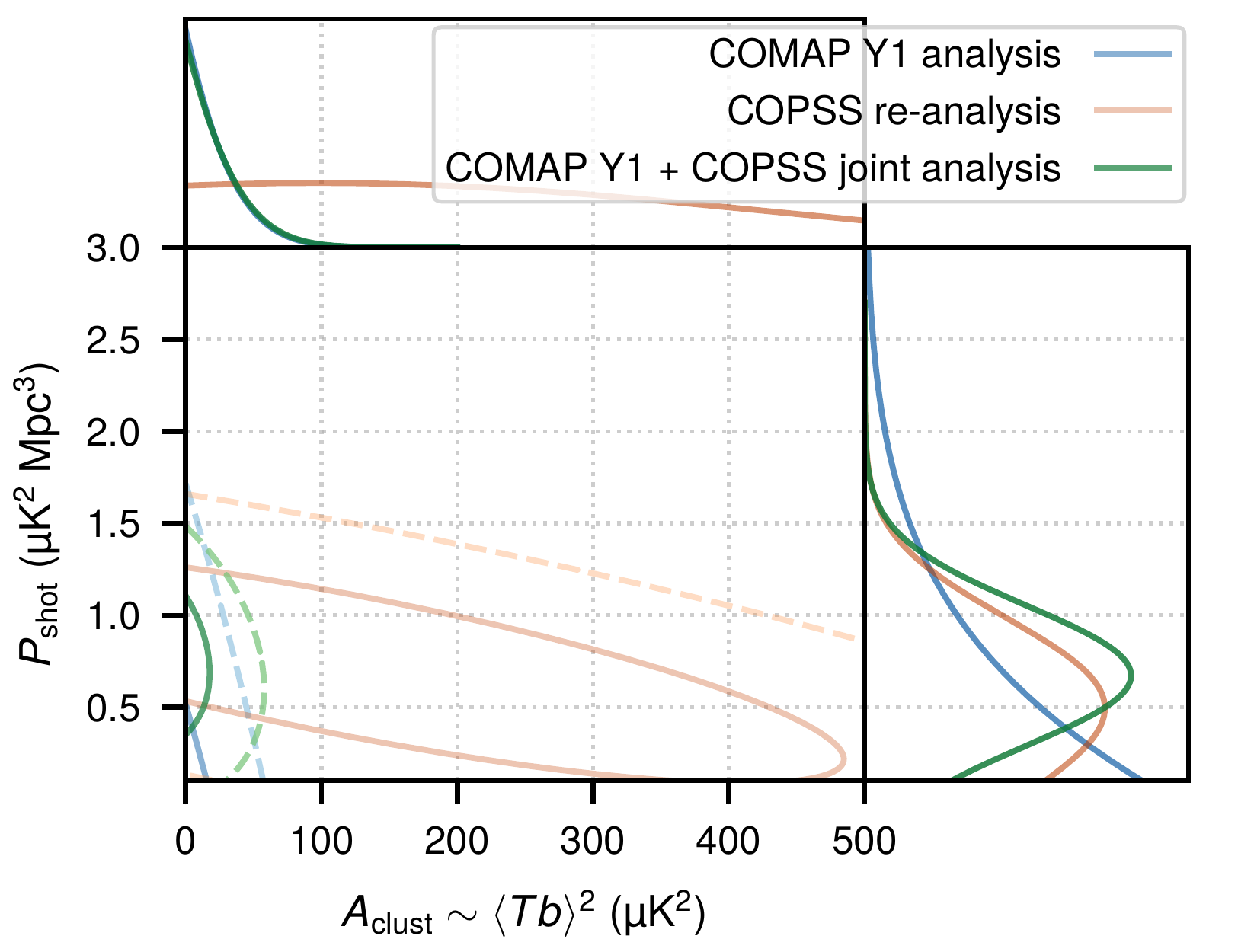}
    \includegraphics[width=0.96\linewidth]{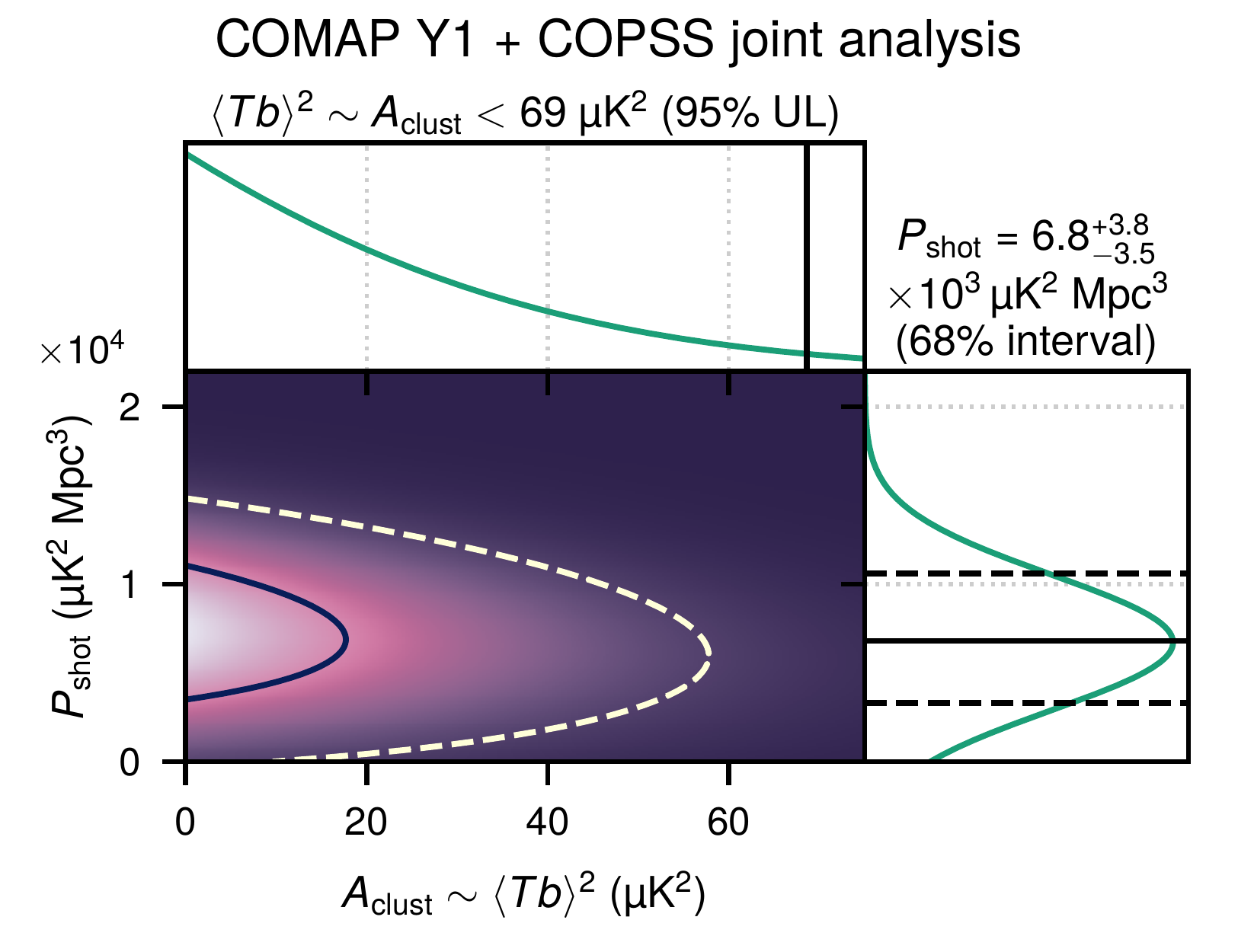}
    \caption{\emph{Upper panel\added{s}:} Likelihood contours\added{ (largest panel) and marginalized probability distributions (smaller panels)} for the clustering and shot-noise amplitudes of the CO power spectrum, based on different datasets. The solid and dashed \added{2D }contours respectively represent $\Delta\chi^2=\{1,4\}$ relative to the minimum $\chi^2$ obtained in the parameter space, corresponding to $1\sigma$ and $2\sigma$ for 2D Gaussians. \emph{Lower panel\added{s}:} Likelihood distribution conditioned jointly on COMAP and COPSS $P(k)$ measurements, with the corresponding contours reproduced from the upper panel and marginalized constraints on each of $A_\text{clust}$ and $P_\text{shot}$ shown.}
    \label{fig:Tbshot_agnostic}
\end{figure}

\begin{figure}[t!]
    \centering
    \includegraphics[width=0.96\linewidth]{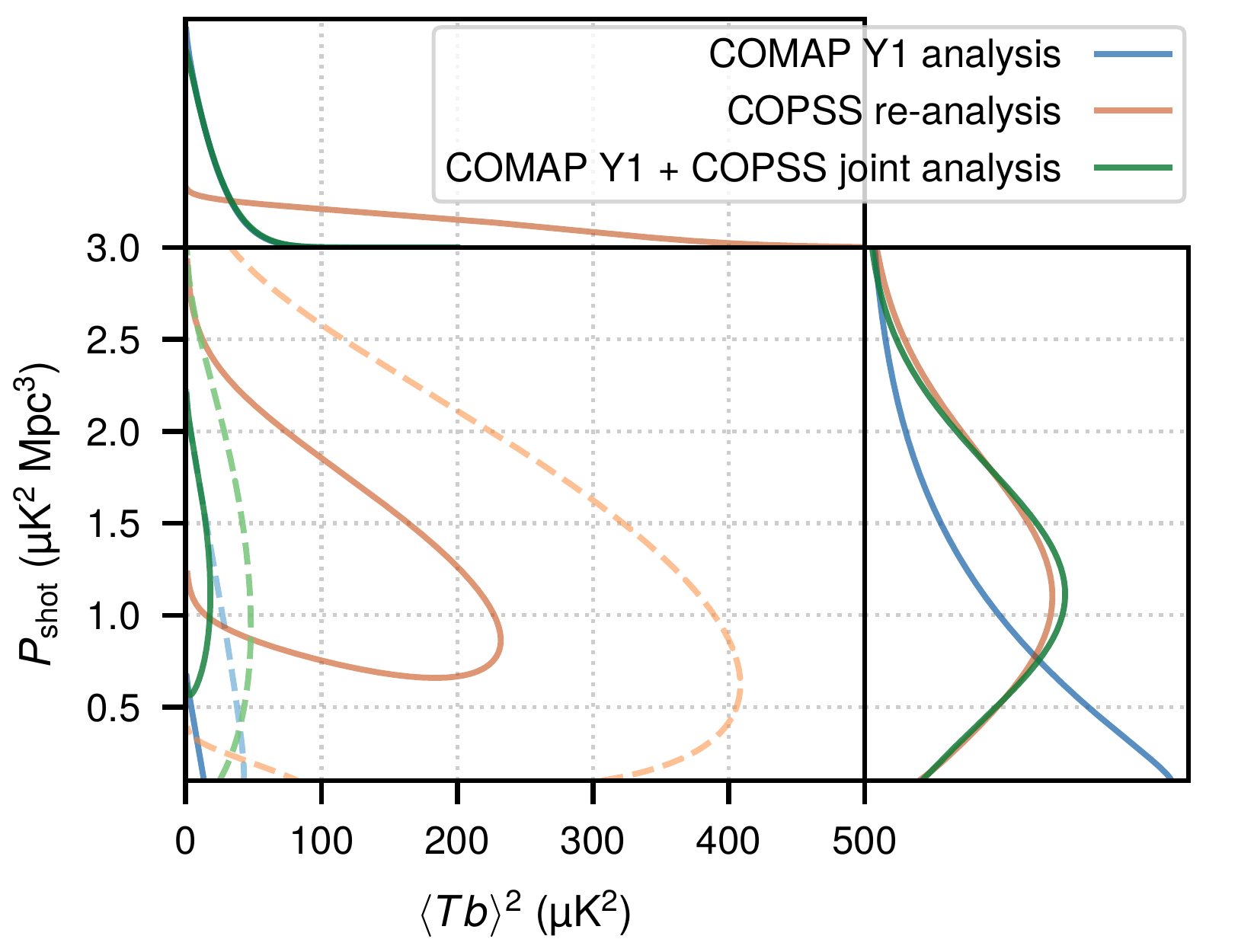}
    \includegraphics[width=0.96\linewidth]{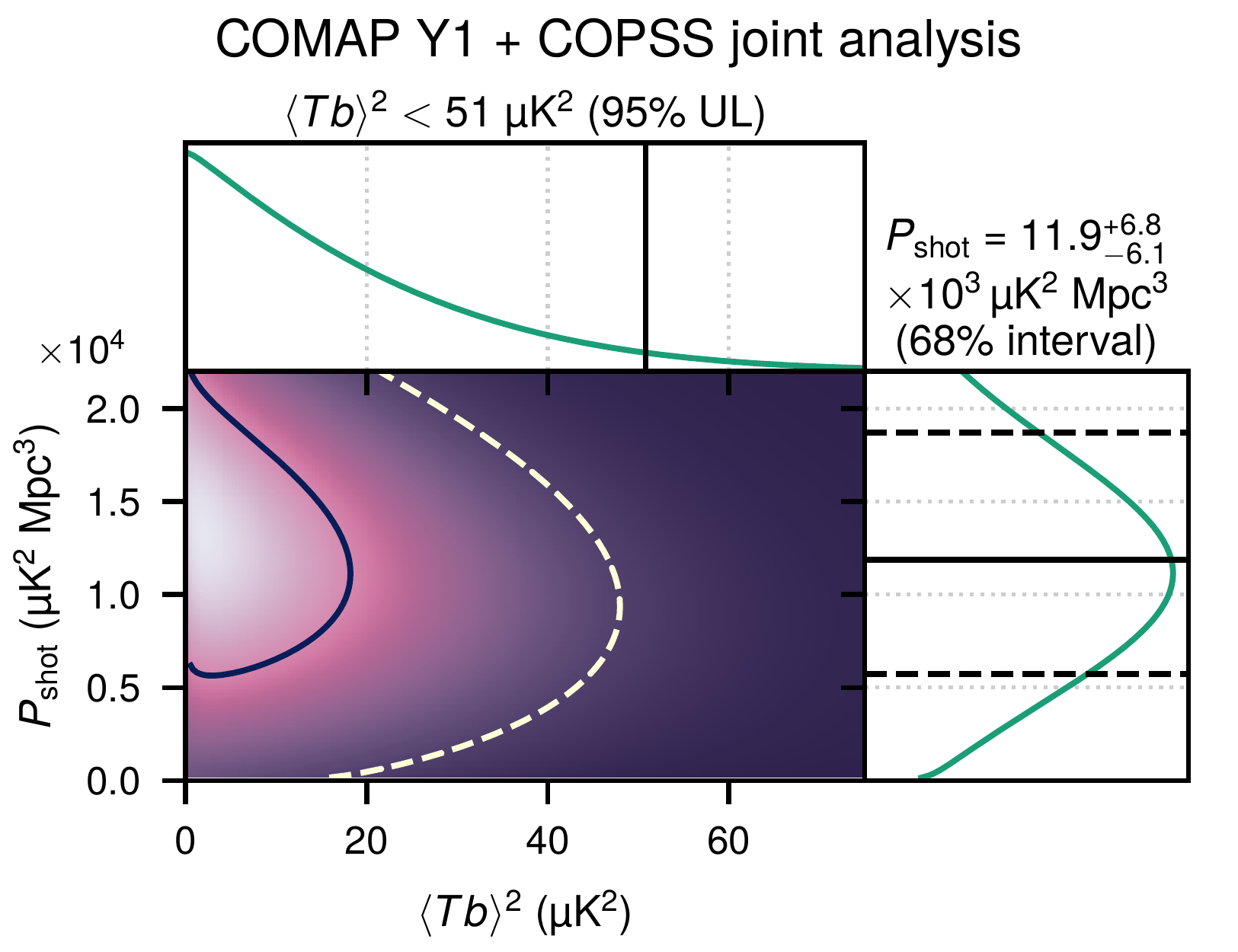}
    \caption{Same as~\autoref{fig:Tbshot_agnostic}, but with strong assumptions around line bias and line broadening as discussed in the main text. These assumptions also allow us to claim a constraint on $\avg{Tb}^2$ itself rather than the redshift-space clustering amplitude $A_\text{clust}$ as considered near the start of~\autoref{sec:ESconstraints_Aclust}.}
    \label{fig:Tbshot_informed}
\end{figure}

\begin{deluxetable*}{ccccccccc}
    \tablecaption{Constraints on the CO Clustering and Shot-noise Amplitudes, and on Derived Quantities\label{tab:alltheconstraints}}
    \tablehead{& \multicolumn{2}{c}{$b$- and $v_\text{eff}$-agnostic:}  & \multicolumn{2}{c}{$b$- and $v_\text{eff}$-informed:}& \multicolumn{2}{c}{$b$- and $v_\text{eff}$-agnostic:}  & \multicolumn{2}{c}{$b$- and $v_\text{eff}$-informed:} \\
        &\colhead{$A_\text{clust}$}&\colhead{$P_\text{shot}$}&\colhead{$\avg{Tb}^2$}&\colhead{$P_\text{shot}$}&\colhead{$\avg{T}$}&\colhead{$\rho_\text{H2}$}&\colhead{$\avg{T}$}&\colhead{$\rho_\text{H2}$}\\\colhead{Data}& \colhead{($\mu$K$^2$)}& \colhead{($10^3\,\mu$K$^2$\,Mpc$^3$)}&\colhead{($\mu$K$^2$)}&\colhead{($10^3\,\mu$K$^2$\,Mpc$^3$)}&\colhead{($\mu$K)}&\colhead{($10^8\,M_\odot\,$Mpc$^{-3}$)}&\colhead{($\mu$K)}&\colhead{($10^8\,M_\odot\,$Mpc$^{-3}$)}}
    \startdata
        COPSS\tablenotemark{a} & $<630$ & $5.7^{+4.2}_{-3.6}$ & $<345$ & $12.1^{+7.5}_{-6.4}$ & $<11.$ & $<7.4$ & $<9.3$ & $<6.4$\\
        COMAP Y1 & $<66$ & $<19$ & $<49$ & $<24$ & $<3.5$ & $<2.4$ & $<3.5$ & $<2.5$\\
        COMAP Y1+COPSS & $<69$ & $6.8^{+3.8}_{-3.5}$ & $<51$ & $11.9^{+6.8}_{-6.1}$ & $<3.5$ & $<2.5$ & $<3.6$ & $<2.5$
    \enddata
    \tablecomments{We use the terms ``$b$- and $v_\text{eff}$-agnostic/informed'' to denote one of two methods used to infer and present constraints on the power spectrum component amplitudes, as discussed in~\autoref{sec:ESconstraints_Aclust}. Bounds on the derived quantities $\avg{T}$ and $\rho_\text{H2}$ depend on a priors-based assumption of $b>2$ and other conversions discussed in~\autoref{sec:ESconstraints_Aclust} and~\autoref{sec:ESconstraints_rhoH2}. Upper limits are 95\% confidence; bounded intervals are 68\% confidence.}
    \tablenotetext{a}{The $A_\text{clust}$ constraint differs somewhat in our re-analysis from the re-analysis of~\cite{mmIME-ACA}, which found a 95\% upper limit of $420\,\mu$K$^2$. We ascribe the discrepancy to differences in assumed cosmology, including in parameters not enumerated by~\cite{mmIME-ACA} that determine $P_m(k)$. Our COPSS-based $P_\text{shot}$ estimate uncorrected for line broadening, which does not depend on such parameters, corresponds to $2.0^{+1.4}_{-1.2}\times10^3h^{-3}\,\mu$K$^2$\,Mpc$^3$ and is entirely consistent with the~\cite{mmIME-ACA} estimate of $2000^{+1100}_{-1200}h^{-3}\,\mu$K$^2$\,Mpc$^3$.}
\end{deluxetable*}
We show results from both methods, using COMAP data and/or COPSS data, in~\autoref{tab:alltheconstraints}. We also illustrate the COMAP--COPSS joint constraints graphically in~\autoref{fig:Tbshot_agnostic} for the $b$- and $v_\text{eff}$-agnostic method and, and in~\autoref{fig:Tbshot_informed} for the $b$- and $v_\text{eff}$-informed method.

With the agnostic method, COMAP data by themselves constrain $A_\text{clust}\lesssim70$\,$\mu$K$^2$ at 95\% confidence, with or without the COPSS data. For the COMAP--COPSS joint analysis, the accompanying shot-noise constraint is $P_\text{shot}=6.8^{+3.8}_{-3.5}\times10^3$\,$\mu$K$^2$\,Mpc$^3$, around 78\% of the total COPSS $P(k)$ measurement. Comparing this to the COPSS re-analysis by~\cite{mmIME-ACA}, which yielded a best estimate of $P_\text{shot}=2.0^{+1.1}_{-1.2}\times10^3h^{-3}\,\mu$K$^2\,$Mpc$^3=5.8^{+3.2}_{-3.5}\times10^3\,\mu$K$^2\,$Mpc$^3$ (around 66\% of the total $P(k)$ measurement), shows that the limit placed by COMAP data on $A_\text{clust}$ constrains how much of the COPSS signal could be ascribed to measuring clustering versus measuring shot noise.

\begin{figure}
    \centering
    \includegraphics[width=0.96\linewidth]{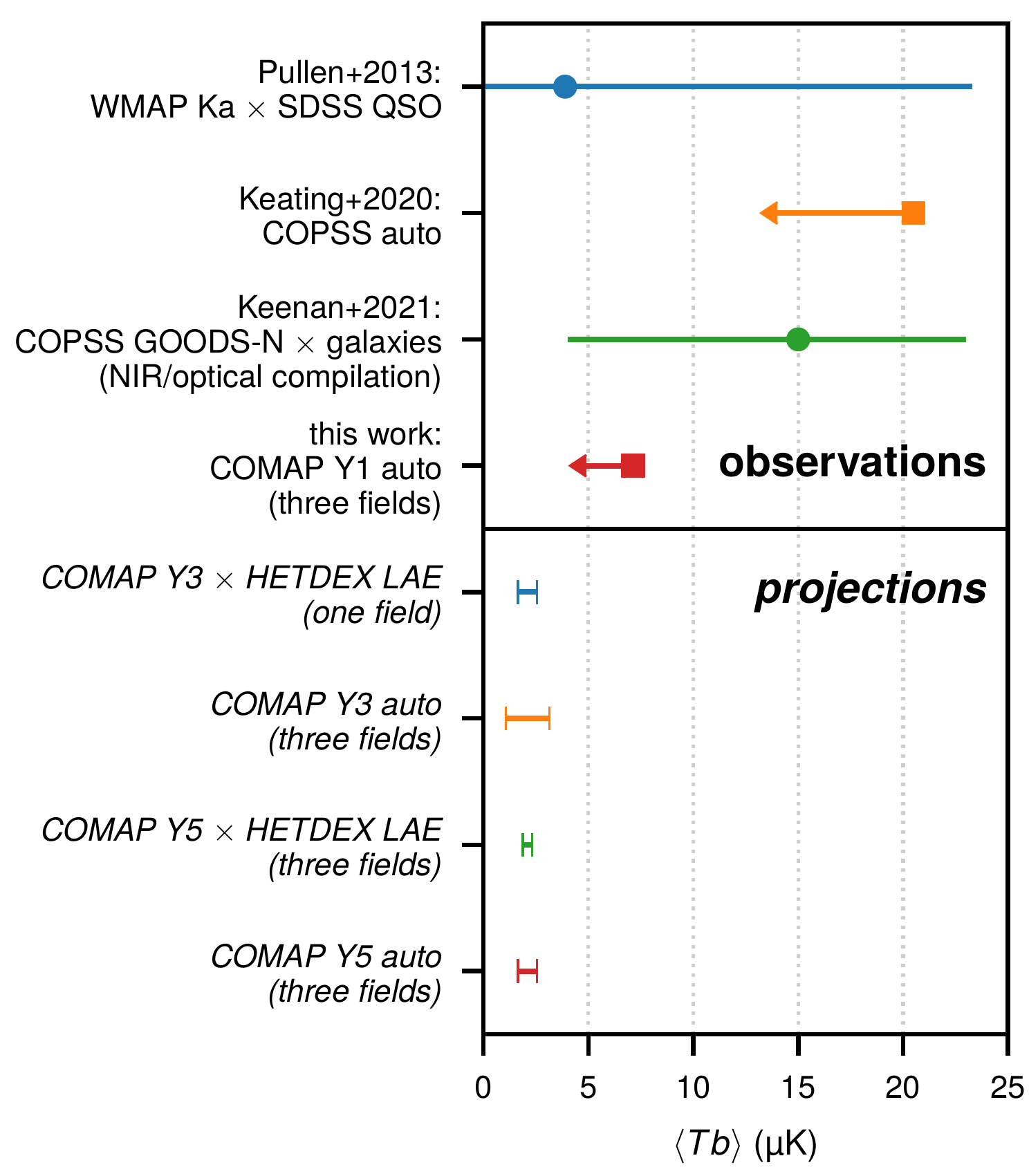}
    \caption{$\avg{Tb}$ constraints from previous observational analyses---the broadband cross-correlation of~\cite{Pullen13}, the auto-spectrum of~\cite{keating_etal_16}, and the 3D cross-correlation of~\cite{Keenan21}---alongside our current upper limit. We also show projections for future results based on COMAP auto- and CO--galaxy cross-spectra in subsequent years (cf.~\autoref{sec:hetdex_maintext}).}
    \label{fig:Tbconstraints}
\end{figure}

\begin{figure}[t!]
    \centering
    \includegraphics[width=0.96\linewidth]{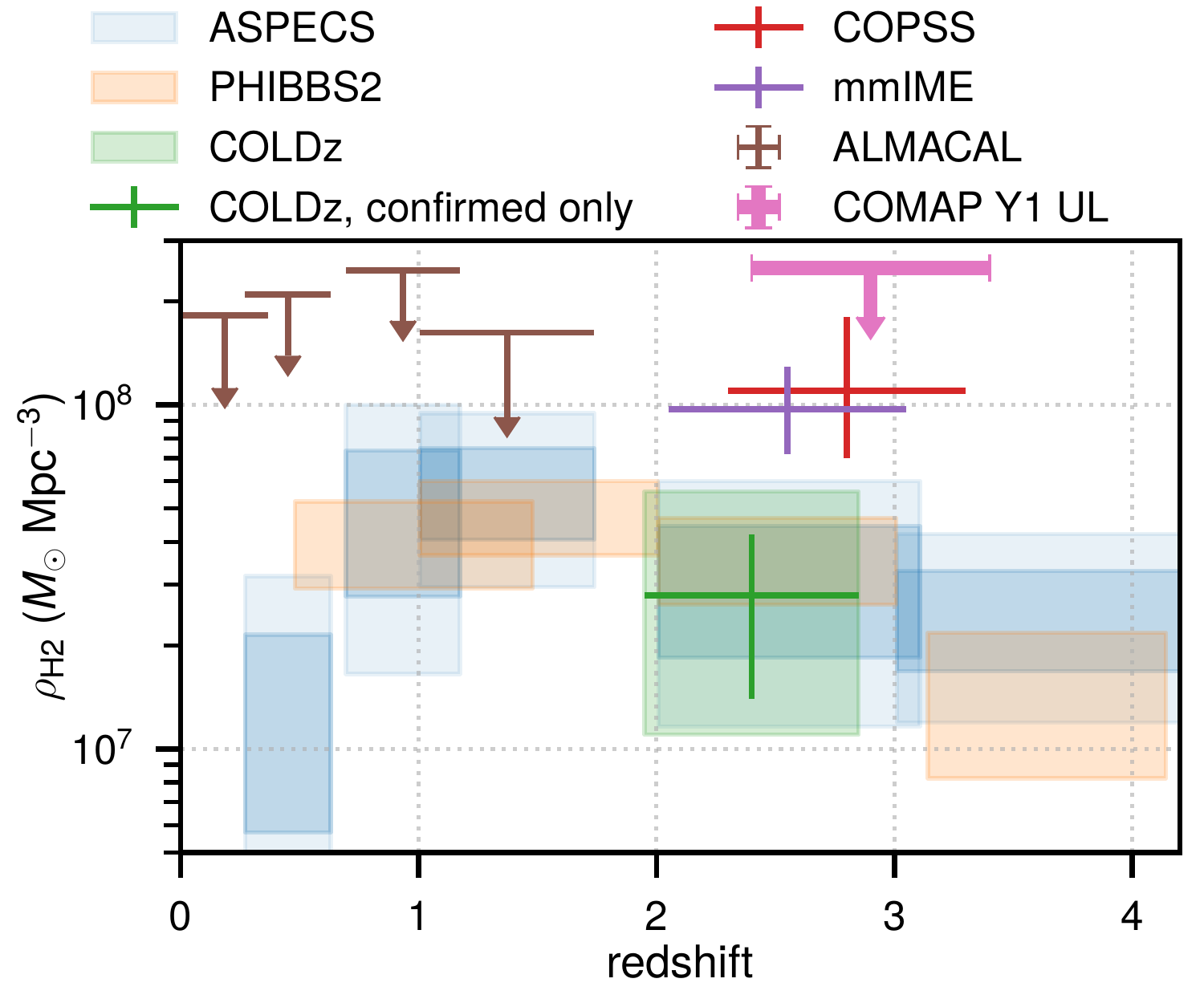}
    \caption{Current COMAP constraint on $\rho_\text{H2}$ (thick bar with downward arrow) in relation to past CO-based results from ASPECS~\citep{ASPECS-LPLF2}, PHIBBS2~\citep{PHIBBS2Lenkic}, COLDz~\citep[from which we show results based on either all line candidates or only those that have confirmatory independent spectroscopic measurements]{COLDzLF}, COPSS~\citep{keating_etal_16}, mmIME~\citep{mmIME-ACA}, and ALMACAL~\citep{Klitsch19}. All results use $\alpha_\text{CO}=3.6\,M_\odot\,($K\,km\,s$^{-1}$\,pc$^{-2})^{-1}$ except COPSS, which uses a conversion of $\alpha_\text{CO}=4.3\,M_\odot\,($K\,km\,s$^{-1}$\,pc$^{-2})^{-1}$.}
    \label{fig:rhoH2UL}
\end{figure}

We show COMAP--COPSS joint constraints from the informed method in~\autoref{fig:Tbshot_informed}; with this model the COMAP data also drive a clustering constraint of $\avg{Tb}^2\lesssim50\,\mu$K with or without COPSS data. The inferred actual $P_\text{shot}$ value\footnote{Unlike with the agnostic method, this value does not change significantly with the incorporation of COMAP data. The incorporation of line broadening into the informed method likely accounts for this fact. The COMAP data exclude very high values of $P_\text{shot}$ that would be consistent with COPSS data on account of attenuation from line broadening at COPSS wavenumbers, but not with the COMAP data at lower $k$. This exclusion suppresses the inferred $P_\text{shot}$ and cancels out the increase in inferred $P_\text{shot}$ from clustering amplitude limits (which was the sole effect of COMAP data on COPSS interpretation with the agnostic method).} of $1.2^{+0.7}_{-0.6}\times10^4\,\mu$K$^2\,$Mpc$^3$ is significantly higher than from our first method, and suggests that line broadening attenuates the COPSS measurement of shot noise by $\approx40\%$; this is entirely consistent with the median expectation for the CO(1--0) $P(k)$ around $k\sim1$ Mpc$^{-1}$ from the simulations of~\cite{linebroad}. That said, the upward correction merely reflects additional assumptions about line broadening rather than any added direct information.\added{ As a result, outside of this particular analysis we will not use this 40\% difference to correct any measurements or derived constraints from either~\cite{keating_etal_16} or~\cite{mmIME-ACA} (e.g., as shown in Figures~\ref{fig:Tbconstraints} through~\ref{fig:rhoH2future}).}

We also note a lack of sufficient sensitivity to further narrow our data-driven priors, even considering the clustering amplitude in isolation. Our upper limit for $\avg{Tb}^2$ corresponds to 13 times the value for the UM+COLDz+COPSS point estimate model, whereas the 68\% credibility interval for $\avg{Tb}^2$ for any of our data-driven priors already spans less than an order of magnitude, as~\autoref{fig:posteriorPk} suggests.

Our two analyses arrive at either an $A_\text{clust}$ constraint or a $\avg{Tb}^2$ constraint, but the two constraints are consistent with each other. Comparing the lower panel of~\autoref{fig:Tbshot_informed} with the estimate of $b$ as a function of $\avg{Tb}^2$ and $P_\text{shot}$ in~\autoref{sec:vbfits}, we can see that the parameter space preferred by the data tends to be associated with luminosity-averaged bias values of $b\sim3$ (although specific points in that space, like our point estimate models, may have even higher $b$). Then the COMAP--COPSS joint 95\% upper limit from~\autoref{tab:alltheconstraints} of $\avg{Tb}^2<51\,\mu$K should translate to an upper limit on the redshift-space clustering amplitude of $A_\text{clust}\approx\avg{T}^2(b^2+2b/3+1/5)\approx63$\,$\mu$K$^2$. This is within 10\% of the $A_\text{clust}$ upper limit obtained from our previous method, with differences likely arising from our simplified treatment of line bias and signal distortions.

We are more conservative about $b$ in deriving an upper limit on $\avg{T}$. For all of our priors, the sampled parameter sets all result almost entirely in $b>2$; a value of $b=2$ would be under the 3rd percentile for ``flat+COLDz'' and under the 1st percentile for the others. Most models in the literature also favour super-linear $L(M_h)$ relations at lower mass (with possible exceptions being older models like those of~\cite{Lidz11} and~\cite{Pullen13}, which had $L\propto M_h$) and thus fairly high values of $b$.

Combining the priors-based constraint of $b>2$ with our first method's limit on $A_\text{clust}=\avg{T}^2(b^2+2b/3+1/5)$, we would obtain $\avg{T}<3.5$\,$\mu$K. Combining $b>2$ with our second method's limit of $\avg{Tb}^2<51$\,$\mu$K$^2$ yields essentially the same limit (within 1\%) of $\avg{T}<3.6$\,$\mu$K. In either case, this result---which the COMAP data primarily drive---is currently the best LIM clustering constraint on the CO(1--0) $\avg{T}$ at $z\sim3$, outperforming by a factor of 3 the joint COPSS auto- and COPSS--galaxy cross-spectra analysis result of $\avg{T}<10.9\,\mu$K from~\cite{Keenan21}. We illustrate this improvement as well as the general history of constraints on $\avg{Tb}$---either from the CO auto-spectrum (via $\avg{Tb}^2$) or from a CO--galaxy cross-spectrum---in~\autoref{fig:Tbconstraints}.

\subsection{Derived Constraints on Molecular Gas Abundance}
\label{sec:ESconstraints_rhoH2}
The constraint on $\avg{T}$ directly translates into a constraint on the cosmic H$_2$ mass density $\rho_\text{H2}$. The conversion $\alpha_\text{CO}$ between H$_2$ mass (noting that here we do not deal with a gas mass density that includes heavier elements or atomic hydrogen) and CO luminosity is typically quoted with H$_2$ mass in intrinsic units of $M_\odot$ and CO luminosity in observer units of K\,km\,s$^{-1}$\,pc$^2$. Then at redshift $z$, given $\alpha_\text{CO}$ and the Hubble parameter $H(z)$,
\begin{equation}
    \rho_\text{H2} = \frac{\alpha_\text{CO}\avg{T}H(z)}{(1+z)^2}.
\end{equation}
At the COMAP central redshift of $z\approx2.8$, our upper limit of $\avg{T}<3.6$\,$\mu$K thus translates to an upper limit of $\rho_\text{H2}<2.5\times10^8\,M_\odot$\,Mpc$^{-3}$ given $\alpha_\text{CO}=3.6\,M_\odot\,($K\,km\,s$^{-1}$\,pc$^2)^{-1}$, which we use for easy comparison with other works that use the same conversion, such as~\cite{ASPECS-LPLF2},~\cite{PHIBBS2Lenkic}, and~\cite{COLDzLF}. We show our upper limit alongside these other works in~\autoref{fig:rhoH2UL}.

Since the COMAP upper limit for $\avg{Tb}^2$ was less stringent than our data-driven (COLDz-based) priors for the clustering power spectrum, we do not expect our upper limit on $\rho_\text{H2}$ to be more constraining than our priors either. Indeed the 90\% interval for $\rho_\text{H2}$ for our UM+COLDz+COPSS priors is given by $\log{[\rho_\text{H2}/(M_\odot\,\text{Mpc}^{-3})]}=7.58^{+0.23}_{-0.25}$, slightly higher than the COLDz standalone calculation from~\cite{COLDzLF} due to both UM and COPSS favoring a higher abundance of molecular gas, respectively through preference for a steep faint-end LF slope and through simply a higher measurement as shown in~\autoref{fig:rhoH2UL}. Therefore the 95th percentile value of $\rho_\text{H2}=9.5\times10^7\,M_\odot$\,Mpc$^{-3}$ from our fiducial priors sits at less than one-half of the COMAP upper limit.

On the other hand, the upper limit is still notable in relation to the other constraints shown in~\autoref{fig:rhoH2UL}. For one, we obtained this limit across a much wider area---on the order of square degrees---compared to the other surveys, which all operate across patches of $\sim$\,square arcminutes. The small volumes of these surveys can result in substantial cosmic variance and systematic biases not necessarily presently accounted for by their analyses, although the COPSS survey design (which spans multiple fields distributed widely across the sky, versus ASPECS and COLDz spanning one or two fields) should be less susceptible to these effects~\citep{Keenan20}.

Since the upper limit is within a factor of between two and three of the upper edge of our priors, the final COMAP Pathfinder measurement should indeed have constraining power beyond our priors, which we explore in~\autoref{sec:sim_inf}. By making use of up to 69 times more science-quality integration time (which would correspond to a map noise level lower by more than 8 times) than even the Field 1 CES-only results (which dominate our coadded CES-only sensitivity), five-year results from the COMAP Pathfinder should be on par with the other results shown in~\autoref{fig:rhoH2UL} and should act as an independent check on those measurements of $z\sim2$--3 $\rho_\text{H2}$. We will discuss expected five-year constraint on $\rho_\text{H2}$ in more quantitative detail later in this work (\autoref{sec:forecasts}).

As the present COMAP constraint and future expected constraints derive from directly measuring $\avg{Tb}$ as opposed to reconstructing $\avg{T}$ from individual detections or shot-noise measurements, they will serve the community as a strongly complementary probe of cosmic molecular gas density at $z\sim3$. In particular, we note that the results of~\cite{mmIME-ACA} depend strongly on models of the multiple overlapping CO lines encompassed by ALMA observing frequencies.

Incidentally, our upper limit also compares favorably to the ALMACAL upper limits of~\cite{Klitsch19} derived for $z\sim0$--2, from a blind search for CO absorption lines against background ALMA calibrators. The survey design for ALMACAL enables $>1500$ hours of integration time spanning a wide sky area---unusual for a community instrument and enabled only by the use of calibrator source observations. However, the ALMACAL approach cannot extend beyond $z\sim2$ due to the nature of ALMA calibrators, the majority of which appear to lie below $z\sim1.5$ with a small tail of the redshift distribution stretching out to $z\sim3$~\citep{Bonato18}. Thus, while molecular gas surveys not limited by cosmic variance are possible with ALMA through absorption line searches, these will not be able to survey the same redshifts as LIM or emission-line searches.

\section{Expectations for COMAP Pathfinder Future Science Results}
\label{sec:forecasts}
As~\cite{es_III} note in their Section 4.2, future observing seasons should improve the rate at which we acquire science-quality integration time through a combination of improvements in hardware, observing efficiency, and analysis. This implies that by the end of Year 5 of the Pathfinder campaign (Y5), sensitivity relative to the current Y1 power spectrum results of~\cite{es_IV} will improve not by a factor of 5, but by as much as a factor of 69 over the Field~1 Y1 result (which, as noted above, accounts for much of the current sensitivity). Of interest is how this final Pathfinder sensitivity will enable exclusion or detection not only of our fiducial UM+COLDz+COPSS model but also other models previously considered in the literature.

We first briefly discuss the expected raw detection sensitivity in~\autoref{sec:det_sign}, then simulate how this sensitivity will enable inferences about $z\sim3$ CO in~\autoref{sec:sim_inf}. Finally in~\autoref{sec:hetdex_maintext} we touch on possible science gains between now and Y5 results through cross-correlation with the Hobby-Eberly Telescope Dark Energy eXperiment~\citep[HETDEX;][]{HETDEX2008,HETDEX2021inst,HETDEX2021}.

\subsection{Current Predictions for Detection Significance}
\label{sec:det_sign}

\begin{figure*}
    \centering
    \includegraphics[width=0.96\linewidth]{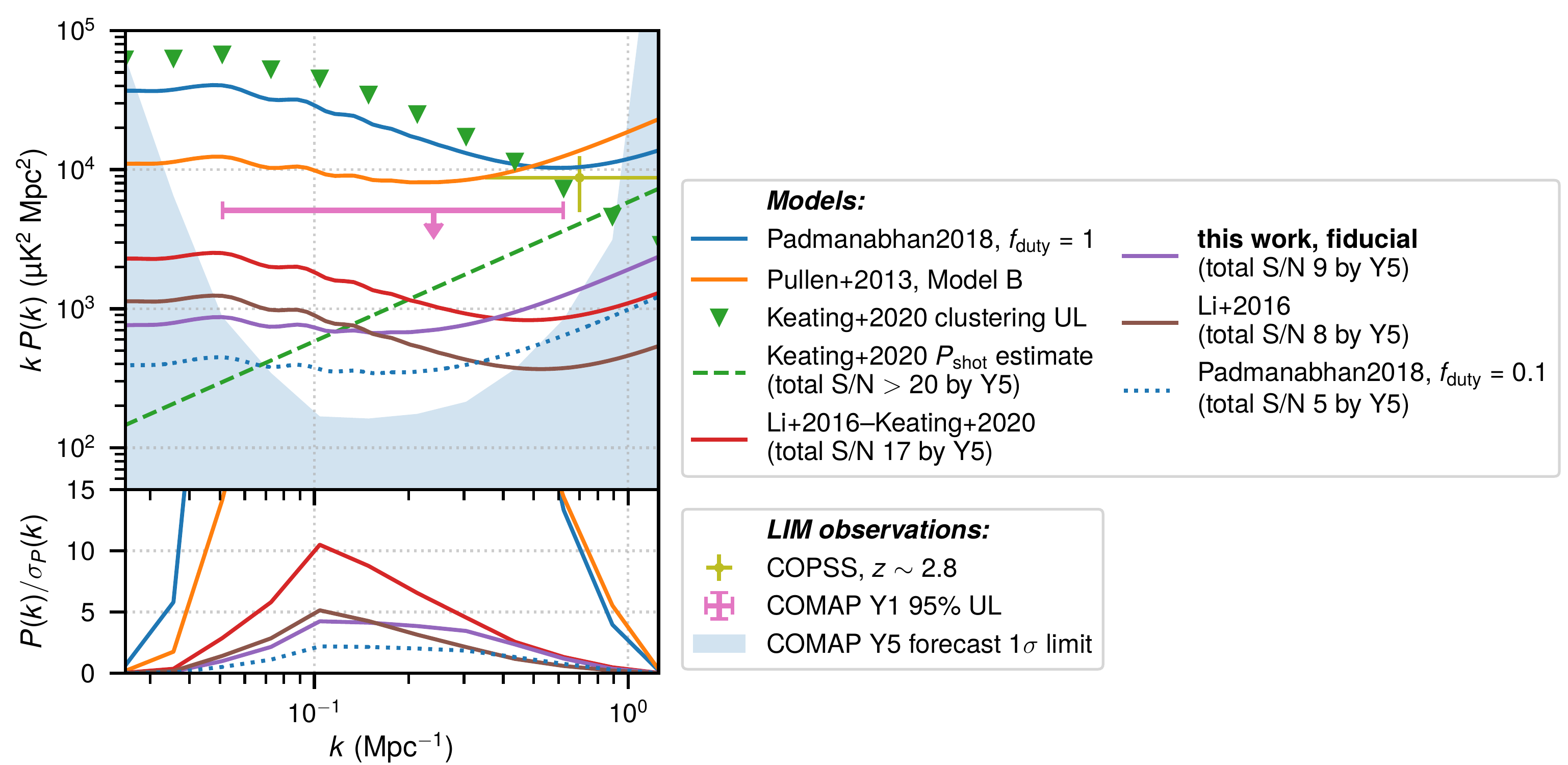}
    \caption{\emph{Upper panel:} The same models and COPSS interpretations from~\autoref{fig:PkSNR_Y1} shown in relation to our Y5 Pathfinder sensitivity forecast (blue shaded area). The legend also indicates the expected S/N with which we would reject the null hypothesis (i.e., excluding sample variance from the calculation). \emph{Lower panel:} S/N per $k$-bin of width $\Delta{[\log{(k\,\text{Mpc})}]}\approx0.16$ for each $P(k)$ prediction shown in the upper panel, accounting for attenuation from line widths in the presence of a beam with FWHM of $4\farcm5$ on sky.}
    \label{fig:PkSNR}
\end{figure*}

We show current and expected Pathfinder sensitivities (with the latter based on the aforementioned improvements forecast by~\citealt{es_III}) in~\autoref{fig:PkSNR}, alongside several models of the CO power spectrum. As our current sensitivity already excludes some of the models shown, as already considered in~\autoref{sec:exclusions}, we will not make signal-to-noise ratio (S/N) forecasts for those models.

We expect Y5 COMAP Pathfinder results to yield confident detections across multiple $k$-bins of other models yet to be excluded, including our own fiducial model, which would be detectable with an all-$k$ S/N of 9 (excluding sample variance). This level of sensitivity will allow COMAP data to discriminate clearly between several of the models shown.

\paragraph{Molecular gas constraints}

In~\autoref{sec:sim_inf} we rigorously consider how this detection, combined with characterisation of the VID, will enable inference of model parameter constraints and of the CO LF. Before we do this, however, we consider a quick Fisher forecast of expected constraints on $\avg{T}$ and thus on $\rho_\text{H2}$.

In addition to our fiducial model, which as we noted towards the end of~\autoref{sec:priors} is a conservative estimate by the very nature of data-driven priors based on direct detection measurements, we also consider the signal estimate derived from the empirical CO model of~\cite{mmIME-ACA}. This model, which we label ``\cite{li_etal_16}--\cite{mmIME-ACA}'' to distinguish it from the COPSS-based shot-noise estimate also calculated by~\citealt{mmIME-ACA}, is also one of the primary models that~\cite{es_VII} use for COMAP forecasts beyond the Pathfinder. The model borrows the general approach of~\cite{li_etal_16}, which composes the simulation- and data-driven halo mass--SFR connection from~\cite{Behroozi13a,Behroozi13b} with an empirical IR--CO luminosity fit, but uses newer (albeit exclusively local) IR--CO correlation fits from~\cite{Kamenetzky16}. The predicted CO(1--0) $\avg{T}$ at the COMAP central redshift of $z=2.8$ is $1.3\,\mu$K, which is several times higher than our fiducial COLDz-driven conservative prediction of $0.5\,\mu$K, owing to significant differences in the faint end of the $L(M_h)$ relation and thus the faint-end slope of the CO LF. Under this model, a Y5 power spectrum analysis would reject the null hypothesis at an all-$k$ S/N of 17.

We run a Fisher analysis across the parameters $\{\avg{T},b,P_\text{shot},v_\text{eff}\}$, imposing loose Gaussian priors around the central line bias and $v_\text{eff}$ values with width $\sigma[b]=1$ and $\sigma[v_\text{eff}]=120$\,km\,s$^{-1}$ (mostly to keep both away from negative values). The applicable central parameter values for our fiducial model are $\{0.52\,\mu\text{K},4.0,1.9\times10^3\,\mu\text{K}^2\,\text{Mpc}^3,330\,\text{km\,s}^{-1}\}$, and the same for the \cite{li_etal_16}--\cite{mmIME-ACA} model are  $\{1.3\,\mu\text{K},2.7,9.7\times10^2\,\mu\text{K}^2\,\text{Mpc}^3,210\,\text{km\,s}^{-1}\}$.

The forecast suggests that the primary parameter being constrained in this exercise is $\avg{T}$, with expectations of $(0.52\pm0.14)\,\mu$K for the conservative fiducial model and $(1.3\pm0.4)\,\mu$K for the more optimistic \cite{li_etal_16}--\cite{mmIME-ACA} model. Through the same conversion as we used in~\autoref{sec:ESconstraints_rhoH2}, these $\avg{T}$ constraints respectively translate into $\rho_\text{H2}$ constraints of $(3.6\pm0.9)\times10^7\,M_\odot$\,Mpc$^{-3}$ and $(9.0\pm2.7)\times10^7\,M_\odot$\,Mpc$^{-3}$, as shown in~\autoref{fig:rhoH2future}.

However, our priors around the CO model are fairly loose, whereas some real-world analyses like those of~\cite{mmIME-ACA} or some Fisher analyses like those of~\cite{es_VII} make stronger assumptions about the shape of the $L(M_h)$ relation---which then completely determines at least the line bias---and constrains only the overall normalisation of $L(M_h)$. In our Fisher forecast's parameter space this would be equivalent to imposing very narrow priors on $b$. If we keep the same prior width for $v_\text{eff}$ but narrow the width for the bias prior to $\sigma[b]=0.1$, we would obtain constraints around the \cite{li_etal_16}--\cite{mmIME-ACA} model of $\avg{T}=(1.3\pm0.07)\,\mu$K and $\rho_\text{H2}=(9.0\pm0.5)\times10^7\,M_\odot$\,Mpc$^{-3}$. We show the latter also in~\autoref{fig:rhoH2future}.

Finally, as these forecasts use the CO power spectrum alone, additional information from the VID and even from cross-correlations would further improve these constraints.

\setcounter{figure}{8}
\begin{figure*}[t!]
    \centering
    \includegraphics[width=0.84\linewidth,clip=True,trim=0 5mm 0 0]{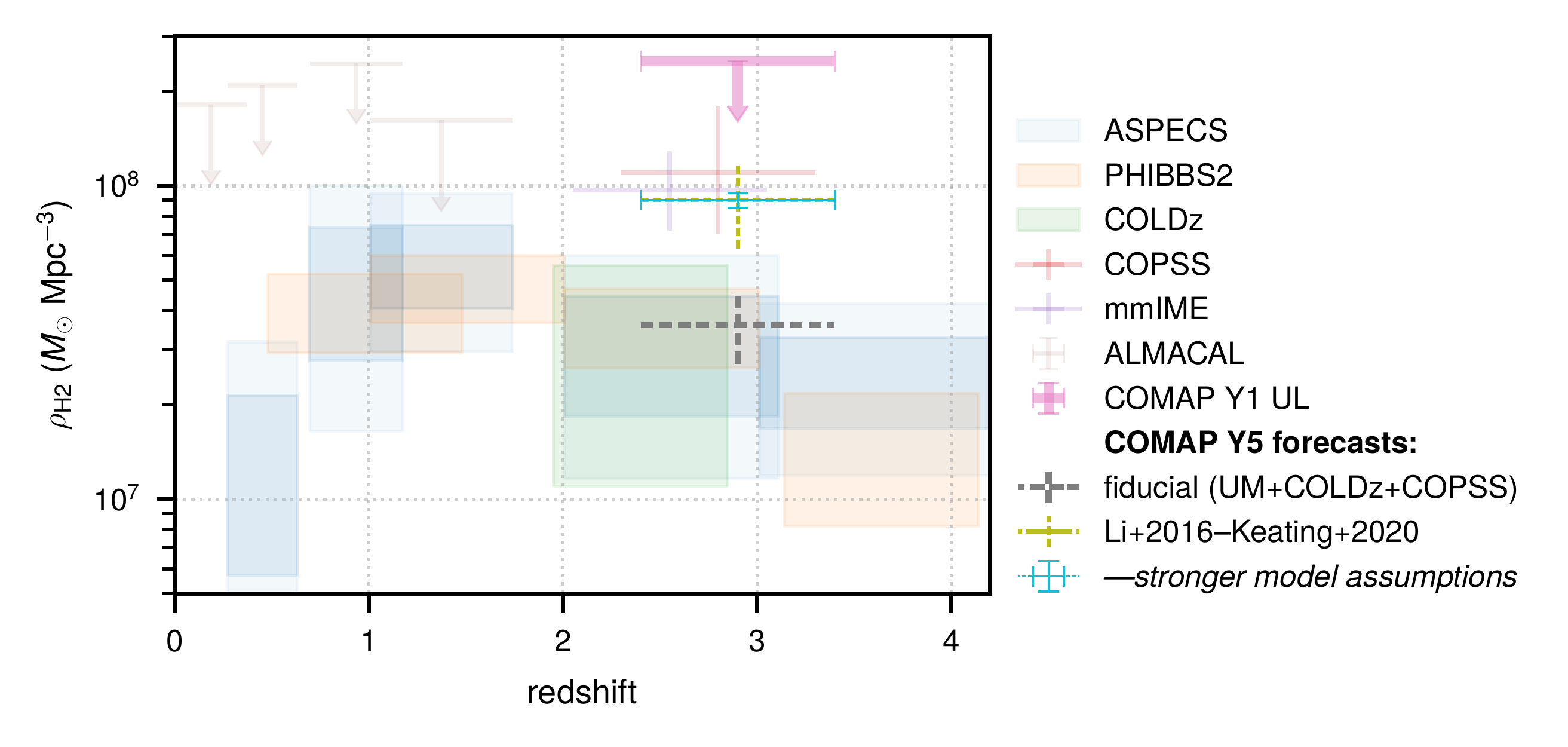}
    \caption{Forecast constraints on $\rho_\text{H2}$ for COMAP Y5 power spectrum analysis, alongside the same constraints from~\autoref{fig:rhoH2UL}. We show forecasts for both the fiducial model in this work and a more optimistic model from~\cite{es_VII}, which is more consistent with current line-intensity mapping measurements. For the latter we also show a forecast constraint with stronger line-bias priors, better mirroring highly informed analyses like that of~\cite{mmIME-ACA}.}
    \label{fig:rhoH2future}
\end{figure*}

\subsection{Simulated Inferences}
\label{sec:sim_inf}

In~\autoref{sec:model} we developed a new parameter set to describe the halo--CO connection, estimated a set of priors for these parameters, and discussed an accurate method to take into account the effect of CO line widths. Now also equipped with predictions for Y5 sensitivities, we can go on to the question of how we could use our model methods to infer constraints from the COMAP experiment. 

Following \citet{Ihle19}, but using the model developed in~\autoref{sec:model}, we run an MCMC inference from simulated data to forecast constraints on astrophysical observables like the LF, $\phi(L)$, as well as posterior distributions of our parameters from~\autoref{sec:model}, $\theta = \{A, B, \log C, \log (M/M_\odot), \sigma\}$. This inference uses both the CO $P(k)$ and the VID in a joint analysis that accounts for covariance between all observables, as first considered by~\cite{Ihle19}.

We focus here on the results of the simulated MCMC inference, but provide further details on the MCMC setup, including the exact priors and survey parameters assumed, in~\autoref{sec:mcmc_appendix}. Broadly speaking, the noise level assumed corresponds to Y5 sensitivity projections already discussed in~\autoref{sec:det_sign}, and the signal simulation uses the fiducial point estimate model (UM+COLDz+COPSS) defined in~\autoref{tab:priorpoints}. The results shown here are from one MCMC run (i.e. one signal and noise realization) and will change somewhat from realization to realization. 

\begin{figure}[t!]
    \centering
    \includegraphics[width=0.96\linewidth]{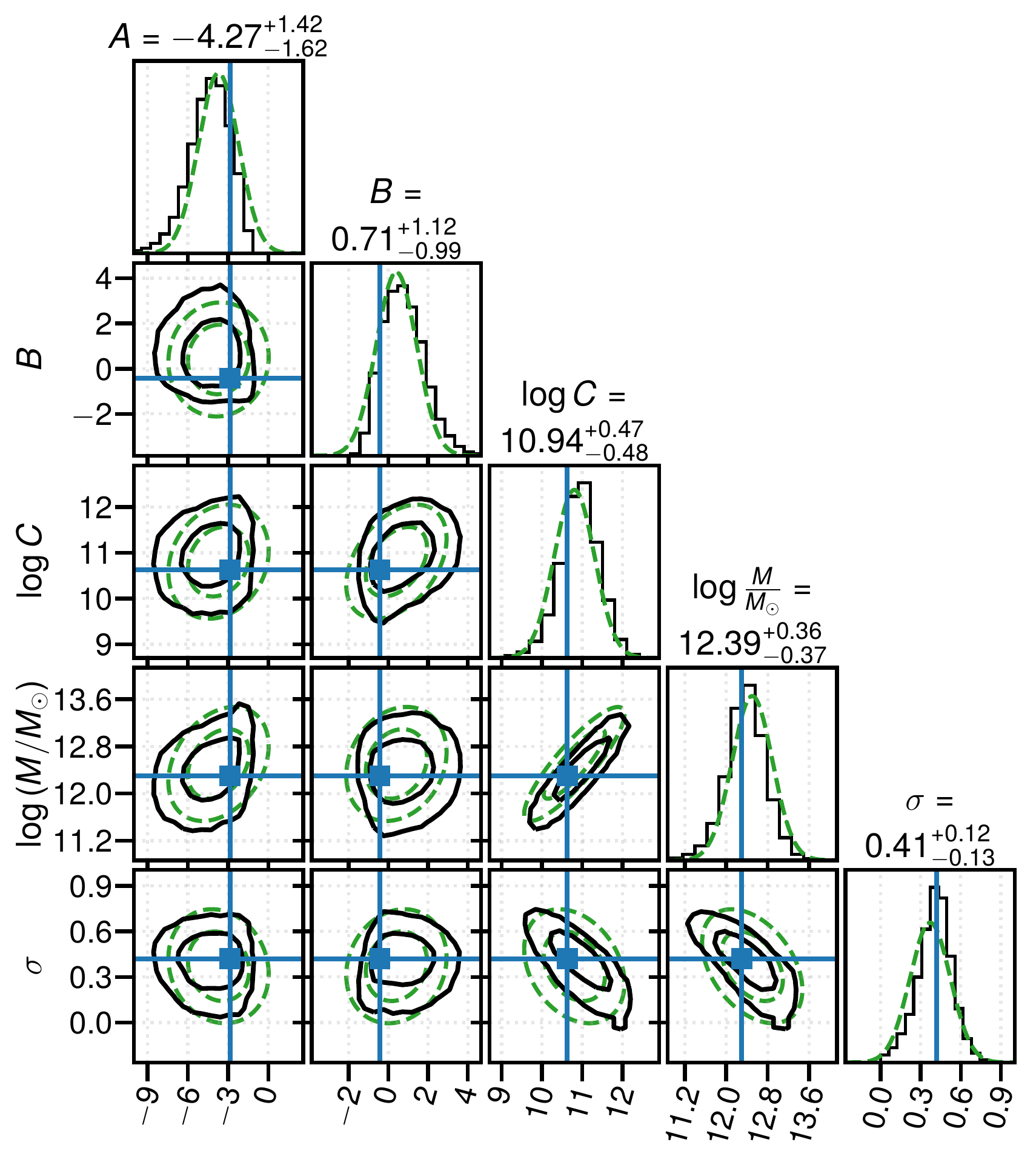}
    \caption{Forecasted posterior distributions for a single realization of the COMAP five-year experiment, given the UM+COLDz+COPSS model parameters from~\autoref{tab:priorpoints}. Blue points in cross hairs show the model point estimates used for the simulated input signal. Black curves outline the posterior distributions, while the green dashed lines correspond to the (Gaussian) priors used for the MCMC. The two curves of each color correspond to 68\% and 95\% credibility regions respectively. The numbers on top of each column correspond to the 68\% credibility interval for each parameter.}
    \label{fig:par_corner}
\end{figure}

\autoref{fig:par_corner} shows the posterior distribution of all the individual model parameters resulting from one MCMC simulated inference run. Comparing the posterior (black curves) to the prior (green dashed curves) we see modest but clear shifts and tightening of the distributions. The simulated COMAP data constrain the power-law slopes, with 95\% limits of $A<-2.1$ and $B>-0.78$, bounding $L(M_h)$ from above in both cases. The data also tightens the probability distributions projected in the $\sigma$--$\log{(M/M_\odot)}$ and in the $\sigma$--$\log C$ planes. This would appear to chiefly reflect information from the VID on the high-luminosity end of the LF, as the anticorrelation of $\sigma$ with both $\log{(M/M_\odot)}$ and $\log{C}$ largely affects predicted abundances of CO emitters beyond the knee of the LF. Overall, the comparison betewen posterior and prior distributions shows that even when including COLDz and COPSS detections in the prior, COMAP improves the constraints on the model.

\begin{figure}[t!]
    \centering
    \includegraphics[width=0.96\linewidth]{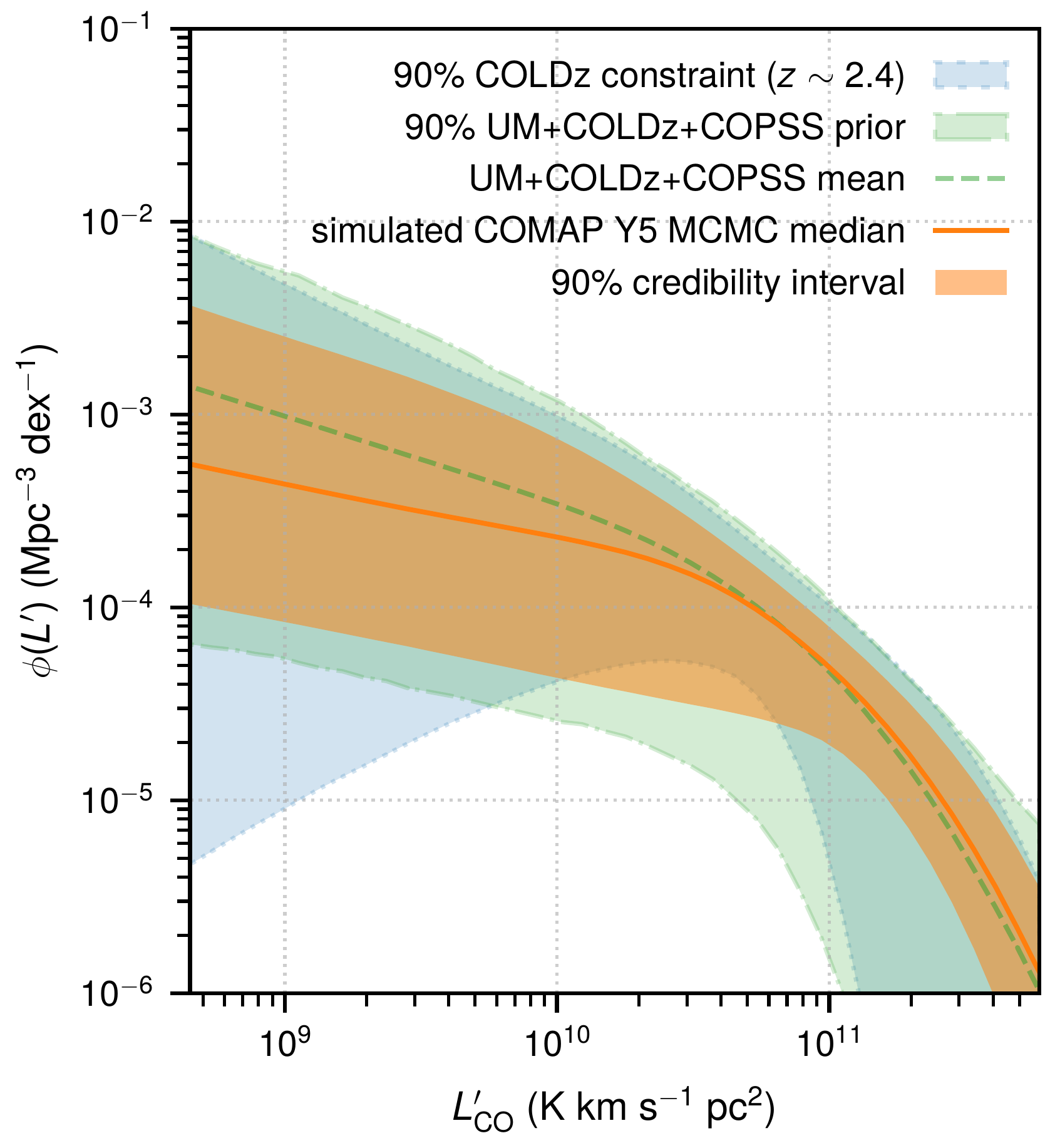}
    \caption{Constraints on the LF from the same single realization of the COMAP 5-year experiment as used for~\autoref{fig:par_corner}. The orange shaded area and solid curve represent the MCMC 90\% credibility interval and median. We also show 90\% intervals given by the COLDz constraints of~\cite{COLDzLF} at $z\sim2.4$ (shaded area bounded by dotted curves), and by the UM+COLDz+COPSS data-driven prior when applied to our $z\sim2.8$ simulations (shaded area bounded by dashed-dotted curves). The green dashed line corresponds to the ensemble mean LF of the UM+COLDz+COPSS model producing the input signal.}
    \label{fig:lum_constr}
\end{figure}

The LF constraints, in~\autoref{fig:lum_constr}, show that even though the improvement of the parameter constraints appeared modest, the LF is significantly more constrained using COMAP compared to the prior (based on COLDz and COPSS), especially at the high-luminosity end. This in turn will correspond to significantly improved measurements of integrated and derived quantities like the previously discussed $\rho_\text{H2}$.

\subsection{HETDEX Cross-correlation Expectations}
\label{sec:hetdex_maintext}

We have considered prospects for cross-correlation between CO intensity maps from COMAP and Lyman-alpha emitter (LAE) data from HETDEX in other works by~\cite{hetdex_selfcite} and~\cite{silva_etal_21}. However,~\cite{hetdex_selfcite} presented cross-spectrum forecasts well before we could characterise real-world performance of the COMAP Pathfinder instrument and data pipeline, and~\cite{silva_etal_21} consider a very detailed LAE model but solely in the context of voxel-level analyses.

Detailed models of the CO--LAE cross-spectrum are beyond the principal scope of our early science papers, which concern themselves with detection and interpretation of CO intensity mapping observations by themselves. However, we note that based on our fiducial CO model and current expectations of LAE bias and number density, we expect to reach an all-$k$ S/N of 7 on the CO--LAE cross-spectrum even with only Year 3 (Y3) data in hand for only Field 1 (whereas we will need all data through Y5 to achieve a similar S/N for the CO auto-spectrum). Fisher forecasts (in the style of, e.g.,~\citealt{BreysseAlexandroff19}) suggest that even with relaxed priors on CO bias and line broadening compared to our assumptions from earlier, this Y3 single-field cross-correlation detection should allow for a constraint of $\avg{Tb}_\text{CO}=2.1\pm0.4$---a $\approx5\sigma$ result. This contrasts with an upper limit of $\avg{Tb}_\text{CO}<5\,\mu$K with only the CO auto $P(k)$ in the same field, or a marginal $2\sigma$ result of $\avg{Tb}_\text{CO}=2.1\pm1.0$ coadding auto $P(k)$ measurements across all three fields.

The constraints from cross and auto $P(k)$ would respectively improve to $\avg{Tb}_\text{CO}=2.1\pm0.2$ and $\avg{Tb}_\text{CO}=2.1\pm0.5$ with the full three-field Y5 data and completely overlapping HETDEX LAE survey coverage in hand. That said, our forecasts suggest that HETDEX data would enable strong constraints on the CO clustering amplitude in advance of Y5, as we illustrate graphically in~\autoref{fig:Tbconstraints} alongside current observational constraints. We refer the reader to~\autoref{sec:hetdex_appendix} for further details on these simple forecasts.

Furthermore, as previous intensity mapping works have shown~\citep{Switzer13,Keenan21}, cross-correlation constraints show strong robustness against systematics present in intensity mapping data. Whether this may relax our data selection requirements in the context of cross-correlation analyses will be the subject of future work, in which we also hope to mirror the more detailed LAE model of~\cite{silva_etal_21} in larger cosmological simulations like the peak--patch simulations used for~\autoref{sec:sim_inf}.
\section{Conclusions}
\label{sec:conclusions}
This paper synthesizes model updates and early COMAP Pathfinder data to answer the following key questions:
\begin{itemize}
    \item \emph{What inferences do our early science verification data enable about the $z\sim3$ CO(1--0) power spectrum, and molecular gas abundance?} Our current result of $\avg{Tb}^2\lesssim50$ $\mu$K$^2$ already excludes certain models directly in the clustering regime, and places a much stronger upper limit on the clustering amplitude of the CO power spectrum than COPSS. In addition, our upper limit is consistent with and readily complements existing constraints on $\rho_\text{H2}$ at $z\sim3$.
    \item \emph{Given early science sensitivities and updated $z\sim3$ models, what are our present expectations for constraints on these same quantities, and others like the CO LF, at the end of five years of COMAP Pathfinder observations?} We expect a detection of the $z\sim3$ CO power spectrum to enable clear discrimination between different models from existing literature that predict different degrees of contribution of faint emitters to the total $P(k)$. For our conservative fiducial data-driven model we forecast an all-scale S/N of 9. Such a firm detection would also enable significant constraining power on the CO LF beyond our priors that conventional direct-detection surveys have not been able to offer, and a measurement of cosmic molecular gas abundance that will be a strong independent check on results from other surveys.
\end{itemize}

These promising early results are possible due to the quality of the COMAP Pathfinder data at the present time, which are entirely consistent with uncorrelated white noise with any systematics successfully suppressed below white noise through data cuts. With further integration time we fully expect the COMAP Pathfinder to detect \emph{an} excess power spectrum over white noise. The key question is whether this excess will be uncharacterized contamination or we will be able to attribute it to the CO signal we are targeting, which Pathfinder Y5 sensitivities should be sufficient to detect and even distinguish between many currently viable CO models as shown in~\autoref{sec:forecasts}. We can only be confident in the interpretation of such an excess through continued technical improvements, not only in mapmaking and power spectrum derivation but also in forward models of the signal.

With future work we also hope to present significant improvements not only in confidence of interpretation of the COMAP data, but also in qualitative range of possible constraints through cross-correlation with external datasets, through both simple power spectrum cross-correlation and voxel-level analyses that will provide high information content around redshift evolution of CO emission and molecular gas content~\citep{silva_etal_21}.

\begin{acknowledgements}

This material is based upon work supported by the National Science Foundation under Grant Nos.~1518282, 1517108, 1517598, 1517288, and 1910999; by the Keck Institute for Space Studies under ``The First Billion Years: A Technical Development Program for Spectral Line Observations''; and by a seed grant from the Kavli Institute for Particle Astrophysics and Cosmology. 

DTC is supported by a CITA/Dunlap Institute postdoctoral fellowship. The Dunlap Institute is funded through an endowment established by the David Dunlap family and the University of Toronto. The University of Toronto operates on the traditional land of the Huron-Wendat, the Seneca, and most recently, the Mississaugas of the Credit River; DTC is grateful to have the opportunity to work on this land. PCB is supported by the James Arthur Postdoctoral Fellowship at New York University. KC, JWL, ACSR, BDU, and DPW acknowledge support from NSF Awards 1518282 and 1910999. Work at the University of Oslo is supported by the Research Council of Norway through grants 251328 and 274990, and from the European Research Council (ERC) under the Horizon 2020 Research and Innovation Program (Grant agreement No.\ 819478, \textsc{Cosmoglobe}). HP acknowledges support from the Swiss National Science Foundation through Ambizione Grant PZ00P2\_179934. JOG acknowledges support from the Keck Institute for Space Science, NSF AST-1517108, and the University of Miami. SEH acknowledges support from an STFC Consolidated Grant (ST/P000649/1). LCK was supported by the European Union's Horizon 2020 research and innovation program under the Marie Sk\l{}odowska-Curie grant agreement No.~885990. JK is supported by a Robert A.~Millikan Fellowship from Caltech.

We thank Riccardo Pavesi for access to the COLDz ABC posterior sample used in this work. This research made use of NASA's Astrophysics Data System Bibliographic Services. Some of the computing for this project was performed on the Sherlock cluster. DTC would like to thank Stanford University and the Stanford Research Computing Center for providing computational resources and support that contributed to these research results. The Scientific color maps \texttt{acton} and \texttt{tokyo}~\citep{Crameri2018} are used in this study to prevent visual distortion of the data and exclusion of readers with color-vision deficiencies~\citep{Crameri2020}.\added{

Finally, we would like to thank an anonymous referee whose comments and suggestions significantly improved this manuscript.}
\end{acknowledgements}
\software{\texttt{hmf}~\citep{hmf}; Matplotlib~\citep{matplotlib}; \added{\texttt{corner.py}~\citep{cornerpy}; }Astropy, a community-developed core Python package for astronomy~\citep{astropy}.}

\appendix
\section{Details of CO Model Prior Formulation}
\label{sec:model_appendix}
Throughout this section, we examine potential ways to inform our CO model priors. First we consider what information we can incorporate from the model papers of~\cite{li_etal_16} and~\cite{Behroozi19}; then we consider $z\sim2$--3 CO(1--0) observations in the past several years and how they can further refine our priors.

Note that for this section only, we use a slightly different cosmology for consistency with~\cite{Behroozi19}, which uses the cosmology of~\cite{Planck15}, such that $\Omega_m = 0.307$, $\Omega_\Lambda = 0.693$, $\Omega_b =0.047$, $h=0.678$, $\sigma_8 =0.823$, and $n_s=0.96$. Differences in cosmological quantities like $H(z)$ and comoving distance are around or less than 1\% at COMAP redshifts, and while the higher $\Omega_m$ will likely result in a $\sim10$\% difference in predicted halo abundances versus our fiducial cosmology, this is a much smaller relative uncertainty than many of our other model uncertainties, including the uncertainties surrounding some observational constraints.

\subsection{Initial Prior Setup from Previous Models}
\label{sec:initialpriors}
The new parameters $\{A,B,C,M\}$ in the parameterization of~\autoref{sec:paramet} are expressible in terms of the parameters used in the scaling relations that have gone into this functional form (again, under the approximation of $\alpha\approx1$\added{ or at least $\alpha\not\ll1$ and $\alpha\not\gg1$}):\begingroup\allowdisplaybreaks
\begin{align}A &= 0.3\alpha_\text{UM}/\alpha;\\
B &= 0.3\beta_\text{UM}/\alpha;\\
\log{C} &= ({10-\log\delta_\text{MF}-\beta+\log\epsilon})/{\alpha};\\
\log{(M/M_\odot)}&=\log{M_\text{200 km/s}}+({10}/{3})\log{\left({V}/{200}\right)}.
\end{align}\endgroup
Then we can propagate through the above equations the priors on $\alpha$, $\beta$, and $\delta_\text{MF}$ from~\citet{li_etal_16}---$\alpha=1.17\pm0.37$, $\beta=0.21\pm3.74$, and $\log\delta_\text{MF}=0.0\pm0.5$---and the 68\% interval around the best-fit values of the other parameters from~\citet{Behroozi19}. (We used the best-fit model from the Early Data Release; we do not consider the changes between this and the official Data Release 1 large enough to recalculate our priors.) The model of~\citet{Behroozi19} is redshift-dependent, but here we fix $z=2.4$, to match the median redshift of the COLDz survey. There should be relatively little evolution in CO abundances and thus the power spectrum between $z=2.4$ and the COMAP central redshift of $z=2.8$ (certainly little more than a factor of 2 or so, less than the current level of uncertainty in models of the signal).

The resulting initial priors on $\{A,B,C,M\}$ are
\begin{align}A&= -1.66\pm2.33,\\B&= 0.04\pm1.26,\\\log{C}&= 10.25\pm5.29,\\\log{(M/M_\odot)}&= 12.41\pm1.77.\end{align}
The central values for these priors do not change significantly across the COMAP redshift range, at least compared to the widths of the priors. We also set an initial prior of $\sigma=0.4\pm0.2$ (dex), which takes the central value from the 0.37 dex total scatter in the~\citet{li_etal_16} fiducial model, and assumes a slightly broader prior than that model would have prescribed.

We consider several alternate sets of initial priors on the model parameters, depending on how confident we think we can be in various pieces of information in the literature. Thus we have, as in~\autoref{tab:allthepriors},
\begin{itemize}
    \item a conservative set of ``flat'', uninformative priors;
    \item an informed set of priors (used for the fiducial model) deriving from empirical models of the galaxy--halo connection as described above;
    \item and an extrapolation-heavy set of priors that derive from calculating the best-fit parameters and errors of the~\cite{Padmanabhan18} model at $z=2.4$ (``P18''), which builds in a range of $z=0$--3 data including LF constraints from COPSS.
\end{itemize}
The names of these initial priors act as prefixes for our data-driven priors, as they represent information unconditioned on observational data.

\begin{deluxetable}{rccccc}
    \tablecaption{Initial Parameter Priors for the CO(1--0) Model\label{tab:allthepriors}}
    \tablehead{\colhead{Prior Prefix} & \multicolumn{5}{c}{Initial Priors on:} \\
        \colhead{}&\colhead{$A$}&\colhead{$B$}&\colhead{$\log{C}$}&\colhead{$\log{(M/M_\odot)}$}&\colhead{$\sigma$}}
    \startdata
        ``flat'' & $\mathcal{U}(-18,9)$ & $\mathcal{U}(-18,9)$ & $\mathcal{U}(5,15)$ & $\mathcal{U}(10,15)$ & $\mathcal{U}(0,1)$\\
        \textbf{``UM''} & $\mathcal{N}(-1.66,2.33)$ & $\mathcal{N}(0.04,1.26)$ & $\mathcal{N}(10.25,5.29)$ & $\mathcal{N}(12.41,1.77)$ & $\mathcal{N}(0.4,0.2)$\\
        ``P18'' & $\mathcal{N}(-2.29,0.52)$ & $\mathcal{N}(-0.57,0.36)$ & $\mathcal{N}(10.59,0.70)$ & $\mathcal{N}(11.79,0.64)$ & $\mathcal{N}(0.4,0.2)$
    \enddata
\end{deluxetable}

\subsection{Observational Constraints on High-redshift CO(1--0)}
\label{sec:priorobs}
As reviewed by~\cite{CW13}, CO observations at high-redshift in general are not especially novel, with hundreds of detections at $z\gtrsim1$. However, a complication is that many of these detections---certainly the ``main sequence'' or ``normal'' star-forming galaxies surveyed by~\cite{Daddi10} and~\cite{Tacconi10}---are in CO(2--1) or CO(3--2) (if not higher-$J$ CO lines), whereas we want to specifically consider CO(1--0) emission. In any case, we have already folded information from all the detections reviewed by~\cite{CW13} owing to the fact that their values for $\alpha$ and $\beta$ are one of four results incorporated into the~\cite{li_etal_16} priors on these parameters.

While the CO LF was not constrained beyond $z=0$ at the time of the~\cite{CW13} review, several major projects have taken place to directly measure the CO LF at redshifts that COMAP will survey. We consider each of these and our rationale for incorporating or not incorporating them into our priors.

\paragraph{ASPECS}
As a molecular line scan survey, ASPECS searches for CO line emitters in a deep interferometric data cube without external pre-selection. The latest iteration is a Large Programme (LP) on ALMA~\citep{ASPECS-LP} covering 4.6 square arcminutes---roughly five times the area of its pilot precursor~\citep{ASPECSPilot}---and the observations in ALMA Band 3 (84--115 GHz) cover CO(3--2) emission at $z\sim2.0$--3.1 as well as lower-$J$ (or higher-$J$) CO lines at lower (or higher) redshift.

While ASPECS LP does constrain the CO LF at COMAP redshifts, we choose not to incorporate these results into our priors for the simple reason that the observations at COMAP redshifts are in CO(3--2) and not CO(1--0). Initial inferred CO(1--0) LF estimates presented by~\cite{ASPECS-LPLF} relied on specific assumptions about CO line excitation, including a line luminosity ratio of $L'_\text{CO(3--2)}/L'_\text{CO(1--0)}=0.42\pm0.07$ taken from~\cite{Daddi15}, which averaged line ratios from three near-IR selected ``normal'' star-forming galaxies at $z=1.5$. While the uncertainties around this ratio were incorporated into the inference of the CO(1--0) LF, in hindsight the quoted uncertainties are severe underestimates of the probable error of the nominal value with respect to the global ratio at $z\sim2.5$. Of the four CO(3--2) detections from~\cite{ASPECS-LP}, three were observed robustly in CO(1--0) in VLA data by~\cite{VLASPECS}, and the line ratios were found to be closer to 0.8--1.1. Further work by~\cite{ASPECS-LPCO} yielded CO excitation models that favoured an average line luminosity ratio of $0.80\pm0.14$---almost twice the original fiducial value used---that was then used for updated LF constraints by~\cite{ASPECS-LPLF2}. The revised value resulted in estimates of luminosity densities and thus molecular gas abundances at roughly half of what was presented by~\cite{ASPECS-LPLF}.

Such significant changes in the presentation of the ASPECS LP results in the span of two years strongly demonstrate both the uncertainty and possible variance in CO excitation across the population of high-redshift galaxies being surveyed. Due to this large uncertainty, we forgo using inferred constraints on $z\sim2$--3 CO(1--0) from ASPECS.
\paragraph{COLDz}
The CO Luminosity Density at High-$z$ (COLDz) survey~\citep{COLDz} is also a molecular line survey, but is in the COSMOS and GOODS-N fields, and uses Ka-band VLA observations at $z=2.0$--2.9 altogether covering almost 60 square arcminutes. The measurement is more directly applicable to our context, as it measures CO(1--0) line emission rather than a higher-$J$ CO line. While the survey only identifies four secure (independently confirmed) line candidates across both fields at $z\sim2$--3, the LF calculation also incorporates a catalogue of line candidates that have not been independently confirmed, many of which do not have a spatially coincident counterpart in optical or near-infrared imagery.

The possibility of spurious line detections should in principle only discourage interpreting each line candidate individually (which~\citealt{COLDz} explicitly do when presenting their non-secure line candidates). However, the understanding of what line candidates should be considered ``reliable'' and which should not continues to evolve. For instance, in the case of ASPECS, between the pilot and large surveys, the requirement on the fidelity of a source (essentially the probability that the source is a genuine line detection rather than a noise peak) to be considered for analysis evolved from 60\% to 90\%. However, of the eight sources (across all CO lines and redshifts) identified by the pilot survey~\citep{ASPECSPilot} in the overlapping area between the pilot and large surveys, only the four sources with identified optical or near-IR counterparts had above 90\% fidelity (simply because having a counterpart meant the fidelity was 100\%). The other four sources had no counterparts, had below 90\% fidelity, and were not recovered by the large survey. Therefore, whether 90\% fidelity is a sufficient threshold to exclude spurious detections remains an open question.

Given the complexities in understanding which sources identified by a molecular line scan like ASPECS or COLDz are spurious, this might discourage using even statistical LF constraints from these surveys. However, it is worth noting that even if the ASPECS-Pilot analysis incorporated spurious sources as a significant fraction of its statistical sample, its $z\sim2.6$ CO(3--2) LF measurement~\citep{ASPECSPilotLF} is actually largely consistent with the ASPECS LP measurement~\citep{ASPECS-LPLF}. Therefore, as purity (along with completeness and other various sources of error and uncertainty) is given due accounting in these analyses, we treat the COLDz measurement of the LF~\citep{COLDzLF} as a reliable one, even if not all of the individual sources in the statistical sample are individually reliable.
\paragraph{COPSS} The work of \cite{keating_etal_16}) represents the first attempt at a dedicated CO(1--0) LIM survey, targeting the same redshifts as COMAP. Following an analysis of Sunyaev-Zel'dovich Array (SZA) archival data~\citep{COPSSPilot}, the same interferometer carried out observations specifically designed to measure the CO power spectrum at $z\sim3$. The result was a constraint of $P(k)=(3.0\pm1.3)\times10^3h^{-3}$ $\mu$K$^2$ Mpc$^3$, or $(8.7\pm3.8)\times10^3$ $\mu$K$^2$ Mpc$^3$, at $k\sim1h$\,Mpc$^{-1}=0.7\,$Mpc$^{-1}$. Theoretical models, including our own, suggest that this should predominantly be a measurement of the shot-noise component of the power spectrum.

\cite{mmIME-ACA} recently re-interpreted the COPSS results to allow for the possibility that the clustering component contributes to the COPSS $P(k)$ value, reporting an estimate of $P_\text{shot}=2.0^{+1.1}_{-1.2}\times10^3h^{-3}$ $\mu$K$^2$ Mpc$^3$. However, significant modification of $P_\text{shot}$ away from the original COPSS value requires $\avg{Tb}^2\gg10$ $\mu$K$^2$, which we consider to be unlikely based on our models; we thus use the original constraint from~\cite{keating_etal_16}, rather than the revised constraint from~\cite{mmIME-ACA}. 

\paragraph{mmIME} The design of mmIME combines archival data and LIM observations on community instruments across a wide range of frequencies to probe CO line emission at high redshift, with~\cite{mmIME-ACA} announcing results from ALMA observations. Using a combination of ASPECS data and ALMA Compact Array observations,~\cite{mmIME-ACA} find a non-zero shot power which they attribute to a combination of CO lines from different redshifts. Based on a CO model consistent with (although not constrained by) the total shot power measured, they expect CO(2--1) at $z\sim1.3$ and CO(3--2) at $z\sim2.5$ to contribute the bulk of this; using assumed line luminosity ratios (again from~\citealt{Daddi15}), the decomposition can be translated into an estimate of the CO(1--0) shot-noise power spectrum at $z\sim2.5$.

We do not incorporate this measurement into our priors because, in addition to the complications reviewed previously surrounding CO line ratios and excitation, the mmIME estimate of CO at $z\sim2.5$ relies on decomposing the total shot power appropriately into the contributions from different CO lines. Since~\cite{mmIME-ACA} assume a specific model to do this, the $z\sim2.5$ CO(1--0) $P_\text{shot}$ estimate could change significantly depending on the model parameters; accounting for these additional uncertainties is beyond the scope of this work.

\paragraph{PHIBBS2} The principal design of PHIBBS2~\citep{PHIBBS2}) is not as a molecular line scan survey, but as targeted observations of CO(2--1), CO(3--2), and CO(6--5) emission from $z=0.5$--0.8, $z=1$--1.6, and $z=2$--3 ``main sequence'' star-forming galaxies. However,~\cite{PHIBBS2Lenkic} were able to identify serendipitous CO line emission from secondary sources in 110 observations of primary PHIBBS2 targets, and constrain the CO LF across $z\sim0.6$--3.6.

As with ASPECS, the measurements at COMAP redshifts are of CO(3--2) or higher-$J$ CO lines. While we thus do not incorporate PHIBBS2 results into our priors either, we note that the ASPECS LP, COLDz, and PHIBBS2 results are all reasonably consistent with each other---at worst in slight tension---when translated to CO(1--0) LF constraints. 
\subsection{Data-driven Priors Constrained by Observational Results}
\label{sec:finalpriors}

\begin{figure}[t!]
    \centering
    \includegraphics[width=0.96\linewidth]{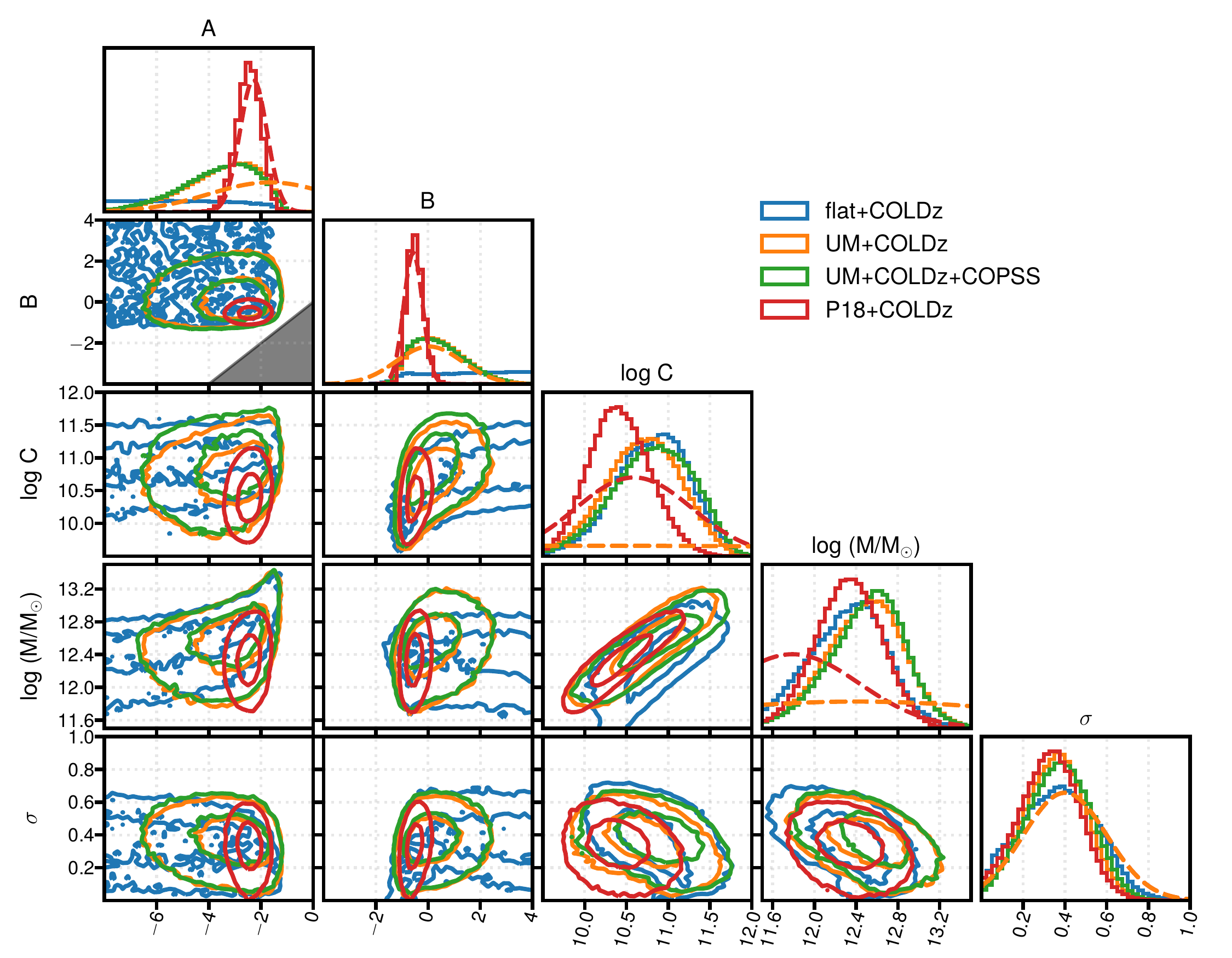}
    \caption{Parameter posterior (or data-driven prior) distributions from the MCMC combining our initial priors (dashed lines in marginalized posterior plots) with a likelihood based on the COLDz ABC constraints. Contours represent 39\% and 86\% mass levels, corresponding to $1\sigma$ and $2\sigma$ levels for 2D Gaussians. The legend indicates the colours for both initial and data-driven prior distribution curves. The dark grey triangle in the $A$-versus-$B$ plot indicates the forbidden parameter space where $A>B$.}
    \label{fig:alltheposteriors}
\end{figure}
To incorporate information from COLDz into our priors, and thus generate refined ``flat/UM/P18+COLDz'' priors for each set of initial priors, we run an MCMC with initial priors on the five parameters $\{A,B,\log{C},\log{(M/M_\odot)},\sigma\}$ as outlined above. At each step of the MCMC, we convert halo masses from a snapshot of the Bolshoi--Planck simulation (as used by~\citealt{Behroozi19}) at $a=0.293560$ into CO luminosities given the sampled model parameters, and calculate the resulting CO LF. Then, to determine the likelihood, we fit a Schechter function to the CO LF and compare the resulting Schechter parameter values to the posterior distribution of Schechter parameters from the COLDz approximate Bayesian computation (ABC). (In a minority of cases, the fitting procedure fails to produce a reasonable result; we find that including or excluding these cases does not significantly influence the posterior distribution.)

We also run an MCMC using UM priors that incorporates the COPSS power spectrum measurement into the likelihood as well. This is done by calculating the expected shot noise power spectrum from the LF as
    \begin{equation}P_\text{shot}=\biggl[\underbrace{\frac{c^3(1+z)^2}{8\pi k_B\nu_\text{rest}^3H(z)}}_{\equiv C_{LT}}\biggr]^2\int dL\,\frac{dn}{dL}\,L^2,\label{eq:Pshot_dL}\end{equation}
which is then compared to the COPSS measurement of $(3.0\pm1.3)\times10^{3}\,h^{-3}$ Mpc$^3$ $\mu$K$^2$. This UM+COLDz+COPSS MCMC will provide our fiducial data-driven prior.

The posterior distribution of this MCMC should then incorporate both our initial priors of~\autoref{tab:allthepriors} and the constraints from COLDz as well as COPSS. Thus, this distribution (``UM+COLDz+COPSS'' in particular) should be a suitable prior distribution for COMAP analysis, and one that provides a new fiducial model for the CO(1--0) power spectrum at the COMAP redshifts.
    
In all MCMCs, we do not force $A<B$ while the chain is run, but we do apply the prior for $A$ to the smaller of the two and the prior for $B$ to the larger, and in analysing the chain after completion, we always take the smaller value of the two at each sample to be $A$, and the larger to be $B$.

While the resulting posterior distributions are highly complex with all kinds of degeneracies, we show them in~\autoref{fig:alltheposteriors}. When using these as data-driven priors for COMAP analysis, we approximate them as multivariate Gaussian distributions based on the means and covariances.

\begin{figure}[t!]
    \centering
    \includegraphics[width=0.48\linewidth]{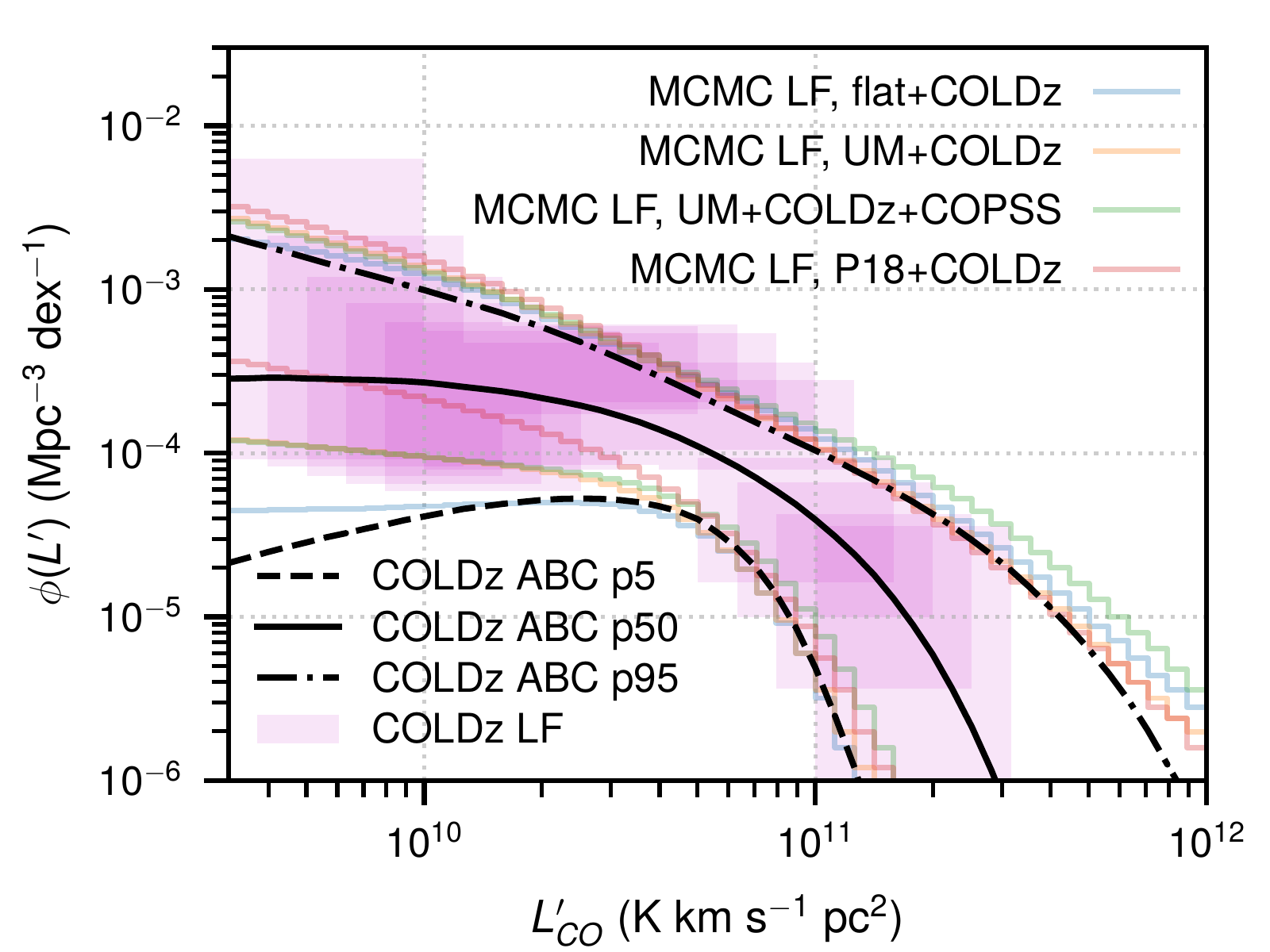}
    \includegraphics[width=0.48\linewidth]{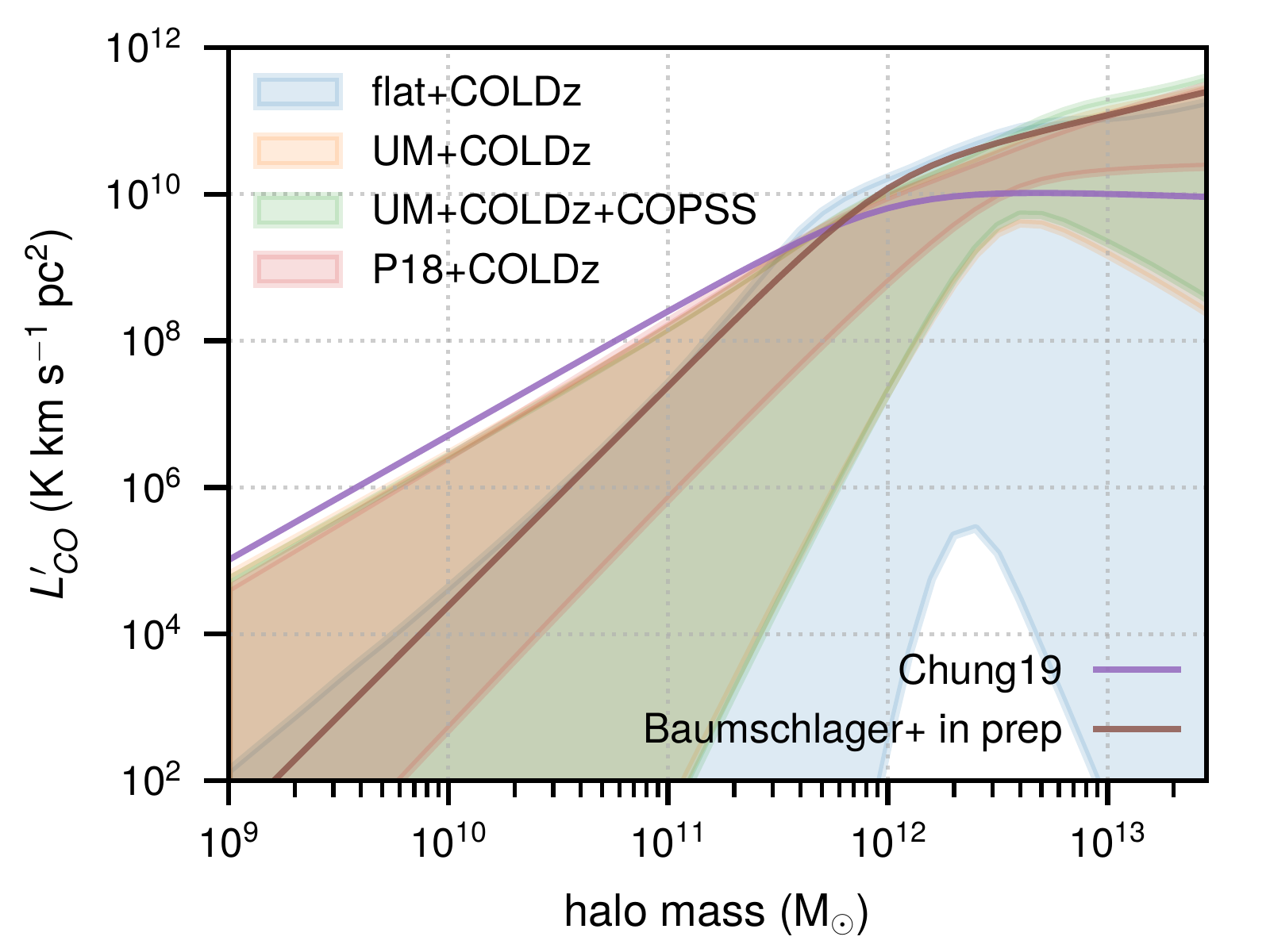}
    \caption{\emph{Left:} $z\sim2.4$ CO LF posterior distributions calculated from the MCMC results. We show 90\% intervals for the MCMC (step plots indicated in legend), the COLDz direct constraints (shaded rectangles), and the COLDz ABC constraints on a Schechter LF (black dashed, solid, and dash-dotted showing 5\%, 50\%, and 95\% percentiles). \emph{Right:} 90\% intervals from each of the MCMC results for the $L(M_h)$ relation. For reference we also show $L(M_h)$ relations from~\cite{anisotropies}---which closely ties to the~\cite{li_etal_16} model---as well as the \texttt{TNG300\_2} model from Baumschlager et al.~(in prep.).}
    \label{fig:posteriorLF}
\end{figure}

Looking at the posterior distributions of the predicted LFs plotted in~\autoref{fig:posteriorLF}, we find they are largely consistent with COLDz constraints, which is exactly as expected. However, one quirk is that the LFs from our MCMC runs tend to have negative faint-end slope, whereas the COLDz constraints do not favour either negative or positive faint-end slope values. This is to be expected based on the fact that the procedure of~\cite{COLDzLF} makes no assumptions about the CO emitters beyond the statistical sample from the survey, whereas we have the implicit assumption of the halo mass function, which approximately follows $dn/dM_h\sim M_h^{-2}$ at the low-mass end. Thus at the faint end of the LF, we expect $\phi(L')=(dn/dM_h)/(d(\log{L'})/dM_h)\sim L^{1/A}$, and $A<0$ being strongly favoured means a negative power-law slope at the faint end is also strongly favoured.

The posterior distributions of the model parameters can be summarized as a posterior distribution of the $L(M_h)$ relation, as shown in the right panel of~\autoref{fig:posteriorLF}. Our ``flat+COLDz'' prior-likelihood combination does not meaningfully constrain anything other than the turnabout scale, but the other data-driven priors tend to additionally favour a relatively flat bright-end slope, and a faint-end power law in the $M_h^2$--$M_h^6$ range.
\section{Average Values of Line Bias and Effective Line Width for the UM+COLDz and UM+COLDz+COPSS MCMC Posterior Distributions}
\label{sec:vbfits}
We find that the behaviour of $b$ and $v_\text{eff}$ with changing $\avg{Tb}$ and $P_\text{shot}$ is relatively smooth across the UM+COLDz MCMC distribution. This allows us to devise the following fits:
\begin{align}
    b&\approx-2.85 -2.30\log{\left(\frac{\avg{Tb}}{\mu\text{K}}\right)}+4.16\log{\left(\frac{P_\text{shot}}{\mu\text{K}^2\,\text{Mpc}^3}\right)}+2.18\log^2{\left(\frac{\avg{Tb}}{\mu\text{K}}\right)}\nonumber\\&\qquad-0.888\log^2{\left(\frac{\avg{Tb}}{\mu\text{K}}\right)}\log{\left(\frac{P_\text{shot}}{\mu\text{K}^2\,\text{Mpc}^3}\right)}+0.0144\log^2{\left(\frac{\avg{Tb}}{\mu\text{K}}\right)}\log^2{\left(\frac{P_\text{shot}}{\mu\text{K}^2\,\text{Mpc}^3}\right)}-0.557\log^2{\left(\frac{P_\text{shot}}{\mu\text{K}^2\,\text{Mpc}^3}\right)}\nonumber\\&\qquad+0.421\log{\left(\frac{\avg{Tb}}{\mu\text{K}}\right)}\log^2{\left(\frac{P_\text{shot}}{\mu\text{K}^2\,\text{Mpc}^3}\right)}-1.42\log{\left(\frac{\avg{Tb}}{\mu\text{K}}\right)}\log{\left(\frac{P_\text{shot}}{\mu\text{K}^2\,\text{Mpc}^3}\right)};\\
    \frac{v_\text{eff}}{\text{km\,s}^{-1}}&\approx241.-219.\log{\left(\frac{\avg{Tb}}{\mu\text{K}}\right)}+32.5\log{\left(\frac{P_\text{shot}}{\mu\text{K}^2\,\text{Mpc}^3}\right)}+21.8\log{\left(\frac{\avg{Tb}}{\mu\text{K}}\right)}\log{\left(\frac{P_\text{shot}}{\mu\text{K}^2\,\text{Mpc}^3}\right)}.
\end{align}

\begin{figure}[t!]
    \centering
    \includegraphics[width=0.48\linewidth]{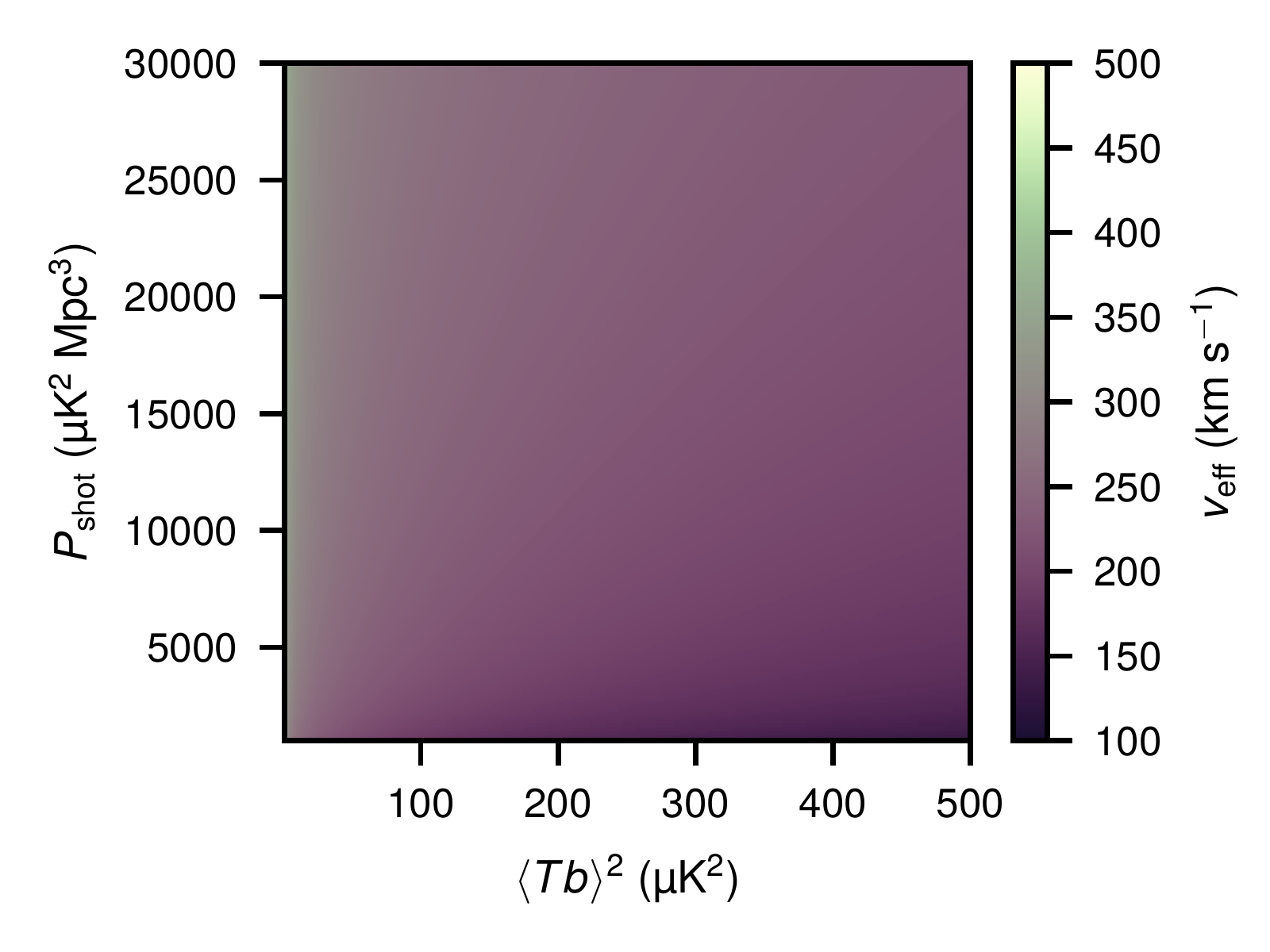}\includegraphics[width=0.48\linewidth]{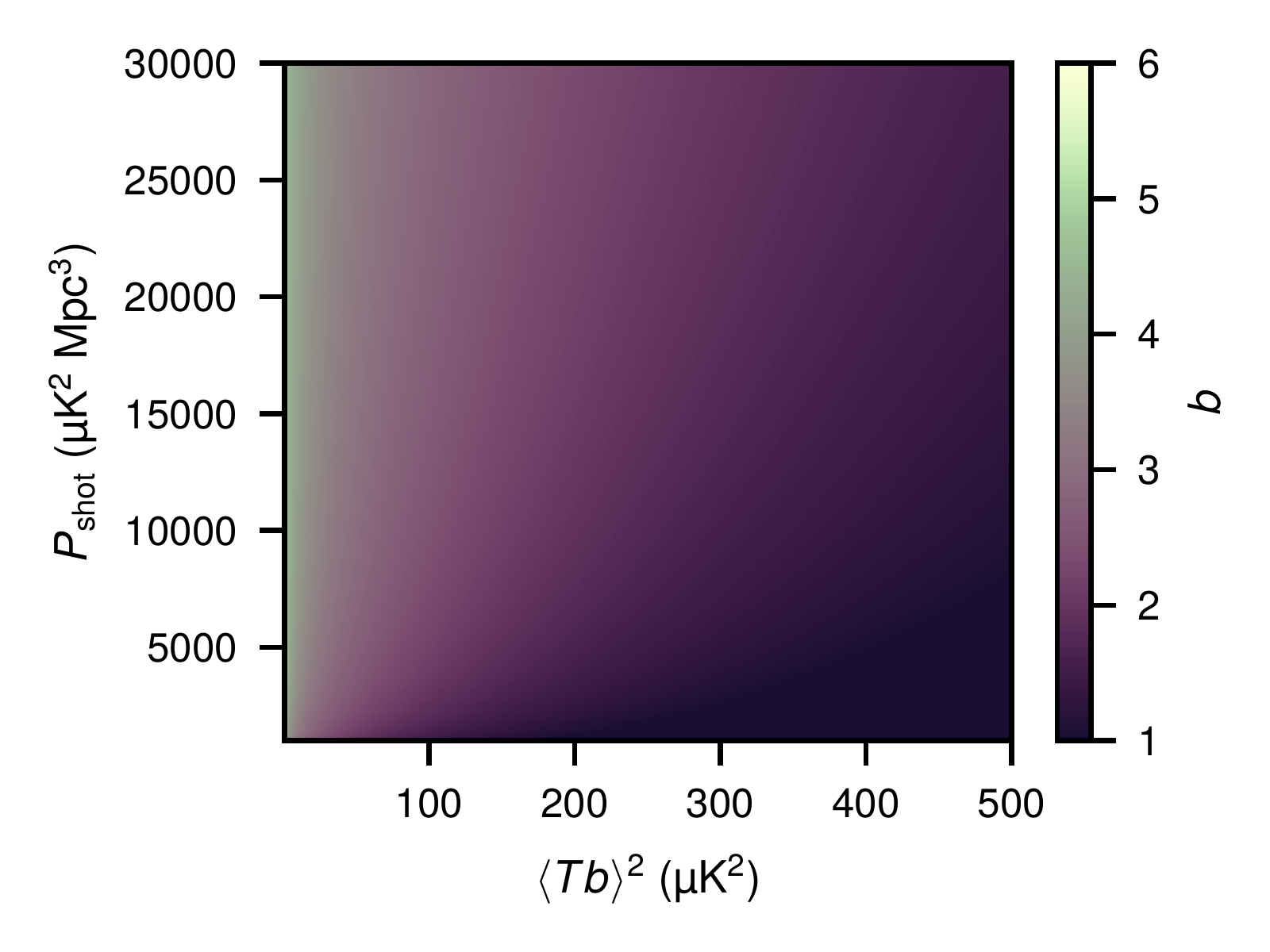}
    \caption{Polynomial fits of $v_\text{eff}$ (left) and $b$ (right) with respect to $\avg{Tb}^2$ and $P_\text{shot}$, as detailed in the main text. Note the general trends of co-correlation of both variables with $P_\text{shot}$ as well as anti-correlation of both with $\avg{Tb}^2$.}
    \label{fig:vbfits}
\end{figure}

Residuals versus these fits are mostly confined to 10--30\% relative error, against both the UM+COLDz and UM+COLDz+COPSS samples. This level of error is sufficient for our purposes given the large uncertainties associated with the observational data. We plot these fits in~\autoref{fig:vbfits}, although note that the MCMC posterior samples only span parts of the parameter space being plotted, largely towards lower values of both $P_\text{shot}$ and $\avg{Tb}$.

\section{Details of MCMC Inference Simulations}
\label{sec:mcmc_appendix}
\subsection{Survey Simulations}
\label{sec:mcmc_methods}

We use a simplified COMAP experimental setup with a sensitivity corresponding roughly to the Y5 sensitivity forecast previously mentioned in~\autoref{sec:det_sign}. The experimental parameters are summarised in~\autoref{tab:exp_setup}. We assume a uniform noise distribution in three cosmological fields each covering four square degrees, with 256 frequency bins covering the full 26--34\,GHz range. Following \cite{Ihle19} we choose a pixel size ($4\times4$ arcmin${}^2$) comparable to the instrumental beam width of 4.5 arcmin (FWHM), which gives us a $30 \times 30$ pixel grid for each field. 

\begin{deluxetable}{rc}
    \tablecaption{Simplified COMAP Experimental Parameters for MCMC Simulated Inference\label{tab:exp_setup}}
    \tablehead{\colhead{Parameter} & \colhead{Value}}
    \startdata
        System temperature [K] & $44$\\
        Number of feeds & $19$ \\
        Beam FWHM [arcmin] & $4.5$ \\
        Frequency band [GHz] & $26$--$34$\\
        Channel width [MHz] & $31.25$ \\
        Number of fields  & $3$ \\
        Field size [deg${}^2$] & $4$ \\
        Number of pixels per field & $30 \times 30$ \\
        Noise per voxel\tablenotemark{a} [$\mu$K] & 17.8
    \enddata
    \tablenotetext{a}{This value corresponds to the Y5 sensitivity forecast discussed at the start of~\autoref{sec:forecasts}.}
\end{deluxetable}

Our signal simulations are based on mock dark matter (DM) halo catalogues generated using the peak patch approach \citep{Bond96,Stein18}. We associate CO luminosities with each of the DM halos using the model presented above. Luminosities are converted to equivalent brightness temperature and then separated by virial velocity before adding up the contributions to each voxel in a high resolution comoving grid. The maps corresponding to the different virial velocity are convolved with the appropriate Gaussian linewidth, as discussed above, before they are added together and convolved with the angular beam. Finally we degrade the map to the low resolution  used for the main analysis. 

We use 161 independent lightcones each covering 9.6$\times$9.6 deg${}^2$ and divide them into smaller angular pieces to correspond to the size of our cosmological fields. This way we get a large number of semi-independent cosmological realizations to use for generating covariance matrices.

\subsection{Observables and Covariances}
\label{sec:mcmc_methods_add}
\citet{Ihle19} showed that using a combination of the power spectrum, $P(k)$, and the VID, $\mathcal{P}(T)$, is a good way to capture different parts of the information in a set of line intensity maps in an efficient manner. We use the same approach here. 

The spherically averaged power spectrum, $P(k)$ is calculated from the (discrete) 3D Fourier components, $f_{\bf k}$, of the temperature map
\begin{equation}
    P(k) = \frac{V_\mathrm{vox}}{N_\mathrm{vox}}\langle |f_{\bf k}|^2 \rangle \approx \frac{V_\mathrm{vox}}{ N_\mathrm{vox}N_\mathrm{modes}} \sum_{j=1}^{N_\mathrm{modes}} |f_{\bf k_j}|^2\equiv P_{k_i},
\end{equation}
where $P_{k_i}$ is the estimated power spectrum in bin number $i$, $V_\mathrm{vox}$ is the voxel volume, $N_\mathrm{vox}$ is the total number of voxels in the map and $N_\mathrm{modes}$ is the number of Fourier components with wave number $|{\bf k_j}| \approx k_i$ (i.e. in the bin corresponding to wave number $k=k_i$).

The most natural observable related to the VID, $\mathcal{P}(T)$, is the temperature bin count
\begin{equation}
    \langle B_i \rangle = N_\mathrm{vox} \int_{T_i}^{T_{i+1}} \mathcal{P}(T) dT,
\end{equation}
where $B_i$ is the number of voxels with a temperature in the $i$'th temperature bin. 

We combine both observables into a data vector 
\begin{equation}
    d_i = (P_{k_i}, B_i).
\end{equation}
If all the components of $d_i$ were independent, they would have the following variance, which we denote as the \emph{independent variance}:
\begin{align}
    \mathrm{Var}_\mathrm{ind}(\mathrm{P_{k_i}}) &= \langle P_{k_i} \rangle^2 / N_\mathrm{modes}, \\
    \mathrm{Var}_\mathrm{ind}(\mathrm{B_i}) &= \langle B_i \rangle.
\end{align}
This assumes that the Fourier modes $f_{\bf k}$ of the maps are independent Gaussians, and that the total number of voxels is much larger than $\langle B_i \rangle$.

Since there typically are correlations between the different elements of the data vector, we can take this into account using a full covariance matrix

\begin{equation}
    \xi_{ij} = \mathrm{Cov}(d_i, d_j).
\end{equation}

We now have all the ingredients we need to build up a likelihood. We assume a Gaussian likelihood of the form (up to a constant)
\begin{equation}
    -2 \mathrm{ln} P(d|\theta) = \frac{N_{s}}{N_{s}+1} \sum_{ij} [d_i - \langle d_i \rangle](\xi^{-1})_{ij}[d_j - \langle d_j \rangle] + \mathrm{ln} |\xi|,
\end{equation}
where $\langle d \rangle(\theta)$ and $\xi(\theta)$ are the mean values and covariance matrix of the observables $d_i$ for specific parameters $\theta$. $N_s$ is the number of simulations used to estimate $\langle d \rangle$, and the factor $N_{s}/N_{s}+1$ takes into account the effect of the uncertainty in the estimate of $\langle d \rangle$. We refer the reader to~\cite{Ihle19} for further details on the mock DM catalogues, the simulation procedure, and how the covariance matrices are estimated. 

\subsection{Mock MCMC Setup}
\label{sec:mcmc_methods_mcmc}
The posterior distribution for our model parameters, $\theta=\{A, B, \log C, \log{({M}/{M_\odot})}, \sigma\}$, is given by Bayes' theorem,
\begin{equation}
    P(\theta|d) \propto P(d|\theta) P(\theta),
\end{equation}
where $P(\theta)$ is the prior on the model parameters, $\theta$. As a prior we approximate the constraints derived in~\autoref{sec:finalpriors} as a multidimensional Gaussian distribution, for computational efficiency. The prior parameters are given in~\autoref{tab:cov_prior}.

\begin{deluxetable}{ccccccc}
    \tablecaption{Mean and Covariance Matrix for Gaussianized UM+COLDz+COPSS Priors on Model Parameters\label{tab:cov_prior}}

    \tablehead{\colhead{}&\colhead{} & \multicolumn{5}{c}{Covariance Matrix} \\
        \colhead{Parameter}& \colhead{Mean}&\colhead{$A$}&\colhead{$B$}&\colhead{$\log{C}$}&\colhead{$\log{\frac{M}{M_\odot}}$}&\colhead{$\sigma$}}
    \startdata
    \colhead{$A$} & -3.71 & 2.26 & 0.0651 & 0.143 & 0.185 & -0.0305\\ 
    \colhead{$B$} & 0.41 & 0.0651 & 1.03 & 0.207 & 0.0805 & 0.0239\\ 
    \colhead{$\log{C}$} & 10.8 & 0.143 & 0.207 & 0.251 & 0.17 & -0.0227\\
    \colhead{$\log{\frac{M}{M_\odot}}$} & 12.5 & 0.185 & 0.0805 & 0.17 & 0.151 & -0.0243\\ 
    \colhead{$\sigma$} & 0.371 & -0.0305 & 0.0239 & -0.0227 & -0.0243 & 0.0228
    \enddata
\end{deluxetable}

To sample from the posterior distribution, we use the \texttt{emcee} package \citep{emcee} implementing an affine invariant ensemble MCMC with 60 walkers. As the ``data'', $d$, for the MCMC forecast we use a single (three-field) cosmological realization of the UM+COLDz+COPSS point estimate from~\autoref{tab:priorpoints}, our default model. 
At each step in the MCMC we estimate the mean observables, $\langle d \rangle(\theta)$, using 10 simulations in order to evaluate the posterior at the current point in parameter space. We estimate the mean CO LF from the 10 signal realizations each step in the MCMC and use this as an estimate of the LF at this point in parameter space. This way we obtain a large number of samples of the LF sampled according to the posterior distribution of the model parameters, giving us a simple way to get posterior constraints on the LF. 

We use a burn-in period of 500 samples out of 5940, and treat the subsequent samples as valid samples from the posterior. We use the Gelman--Rubin and Geweke convergence diagnostics as implemented by \texttt{ChainConsumer}~\citep{ChainConsumer}, which both suggest the MCMC converges. The Geweke statistic in particular suggests that we could even shorten the burn-in period considerably to the first 30 samples, but we do not presume these statistics have great sensitivity in identifying chain convergence, although they may have great specificity. Therefore, regardless of the convergence statistics, we take a conservative approach and continue to treat the first 500 MCMC samples as the burn-in period.

\section{Details of COMAP--HETDEX Fisher Forecasts}
\label{sec:hetdex_appendix}
The quantitative forecasts of~\cite{hetdex_selfcite} for COMAP--HETDEX cross-spectra are somewhat esoteric at the present time for \replaced{two}{several} reasons:
\begin{itemize}
    \item COMAP Pathfinder parameters such as field sizes and observation efficiencies have substantially evolved.
    \item The LAE model, while in principle matching LAE counts in the literature, did not correctly account for Lyman-alpha emission duty cycles of high-redshift galaxies, effectively setting the LAE fraction to 100\%. Thus the model of~\cite{hetdex_selfcite} overestimates the mean CO luminosity of LAE samples and thus overestimates the cross shot noise, while also overestimating the LAE bias.
    \item The forecast S/N values were never translated to parameter constraints.
\end{itemize}
It is entirely out of scope for this paper to provide a full-fledged Lyman-alpha emission model in order to forecast COMAP--HETDEX cross-correlation analyses. We can, however, forecast the observable auto- and cross-spectra without having to devise such a model.

\subsection{Observables and Parameters}
Adapting~\cite{BreysseAlexandroff19}, in real (comoving) space we would have
\begin{align}
P_\text{CO}(k) &= \avg{Tb}_\text{CO}^2P_m(k)+P_\text{shot,CO};\\
P_\text{LAE}(k) &= b_\text{LAE}^2P_m(k) + \bar{n}_\text{LAE}^{-1};\\
P_{\text{CO}\times\text{LAE}}(k) &= \avg{Tb}_\text{CO}b_\text{LAE}P_m(k) + P_{\text{shot,CO}\times\text{LAE}}.
\end{align}
Normally forward models would link parameters like our fiducial model's $\{A,B,C,M,\sigma\}$ to quantities like $\avg{Tb}_\text{CO}$ and $P_\text{shot}$. However, our approach for these simple forecasts will focus on directly constraining the ``derived'' quantities, much as we directly constrained $\avg{Tb}$ and $P_\text{shot}$ with the early science data.

Apart from the matter power spectrum $P_m(k)$, five quantities fully define the real-space power spectra: the CO clustering amplitude $\avg{Tb}_\text{CO}$, the CO shot noise $P_\text{shot,CO}$, the LAE bias $b_\text{LAE}$, the LAE number density $\bar{n}_\text{LAE}$, and finally the cross shot noise $P_{\text{shot,CO}\times\text{LAE}}$, which encodes the mean CO luminosity of the LAE population:
\begin{align}
P_{\text{shot,CO}\times\text{LAE}}&=C_{LT}\avg{L_{\text{CO}|\text{LAE}}}
=\frac{C_{LT}}{\bar{n}_\text{LAE}}\int L_\text{CO}\,\left.\frac{dn}{dL_\text{CO}}\right|_\text{LAE}\,dL_\text{CO},
\end{align}
where $C_{LT}$ is the same as defined in~\autoref{eq:Pshot_dL}.

We can define fiducial values easily for four of the five quantities. For CO, given the UM+COLDz+COPSS point model (which is our fiducial model) we have $\avg{Tb}_\text{CO}=2\,\mu$K (with $b_\text{CO}=4$) and $P_\text{shot,CO}=1.9\times10^3\,\mu$K$^2$\,Mpc$^3$. For HETDEX, references in~\cite{HETDEX2021} suggest $b_\text{LAE}=1.8$--$2.2$ so we can take a central value of 2. Taking into account the fact that HETDEX sparsely samples most of its survey footprint at $1/4.6$-fill,~\cite{HETDEX2021} quote an expected number density of $\bar{n}_\text{LAE}=1.1\times10^5\,$Gpc$^{-3}$.

For the cross shot noise, we assume a LAE fraction and give our best estimate for the mean CO luminosity for the LAE population based on that assumption. Using various subsamples of UV-selected galaxies and deep data from the Multi-Unit Spectroscopic Explorer (MUSE), \cite{Kusakabe20} found the LAE fraction to range between 4\% and 30\%, rising weakly with redshift and not evolving significantly with absolute rest-frame UV magnitude. Based on the LAE fractions found in their Table 1, we consider $X_\text{LAE}=0.05$ a reasonable LAE fraction to assume.

Proceeding from this assumption that 5\% of galaxies are LAEs, with no significant dependence on UV luminosity---and thus, one might assume, no significant dependence on star-formation rate or halo mass---we assume that 5\% of dark matter halos host LAEs. To reach a number density of $\bar{n}_\text{LAE}=1.1\times10^5\,$Gpc$^{-3}$ with $1/4.6$-fill, which is to say $\bar{n}_\text{LAE}=5\times10^5\,$Gpc$^{-3}$ without sparse sampling, the number density of halos that could \emph{ever} possibly host a LAE would have to be $\bar{n}_\text{LAE}/X_\text{LAE}=10^7\,$Gpc$^{-3}$. This number density can be achieved by selecting all dark matter halos with a halo mass above $M_h>9.3\times10^{10}\,M_\odot$, a halo population with an average halo bias of 2.2 roughly consistent with the $b_\text{LAE}$ central expectation. For this population the average CO luminosity is $4.9\times10^4\,L_\odot$ under our fiducial model, which we multiply by $C_{LT}$ to obtain an estimate of $P_{\text{shot,CO}\times\text{LAE}}=51\,\mu$K\,Mpc$^3$.

However, the redshift-space observables are more complicated than the real-space power spectra, as the CO intensity field is subject to line broadening and large-scale redshift-space distortions manifest in all observed clustering. As in the early science analyses in the main text, here we are only concerned with the simply spherically-averaged monopole power spectrum, so we approximate the effect of line broadening with a single Gaussian filter with associated effective line width $v_\text{eff}$. On top of these astrophysical redshift-space effects, we must also consider the COMAP transfer function and the VIRUS instrumental resolution.

Although the early science analysis in the main text uses de-biased sensitivities already corrected for the transfer function, for this section we compare the uncorrected signals against raw sensitivities. We can approximate the transfer function for COMAP CES data, in the space of the transverse and line-of-sight wavevector components $k_\perp$ and $k_\parallel$, as a combination of sigmoid and Gaussian functions:
\begin{equation}\mathcal{T}(k_\perp,k_\parallel)=\frac{0.4\exp{[-(3.5\,\text{Mpc}\cdot k_\perp)^2]}\exp{[-(1.8\,\text{Mpc}\cdot k_\parallel)^2]}}{[1+\exp{(5-100\,\text{Mpc}\cdot k_\perp)}][1+\exp{(5-200\,\text{Mpc}\cdot k_\parallel)}]}\end{equation}

We can combine this with an estimated noise power spectrum of $P_N\sim10^6\,\mu$K$^2$\,Mpc$^3$ for Field 1 CES data, as well as Fourier mode counts expected from a $60\times60\times256$ voxel grid spanning $2\degr\times2\degr\times8\,$GHz in angular and spatial extent (mirroring the actual COMAP pixelisation in all dimensions). The resulting noise limit and transfer function are both within $1/3$ of ground truth across a majority of the range of $k$ values. Scaling the resulting noise limit ($P_N$ divided by the number of modes) down by 69 to obtain our Y5 sensitivity estimate and applying the approximate $\mathcal{T}$ and line broadening to the fiducial CO model power spectrum, we obtain a forecast all-$k$ S/N of 8 for the uncorrected signal versus the raw sensitivity. This estimate is not at all far from the value of 9 forecast in the main text comparing the de-biased noise limit against a line-broadened signal.

Working from~\cite{anisotropies} we can define the pseudo-power auto- and cross-spectra as functions of $(k_\perp,k_\parallel)$:
\begin{align}
\tilde{P}_\text{CO}(k_\perp,k_\parallel) &= \mathcal{T}(k_\perp,k_\parallel)\exp{\left(-k_\parallel^2\sigma_\text{eff}^2\right)}\left[\avg{Tb}_\text{CO}^2(1+\beta_\text{CO}k_\parallel^2/(k_\parallel^2+k_\perp^2))^2P_m(k)+P_\text{shot,CO}\right];\\
\tilde{P}_\text{LAE}(k_\perp,k_\parallel) &= \exp{\left(-k_\parallel^2\sigma_\text{VIRUS}^2\right)}\left[(b_\text{LAE}+fk_\parallel^2/(k_\parallel^2+k_\perp^2))^2P_m(k) + \bar{n}_\text{LAE}^{-1}\right];\\
\tilde{P}_{\text{CO}\times\text{LAE}}(k_\perp,k_\parallel) &= \mathcal{T}^{1/2}(k_\perp,k_\parallel)\exp{\left(-\frac{k_\parallel^2(\sigma_\text{eff}^2+\sigma_\text{VIRUS}^2)}{2}\right)}\times\nonumber\\&\hspace{1.5cm}\left[\avg{Tb}_\text{CO}(1+\beta_\text{CO}k_\parallel^2/(k_\parallel^2+k_\perp^2))(b_\text{LAE}+fk_\parallel^2/(k_\parallel^2+k_\perp^2))P_m(k) + P_{\text{shot,CO}\times\text{LAE}}\right].
\end{align}
Note that we define $\sigma_\text{eff}$ in terms of $v_\text{eff}$:
\begin{equation}
    \sigma_\text{eff}=\frac{(1+z)}{H(z)}\frac{v_\text{eff}}{\sqrt{8\ln{2}}}.
\end{equation}

We may then average the pseudo-power spectra in cylindrical $k$-space into $k$-bins, weighting by inverse variance, to yield spherically averaged pseudo-signal spectra $\tilde{P}_i(k)$. For the biased signals this is equivalent to simply weighting by the mode count in each pair of $k_\perp$- and $k_\parallel$-bins, thus ending up with a simple arithmetic average over some number of modes $N_m(k)$ for each $k$-bin. Since the field being auto-correlated or Fourier-transformed is entirely real, only half the Fourier modes are independent; we count modes so that $N_m(k)$ already includes this halving.

\subsection{Fisher Forecasts}

For a Fisher analysis in the style of~\cite{BreysseAlexandroff19}, we first define the covariance matrix of the (distorted) CO intensity and LAE overdensity fields:
\begin{equation}
    C_{ij}(k) = \begin{bmatrix}\tilde{P}_\text{CO}(k)+P_N&\tilde{P}_{\text{CO}\times\text{LAE}}(k)\\\tilde{P}_{\text{CO}\times\text{LAE}}(k)&\tilde{P}_\text{LAE}(k)\end{bmatrix}
\end{equation}
Unlike~\cite{BreysseAlexandroff19}, we should not divide $P_N$ by any kind of window function as we already fold applicable transfer functions into $\tilde{P}_i(k)$.

To forecast constraints for parameters $\{x_i\}$, we would obtain the covariance of those parameters by inverting the Fisher matrix:
\begin{equation}
    F_{ij}=\sum_k N_m(k)\operatorname{Tr}{\left[\pd{C}{x_i}C^{-1}\pd{C}{x_j}C^{-1}\right]}.
\end{equation}
Note that we discard the factor of $1/2$ from Equation 13 of~\cite{BreysseAlexandroff19} as we use $N_m(k)$ to denote the number of \emph{independent} Fourier modes in each $k$-bin, whereas~\cite{BreysseAlexandroff19} appear to write Equation 13 under the assumption that $N_m$ denotes twice this. Corroborating this is the fact that Equation 12 of~\cite{hetdex_selfcite} divides by $2N_m(k)$---using an equivalent definition of $N_m(k)$ as in this section---when calculating cross-spectrum variance, but Equation 19 of~\cite{BreysseAlexandroff19} divides only by $\sqrt{N_m(k)}$ when calculating cross-spectrum error.

We project constraints for six parameters: $\avg{Tb}_\text{CO}$ (in units of $\mu$K), $\beta_\text{CO}$, $p_\text{shot,CO}\equiv P_\text{shot,CO}/(10^3\,\mu$K$^2$\,Mpc$^3)$, $v_\text{eff}$ (in units of km\,s$^{-1}$), $b_\text{LAE}$, and $p_{\text{shot,CO}\times\text{LAE}}\equiv P_{\text{shot,CO}\times\text{LAE}}/(10^2\,\mu$K\,Mpc$^3)$. We do not attempt to project constraints on $\bar{n}_\text{LAE}$ as the HETDEX data by itself will constrain this extremely finely. We also note the addition of two parameters not in our real-space parameter set: $\beta_\text{CO}$, which must be defined separately from $\avg{Tb}_\text{CO}$ to fully describe redshift-space distortions; and $v_\text{eff}$, which we use to describe line broadening. We base the central values for these parameters on our fiducial model, and thus obtain fiducial values for all six parameters in our Fisher forecast:
\begin{align}
\avg{Tb}_\text{CO}&=2.1\,[\mu\text{K}];\\
\beta_\text{CO}&=0.24;\\
p_\text{shot,CO}&=1.9;\\
v_\text{eff}&=330\,[\text{km\,s}^{-1}];\\
b_\text{LAE}&=2;\\
p_{\text{shot,CO}\times\text{LAE}}&=0.51.
\end{align}
We also continue to assume $\bar{n}_\text{LAE}=1.1\times10^{-4}\,\text{Mpc}^{-3}$ to calculate $\tilde{P}_\text{LAE}(k)$.

Since a significant degeneracy exists between $\avg{Tb}_\text{CO}$ and $\beta_\text{CO}$, and also between $p_\text{shot,CO}$ and $v_\text{eff}$, we impose Gaussian prior distributions of $\beta_\text{CO}=0.24\pm0.15$ and $v_\text{eff}=330\pm165$ (again, in units of km\,s$^{-1}$). These priors are conservative; the prior width on $v_\text{eff}$ is defined to keep $v_\text{eff}>0$ in the vast majority of cases, while the prior width on $\beta_\text{CO}$ derives from the main text's assertion that $b_\text{CO}>2$.

We run Fisher analyses for two survey sensitivity scenarios:
\begin{itemize}
    \item \emph{Y3:} We assume that at minimum, by Y3 we will have a (sparsely sampled) HETDEX LAE catalogue that covers Field 1, and that we will have accumulated 15 times the integration time that we currently have in this field (entirely consistent with our Y5 forecast). This means we scale the estimated current $P_N=10^6\,\mu$K$^2$\,Mpc$^3$ down by a factor of 15, but assume we only have Fourier modes available in this one field for cross-correlation. For the CO auto-spectrum we consider sensitivities for both one field and for all three fields, keeping $P_N$ the same but tripling the number of modes available in the survey for the latter case.
    \item \emph{Y5:} We assume that we integrate deep enough to achieve a noise power spectrum of $10^6/40=2.5\times10^4\,\mu$K$^2$\,Mpc$^3$ in all fields, and that a sparsely sampled HETDEX LAE catalogue covers all fields as well. This scenario is designed such that the net auto-spectrum sensitivity gain of $40\sqrt{3}=69$ relative to the current Field 1 limit is consistent with the improvement forecast for Y5 in the main text.
\end{itemize}

 The assumed HETDEX LAE data availability for Y3 and Y5 does not reflect potential proprietary periods for HETDEX data before they are shared with either the general community or the COMAP collaboration specifically. However, given the expectation of full-fill sampling of the HETDEX zero-declination field (which overlaps with COMAP Field 1) and the current estimated HETDEX survey completion date of 2024 quoted by~\cite{HETDEX2021}, we believe we have a reasonable guess of how much data HETDEX would have available internally.
 
\begin{figure}[t!]
    \centering
    \includegraphics[width=0.48\linewidth]{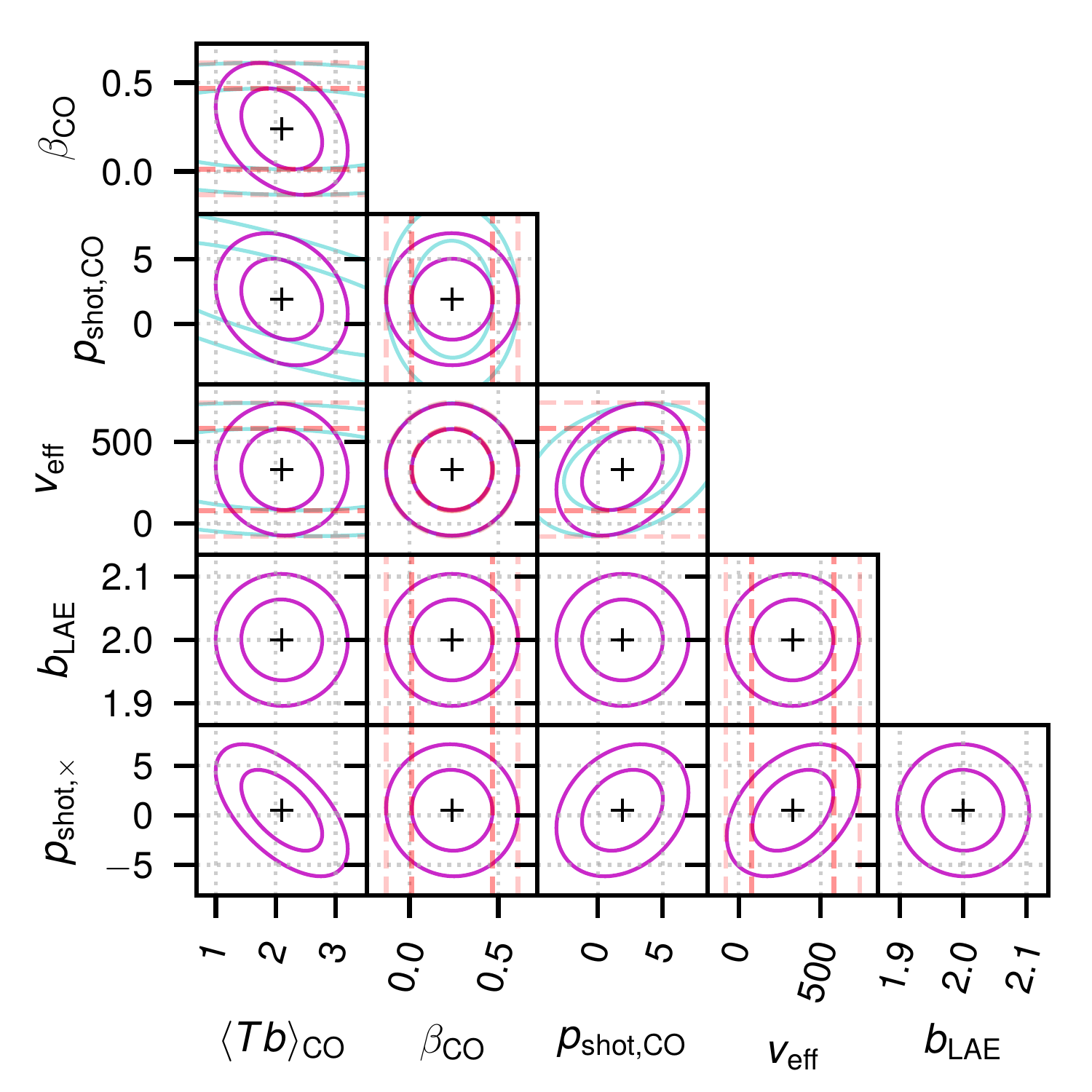}\includegraphics[width=0.48\linewidth]{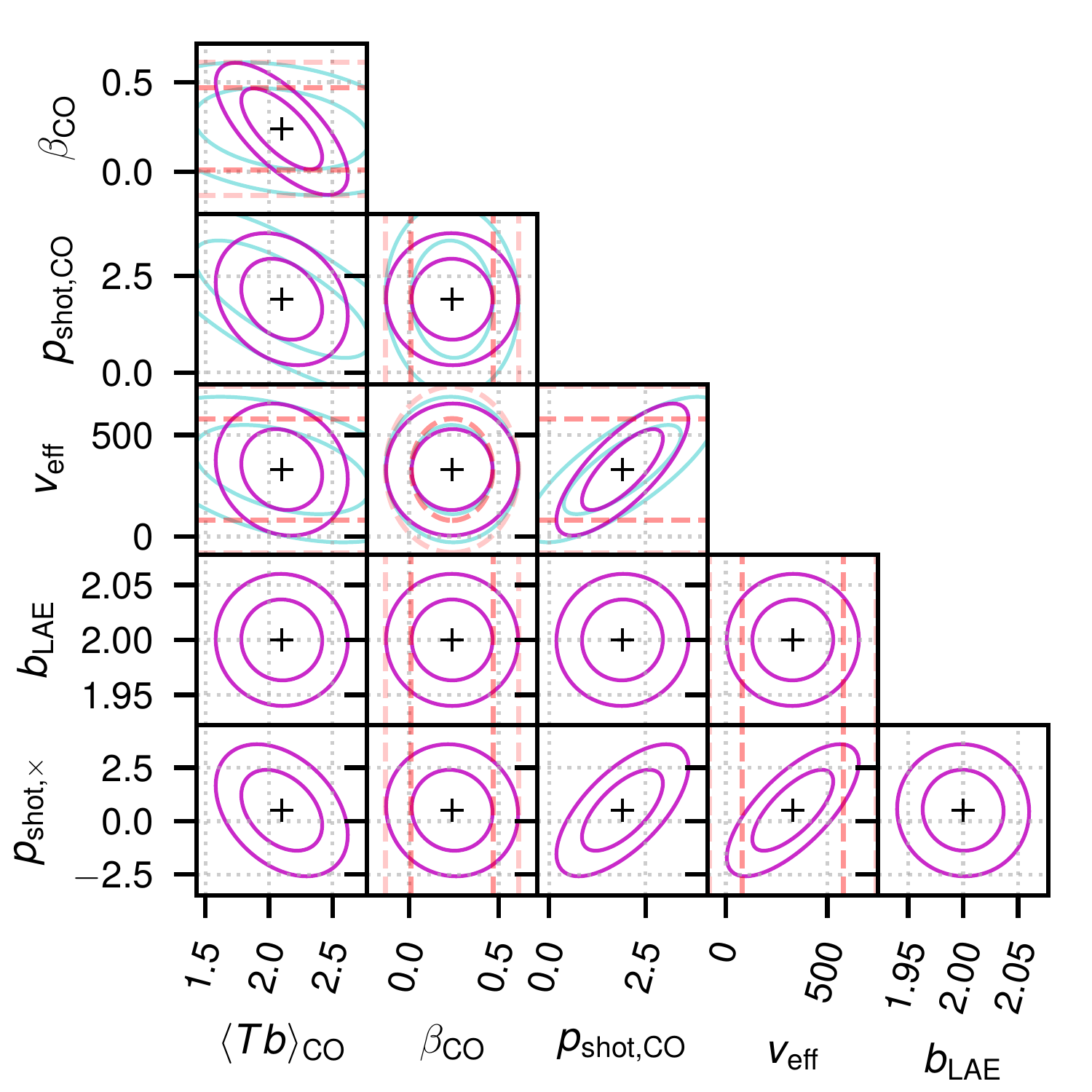}
    \caption{68\% and 95\% ellipses from the Fisher analyses described in the text for the Y3 one-field (left) and Y5 three-field (right) scenarios. Parameters are dimensionless except $\avg{Tb}_\text{CO}$ (in units of $\mu$K) and $v_\text{eff}$ (in units of km\,s$^{-1}$). Faint cyan ellipses show constraints expected from the CO auto-spectrum only, while the solid magenta ellipses show joint constraints expected from CO and LAE data. We also show priors for $\beta_\text{CO}$ and $v_\text{eff}$ (red dashed) applied in the Fisher analyses.}
    \label{fig:hetdex_fisher}
\end{figure}

We show the resulting error ellipses in~\autoref{fig:hetdex_fisher}. While we only resolve some parameter degeneracies through priors, note that we significantly reduce the degeneracy between $\avg{Tb}_\text{CO}$ and $p_\text{shot,CO}$---which is to say that we can better disambiguate CO clustering from CO shot noise---through cross-correlation.

\begin{deluxetable}{ccccccc}
    \tablecaption{Fisher Forecasts for Clustering Constraints from COMAP Auto- and COMAP--HETDEX Cross-Spectra\label{tab:allthefishers}}

    \tablehead{\colhead{} & \colhead{} & \multicolumn{5}{c}{$\avg{Tb}_\text{CO}/\sigma[\avg{Tb}_\text{CO}]$} \\\cline{3-7}
        \colhead{Model}& \colhead{$\avg{Tb}_\text{CO}$} &\colhead{\parbox[b][1.25cm]{2cm}{\centering COMAP Y3 $\times$\\HETDEX LAE\\(One Field)}}&\colhead{\parbox[b]{2.3cm}{\centering COMAP Y3 Auto\\(One Field)}}&\colhead{\parbox[b]{2.3cm}{\centering COMAP Y3 Auto\\(Three Fields)}}&\colhead{\parbox[b]{2cm}{\centering COMAP Y5 $\times$\\HETDEX LAE\\(Three Fields)}}&\colhead{\parbox[b]{2.3cm}{\centering COMAP Y5 Auto\\(Three Fields)}}}
    \startdata
    UM+COLDz+COPSS & 2.1\,$\mu$K & 4.7 & 1.2 & 2.0 & 10. & 4.6 \\ 
    Li+2016--Keating+2020 & 3.5\,$\mu$K & 8.0 & 3.7 & 6.1 & 13. & 12.
    \enddata
    \tablecomments{The model labels ``UM+COLDz+COPSS'' and ``Li+2016--Keating+2020'' respectively denote the fiducial model from this work (derived from the namesake data-driven prior) and the \cite{li_etal_16}--\cite{mmIME-ACA} model, both discussed in the main text.}
\end{deluxetable}

The main parameter we would meaningfully constrain in the Y3 scenarios is $\avg{Tb}_\text{CO}$, so we tabulate the constraining power in~\autoref{tab:allthefishers} as the ratio between the central $\avg{Tb}_\text{CO}$ value and the Fisher forecast uncertainty $\sigma[\avg{Tb}_\text{CO}]$. From Field 1 data alone we expect $\sigma[\avg{Tb}_\text{CO}]=0.445$ from a joint analysis of the CO auto- and CO--LAE cross-spectra (the latter detected with an all-$k$ S/N of 7). This would be a significant improvement over the early science result of $\avg{Tb}_\text{CO}^2<51\,\mu$K (or $\avg{Tb}_\text{CO}<7\,\mu$K), and would still be better than an analysis of the Field 1 CO auto-spectrum by itself which would only yield an upper limit of $\avg{Tb}_\text{CO}<5\,\mu$K. (Note the priors on $v_\text{eff}$ and line bias applied in the Fisher forecasts are much looser than the axiomatic assumptions applied in the main text's analysis, so we should not expect this forecast upper limit to improve on the early science result by a factor of $\sqrt{15}\approx4$.) Even if we multiply the number of modes available by a factor of three to simulate an auto-spectrum-based constraint that uses data from all three COMAP fields, we forecast a marginal $2\sigma$ result as the predicted uncertainty is $\sigma[\avg{Tb}_\text{CO}]=1.04$.

With Y5 data in hand, the CO--LAE cross-spectra detection should continue to improve to an all-$k$ S/N of 19, and the uncertainty on the CO clustering amplitude from joint analysis of auto- and cross-spectra should also improve by approximately a factor of 2 to $\sigma[\avg{Tb}_\text{CO}]=0.209$. The CO auto-spectrum alone will now be securely detected as we have forecast previously in this work, and achieve $\sigma[\avg{Tb}_\text{CO}]=0.455$, on par with the Y3 Field 1 cross-correlation result.

We also repeat these analyses for the \cite{li_etal_16}--\cite{mmIME-ACA} model that we considered in~\autoref{sec:det_sign}, recalculating central values for all parameters except $b_\text{LAE}$:
\begin{align}
\avg{Tb}_\text{CO}&=3.5\,[\mu\text{K}];\\
\beta_\text{CO}&=0.36;\\
p_\text{shot,CO}&=0.97;\\
v_\text{eff}&=210\,[\text{km\,s}^{-1}];\\
p_{\text{shot,CO}\times\text{LAE}}&=1.06.
\end{align}
We continue to use Gaussian priors for $\beta$ and $v_\text{eff}$ with the same widths, though the central values are changed. The stronger auto-spectrum detection forecast for this model, as forecast previously in~\autoref{sec:det_sign}, means that Y5 results are similar between auto- and cross-spectrum analyses. However, cross-correlation still provides a significant advantage in intermediate stages of data acquisition, as we show in tabulations alongside the fiducial predictions in~\autoref{tab:allthefishers}. We also recall the point raised towards the end of~\autoref{sec:hetdex_maintext} about the advantages of LIM--galaxy cross-correlation against systematics as discussed by other works~\citep[e.g.:][]{Switzer13,Keenan21}.

In all cases, HETDEX observations beyond sparse sampling that fully fill in all COMAP patches would benefit S/N by lowering HETDEX shot noise. The quantitative improvement would depend on relative contribution of HETDEX shot noise versus COMAP thermal noise to the uncertainties, but the improvement predicted by~\cite{hetdex_selfcite} was around 50\%.

We do not detect the cross shot noise in any of the scenarios considered above. However, even if the cross-spectrum yields only an upper limit on the mean CO luminosity of LAEs, this can be combined with voxel-level analyses as proposed by~\cite{silva_etal_21} and should still provide key insights into galaxy and IGM properties at $z\sim3$.
\bibliography{CO_bib,correlate_references,COMAP_ESV_refs,early_science}{}
\bibliographystyle{aasjournal}




\end{document}